\documentclass[trackchanges,twocolumn]{aastex701}
\usepackage{newtxtext,newtxmath}
\usepackage{amssymb}
\usepackage{amsmath}
\usepackage{graphicx}
\usepackage{rotating}
\usepackage{soul}

\begin{document}

\title{Mapping Interstellar Ice Inventory toward Class~0 Protostars in Star-forming Region Orion A with JWST Data}

\author[0000-0002-1584-3620]{Igor Petrashkevich}
\affiliation{Research Laboratory for Astrochemistry, Ural Federal University, Kuibysheva St. 48
Yekaterinburg 620026, Russia}
\email[show]{petrashkevich.igor@gmail.com}  

\author[0000-0002-8150-2795]{Yaroslav Pavlyuchenkov}
\affiliation{Institute of Astronomy of the Russian Academy of Sciences, 8 Pyatnitskaya St.,
Moscow 19017, Russia }
\email{petrashkevich.igor@gmail.com}  

\author[0000-0001-6004-875X]{Anna Punanova}
%\affiliation{Onsala Space Observatory, Chalmers University of Technology, Observatoriev\"gen 90, R\aa\"o, Onsala, Sweden}
\affiliation{Independent researcher}
\email{petrashkevich.igor@gmail.com}  

\author[0000-0002-7308-9056]{Maksim Ozhiganov}
\affiliation{Research Laboratory for Astrochemistry, Ural Federal University, Kuibysheva St. 48
Yekaterinburg 620026, Russia}
\email{petrashkevich.igor@gmail.com}  

\author[0009-0004-5420-8824]{Ruslan Nakibov}
\affiliation{Research Laboratory for Astrochemistry, Ural Federal University, Kuibysheva St. 48
Yekaterinburg 620026, Russia}
\email{petrashkevich.igor@gmail.com}  
\author[0009-0007-0109-3439]{Varvara Karteyeva}
\affiliation{Research Laboratory for Astrochemistry, Ural Federal University, Kuibysheva St. 48
Yekaterinburg 620026, Russia}
\email{petrashkevich.igor@gmail.com}

\author[0009-0003-5504-9346]{Svetlana Salii}
\affiliation{Institute of Natural Sciences and Mathematics,  Ural Federal University, Kuibysheva St. 48
Yekaterinburg 620026, Russia}
\email{petrashkevich.igor@gmail.com}  
%\author[0009-0001-3921-604X]{Mikhail Kiskin}
%\affiliation{Research Laboratory for Astrochemistry, \\ Ural Federal University, Kuibysheva St. 48 \\
%Yekaterinburg 620026, Russia}
\author[0000-0001-7575-5254]{Andrej Sobolev}
\affiliation{Xinjiang Astronomical Observatory, Chinese Academy of Sciences, Urumqi 830011, People’s Republic of China}
\email{petrashkevich.igor@gmail.com}  

\author[0009-0001-3921-604X]{Mikhail Medvedev}
\affiliation{Research Laboratory for Astrochemistry, Ural Federal University, Kuibysheva St. 48
Yekaterinburg 620026, Russia}
\email{petrashkevich.igor@gmail.com}  
\author[0000-0003-1684-3355]{Anton Vasyunin}
\affiliation{Research Laboratory for Astrochemistry, Ural Federal University, Kuibysheva St. 48
Yekaterinburg 620026, Russia}
\email{petrashkevich.igor@gmail.com}  



\begin{abstract}

We present a detailed study of the spatial distribution and chemical composition of interstellar ices toward six Class 0 protostars (HOPS-56, HOPS-60, HOPS-73, HOPS-91, HOPS-96, and HOPS-108) in the Orion A molecular cloud. Using high-resolution spectroscopic data from the JWST NIRspec and MIRI MRS instruments (4.3–8.1 $\mu$m), we have constructed the first pixel by pixel absorption maps with a resolution of $\sim$100~AU for key ice species, including $^{13}$CO$_2$, OCN$^-$, CO, H$_2$O, NH$_4^+$, and H$_2$CO. CH$_4$ and OCS were analyzed toward the continuum peaks. The column densities were derived by fitting the observed spectra with laboratory ice analogs. We employed radiative transfer modeling, which confirmed the reliability of our column density estimates within the protostellar envelopes. Our analysis reveals significant variations in ice abundances and distributions, reflecting the physical structure and energetic processes within the envelopes. Specifically, we observe the influence of protostellar heating and outflows on the ice mantles, most notably in HOPS-60. The total ice composition is consistent with astrochemical models and covers $\sim$90\% of observed ice inventory suggesting that ice is primarily formed during the prestellar stage and subsequently inherited by the protostellar envelope. Based on the abundance relative to water, the sources can be categorized into two distinct groups, possibly indicating evolutionary differences or variations in envelope density and temperature profiles.

%%\footnote{Abstracts for Research Notes of the American Astronomical Society (RNAAS) are limited to 150 words}.
\end{abstract}



\section{Introduction}

Volatiles such as water (H$_2$O), carbon monoxide (CO), methane (CH$_4$) and others are essential for the emergence of life. The volatiles on the planets including Earth were formed in molecular clouds prior to star and planet formation and were stored mainly as ices in pre- and protostellar systems \citep[$>$87\% of volatiles except for H$_2$ are in the ice form, e.g.,][]{Boogert2015,Drozdovskaya2015}. Studying the chemical inventory at different stages of star formation is important for understanding chemical evolution of stars and planetary systems. The main constituents of interstellar ices are simple volatile molecules -- H$_2$O, CO, CO$_2$, NH$_3$, CH$_3$OH and semi-volatile salts such as NH$_4^+$OCN$^-$ \citep[e.g.,][]{Boogert2015}. 

Vibrational modes of the molecules allow to study them observing their absorption of background infrared (IR) emission. One can directly obtain their column densities from the optical depths \citep[e.g.,][]{Gibb2004,McClure2023}. This approach requires a bright IR background source or high sensitivity of observations. Previous observations obtained with ground-based, airborne and space-based facilities \citep[such as UKIRT, IRTF, VLT, KAO, SOFIA, ISO, Spitzer, AKARI, see, e.g.,][]{Storey1981,Chiar1995,Whittet1996,Boogert2008,Yamagishi2011,Goto2021,DeWitt2026} shed light on composition of interstellar ices and provided mostly pointing data (one measurement per source), with only few maps of the most abundant ices, H$_2$O, CO and CO$_2$ \citep[e.g.,][]{Pontoppidan2004,Pontoppidan2006,Sonnentrucker2008}. JWST, with over 100 times better sensitivity and about 10 times better angular resolution allows us to reveal spatial distribution of the ices \citep[e.g.,][]{Smith2025,Tyagi_2025,Gramze2025}, crucial for understanding chemistry and benchmark models of chemistry and dynamics, especially in such complex systems as protostellar cores.

Single protostars in nearby star-forming regions are the best sources for studying chemical and physical processes in the ices. We used the open archive spectral maps from the Mikulski Archive Space Telescopes, MAST,\footnote{MAST database: \url{https://mast.stsci.edu/portal/Mashup/Clients/Mast/Portal.html}} database, obtained via JWST NIRspec and MIRI MRS observations, to map the distribution of the most abundant ices in Class~0 protostars. The Class~0 sources should have the thickest icy mantles; they provide both the closest link to prestellar ices and reveal transformation of ice caused by heating. For our study, we chose the JWST program ID: 5804 that targeted 13 protostars in the massive Orion~A molecular cloud. Orion A (L1641) is a region of active star formation located at a distance of 430~pc \citep[][]{Tobin2020}. We avoided complex multi-component systems and selected six single Class~0 protostars: HOPS-56, HOPS-60, HOPS-73, HOPS-91, HOPS-96, and HOPS-108. These protostars have been actively studied with Spitzer and ALMA (Atacama Large Millimeter Array) observations, characterizing their physical parameters as well as their accretion disks and envelopes \citep[][]{Furlan2016,Tobin2019,Tobin2020,Sheehan2022}. HOPS-60 and HOPS-73 have prominent CO outflows that extend for thousands of astronomical units \citep[][]{Matsushita_2021,Huang2024}. Other selected sources do not have characteristic CO outflow features in the IR spectral range.

To fit the complex blends of the ice absorption features, we used the laboratory spectra of the ice mixtures, from public libraries \citep[e.g.,][]{Rocha2022} and those produced in our laboratory \citep{Ozhiganov_2024}. While the IR transmission from distant background stars is straightforward to interpret since the entire flux falls into the aperture, spatially resolved illumination by an embedded protostar may be affected by scattering, multi-scattering and diffusion effects \citep{Ehrenfreund1997,Dartois2024}. Thus we used a radiative transfer model to test if the common approach to convert flux to optical depths of the absorption features is applicable to study IR spectra of ices toward protostars. 

We estimated the optical depths of CO$_2$, OCN$^-$, CO, OCS, H$_2$O, NH$_4^+$, CH$_4$ and H$_2$CO based on absorption bands in the range from 4.3 to 8 $\mu$m in each pixel of the NIRcam and MIRI MRS spectral cubes. This allowed us to construct maps of the column densities for these species and their abundance maps with respect to water.

Section~\ref{sec:observations} presents NIRspec and MIRI MRS JWST observations and ALMA continuum observations from literature. Section~\ref{sec:processing} describes the reduction of the JWST observation data. Section~\ref{sec:lab_spectra} presents the laboratory comparison spectra of the ice mixtures, selected to fit the observation data. Section~\ref{sec:results} presents the studied absorption bands of interstellar ice, obtaining optical depth of the spectral features, the fitting of the absorption bands, conversion to the column density maps, the obtained column densities and the abundances with respect to H$_2$O, and the radiative transfer model. In Section~\ref{sec:discussion} we discuss the results. Section~\ref{sec:conclusions} presents our conclusions.

\section{Observations}\label{sec:observations}
\subsection{MIRI MRS and NIRspec IFU archive observations}

\begin{figure}
\centering
\includegraphics[scale=0.48]{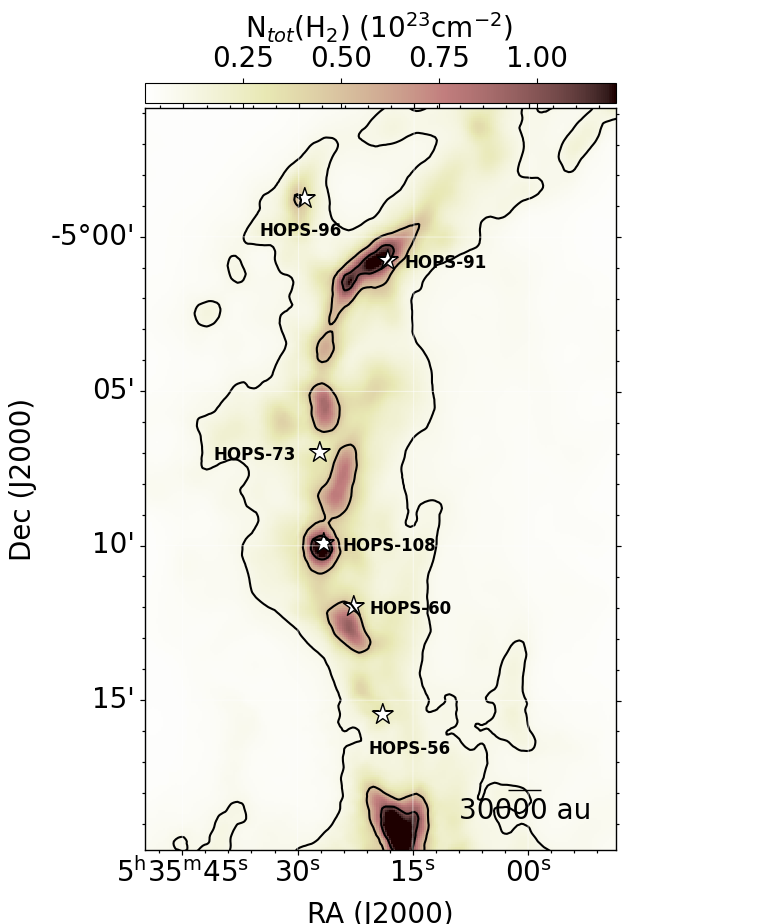}
\caption{Location of observed sources (shown as stars) with the column density of molecular hydrogen toward Orion A \citep[][the Herschel Gould Belt Survey Archive]{Polychroni2013}. The molecular hydrogen column density, $N$(H$_2$), contours start at 1$\times$10$^{22}$~cm$^{-2}$ with a contour step of 5$\times$10$^{22}$~cm$^{-2}$.} 
\label{pic:Orion}
\end{figure}

The JWST NIRspec and MIRI MRS observation data used in this work are part of the observational program that is dedicated to studying protostellar structures in Orion A (PI: Tom Megeath, ID: 5804, Cycle~3). The observations toward six Class~0 protostars HOPS-56, HOPS-60, HOPS-73, HOPS-91, HOPS-96, and HOPS-108 were taken from the MAST %Mikulski Archive Space Telescopes\footnote{MAST database: \url{https://mast.stsci.edu/portal/Mashup/Clients/Mast/Portal.html}} 
database. Table \ref{tab:coord} shows the coordinates of the centers of the observed sources and Fig.~\ref{pic:Orion} shows the location of the sources in the Orion~A filament. NIRspec uses the G395M grating and F290LP filter with the spectral resolution of $\Delta\lambda$/$\lambda$ $\sim$ 1000 and covers the range of 2.87 -- 5.27~$\mu$m. MIRI MRS observations were carried out with readout mode SLOWR1 in three gratings (A, B and C). We selected MIRI MRS channels 1 and 2 with the spectral resolution of $\Delta\lambda$/$\lambda$ = 3750 - 2750 in the range of 4.9--11.7~$\mu$m. NIRspec and MIRI MRS IFU have 2x2 mosaic with a 10\% overlap using the four-point dither mode with spatial resolution of 0.1$^{\prime\prime}$ and 0.13 -- 0.17$^{\prime\prime}$, respectively.

\begin{figure*}
\centering
\includegraphics[scale=0.35]{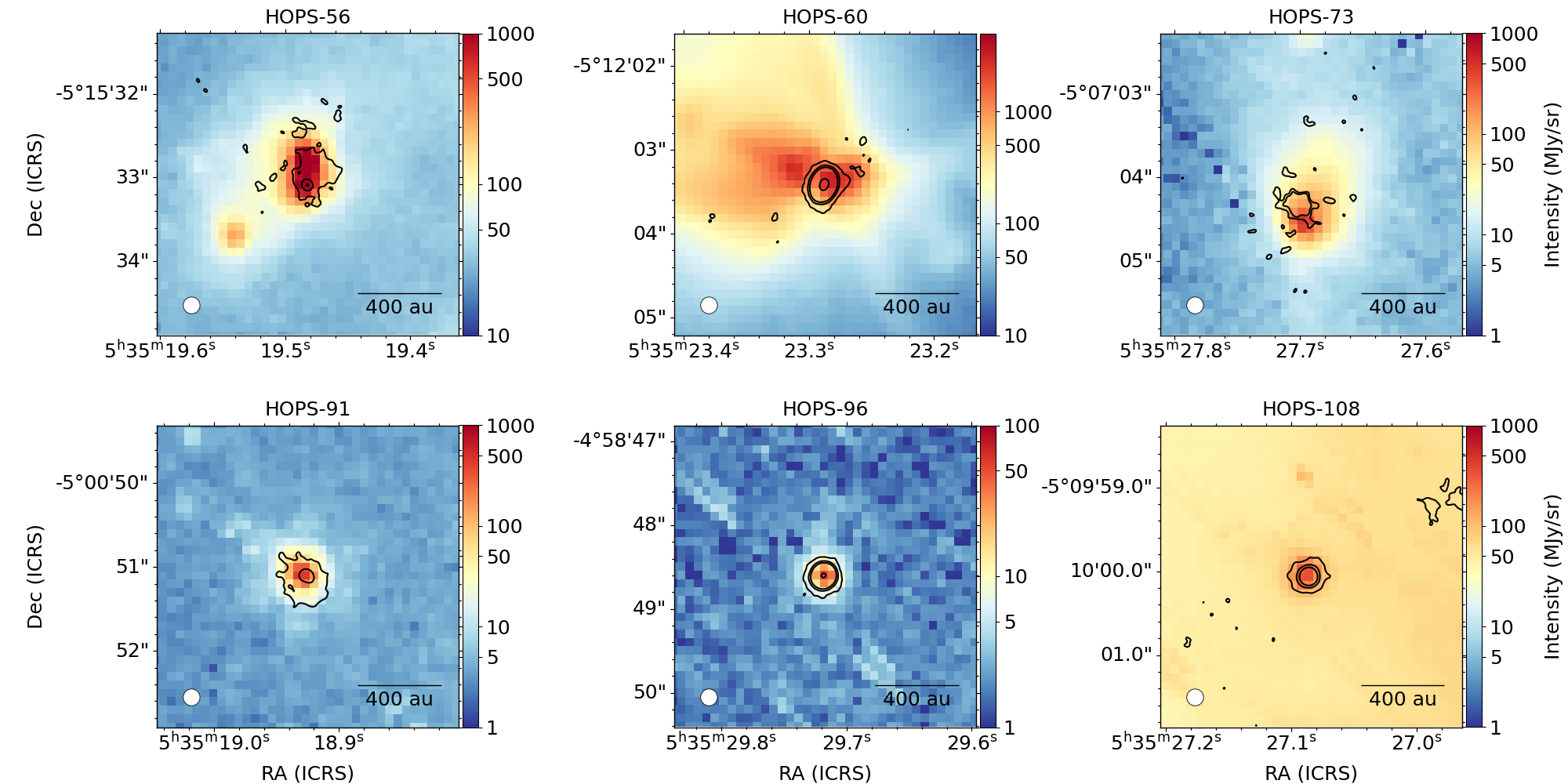}
\caption{NIRspec continuum emission at 4.7 $\mu$m toward HOPS-56, HOPS-60, HOPS-73, HOPS-91, HOPS-96, and HOPS-108. The black contours show the continuum emission at 0.87 mm according to ALMA observations \citep[][]{Tobin2020}. It starts at 1$\times$10$^{-3}$~Jy~beam$^{-1}$ with a step of 10$^{-3+n}$~Jy~beam$^{-1}$ and at 1$\times$10$^{-4}$~Jy~beam$^{-1}$ for HOPS-96.} 
\label{pic:continuum}
\end{figure*}

All observations were processed through three levels of JWST calibration pipeline with the version 1.18.0 and CRDS context jwst\_1364.pmap (CRDS\_VER = `12.1.4') described in \cite{Greenfield2016,Bushouse2024,Gelder2024}. In the final calibration stage for the science-ready data cubes, individually observed background observations were subtracted to account for the telescope background. The science-ready NIRspec and MIRI MRS cubes were additionally processed by the custom astrometry pipeline taking into account ALMA observations to correct the coordinate grid of the maps \citep[geometric distortion of MIRI MRS is described in][]{Patapis2024}. Figure~\ref{pic:continuum} shows the NIRspec continuum emission at 4.7~$\mu$m for all sources and ALMA observations (black contour).

\begin{table*}
\caption{Coordinates of protostars toward accretion disk centers and spectrum energy distribution (SED) data fitted to Spitzer and Herschel observations. Coordinates established with ALMA observations from \cite{Tobin2020}. SED fit results were taken from \cite{Furlan2016}.}\label{tab:coord}

\begin{tabular}{ l  c   c   c  c  c  c  c  c  c  c  c }\hline
Core & $\alpha$(J2000) &$\delta$(J2000) & CO gas line & $L$ & $M_{\rm env}$ & $\theta$ & $A_V$(a)& $A_V$(b)& $T_{\rm bol}$& $R_{\rm disk}$ & $M_{\rm disk}$\\ 
   & (hh:mm:ss)  &(dd:mm:ss)   &   & ($L_\odot$) & ($M_\odot$) &  ($^{\circ}$) & & & (K) & (AU) & $M_{\oplus}$\\
\hline
HOPS-56 & 05:35:19.482  & -05:15:33.08& + & 20.4 & 0.8 & 50 & 52.7& 137 & 48.1 & 30 & 5.4 \\ 
HOPS-60 & 05:35:23.287  &-05:12:03.41& + & 19.2 & 0.1 & 
83 & 8.3& 132 & 54.1 & 62.7& 152.7\\ 
HOPS-73 & 05:35:27.699  &-05:07:04.34& + &1.5 & 0.1 & 
70 & - & -& 43.0 & 63 & 9.8\\ 
HOPS-91 & 05:35:18.925  &-05:00:51.11& - &4.1 & 1.8 & 
57 & 68.8 & 88& 41.7 & 86.7 & 22.9  \\ 
HOPS-96 & 05:35:29.717 &-04:58:48.61& - & 5.4 & 1.8  &
57 & 88.8& 175 & 35.6 & 41.4 & 135.0\\ 
HOPS-108 &  05:35:27.084  &-05:10:00.06& - & 33.5 & 0.7 &  70 & 19 & 134 & 38.2 & 40.4 & 24.4 \\ 

\hline
\end{tabular}
      \small
      \begin{flushleft}
      \item \textbf{Note:} (a) -- visual extinction ($A_V$) from \citet{Furlan2016}, (b) -- visual extinction ($A_V$) from water fitting in Section \ref{col_H2}.
      \end{flushleft}

\end{table*}

\subsection{Observations from literature}
We used ALMA dust continuum observations toward HOPS-56, HOPS-60, HOPS-73, HOPS-91, HOPS-96, and HOPS-108 from VANDAM project \citep[resolving the accretion disk parameters with a spatial resolution of 0.107$^{\prime\prime}$][]{Tobin2020} for astrometrical calibration. We adopted the position of the protostars and disks from \citet{Tobin2020} to interpret our results on the ice composition. We used the maps of the molecular hydrogen column density, $N_{\rm tot}({\rm H_2})$, and dust temperature from \citet{Polychroni2013}\footnote{Herschel Archive: \url{http://www.herschel.fr/cea/gouldbelt/en/index.php}}, with the beam size of 36.3$^{\prime\prime}$ and the pixel size of 3$^{\prime\prime}$, to compare our $N$(H$_2$) obtained via our estimate of the water ice column density in Section \ref{col_H2}.

\section{Processing of the observational spectra}\label{sec:processing}
The analysis of absorption bands in the JWST NIRspec and MIRI MRS IFU observational data is complicated by variable spatial resolution, gas and PAHs emission. We have implemented a pipeline to ensure consistent spectral processing for each pixel in the observational data:
\begin{enumerate}
\item NIRspec: 
      \begin{itemize} 
      \item Detection and filtering of CO gas lines;
      \end{itemize}
 MIRI MRS:
      \begin{itemize} 
      \item spatial regridding to a single spatial resolution; % Resampling the spectrum to a single spatial resolution (only for MIRI MRS);
      \item spectrum interpolation in areas of channel overlap;
      \item subtraction of PAH emission using template spectra from \citet{Chown2024};
      \end{itemize}
\item Continuum fitting, based on the ice absorption-free windows listed in Table \ref{Tab:window};
\item Conversion of the spectrum from intensity to optical depth using the fitted continuum.
\end{enumerate}

These steps are described in detail below.

\subsection{Initial processing}\label{sec:initial_processing}
We chose the 4.3--8.1~$\mu$m wavelength range for the study owing to the following limitations. Class~0 protostars have low emission in the near-IR range ($<4$$\mu$m), which makes it difficult to study the spectra in each pixel. We set the lower limit of the NIRspec range to 4.3 $\mu$m where the emission allows for the study of optically thin bands. MIRI MRS has four channels, each having SHORT, MEDIUM, and LONG bands, covering 4.9--27.9~$\mu$m. The spatial resolution at $\lambda$ $\geq$ 12 $\mu$m (FWHM $\geq$ 0.36$^{\prime\prime}$ and linear resolution $\geq$ 155~AU at 430~pc) does not allow us to resolve protostellar structures in Orion A (disk and inner envelope near disk, see Table \ref{tab:coord}). The field of view of channel 1 of MIRI MRS (the range is 4.9 - 7.4 $\mu$m, frame field of view is 3.2$^{\prime\prime}$ × 3.7$^{\prime\prime}$ and pixel size is 0.13$^{\prime\prime}$) almost completely falls within a few pixels of channel 4 (the range is 17.7 - 27.9 $\mu$m, frame field of view is 6.6$^{\prime\prime}$ × 7.7$^{\prime\prime}$ and pixel size is 0.35$^{\prime\prime}$) and the effect of FWHM overlap in each pixel of channels 3 and 4 becomes much more noticeable (FWHM = 0.314$^{\prime\prime}$$\times\lambda/10~\mu m$). Because of that, we chose only channels 1 and 2 of the MIRI MRS for our study. Silicates absorption ($>$8.1~$\mu$m) and the detector heating limit our options to using only the SHORT range of channel 2 (7.4 - 8.1~$\mu$m).

\subsubsection{NIRspec}\label{NIR_ob}
The NIRspec observations are spectral cubes with the field of view of the mosaic of 5.7$^{\prime\prime}$ x 5.7$^{\prime\prime}$ and the spectral range from 2.8 to 5.0~$\mu$m. HOPS-56 and HOPS-60 have strong emission lines of gaseous CO and HOPS-96 has noticeable absorption lines of gaseous CO at 4.5--5.0~$\mu$m overlapped with the ice features of interest (OCN$^-$, CO and OCS absorption bands). We used percentile, gray erosion and dilation filters to filter the spectra \citep[see, e.g.,][]{Soille2003} in each pixel where CO gas emission was present. As a result, the spectral resolution in these pixels decreased by a factor of 1.5 (by percentile filters), and this did not affect the ice absorption feature profiles. This cleaning allowed us to correctly set continuums which is crucial to explore the absorption features. Figure~\ref{pic:spec_filter} (in Appendix~\ref{Append}) shows the spectrum before and after filtering in a pixel toward the center of HOPS-60. Pixels without CO gas lines were not filtered.

\begin{table*}
\caption{Absorption band peak position and their band strengths.}\label{tab:bands}
\begin{tabular}{ l  c  c  c  c  c }\hline
Band & Peak & Peak& $A'$ & Vibration Mode & Reference  \\ 
     &    (cm$^{-1}$)  &     ($\mu$m)   &  (cm~molecule$^{-1}$)  &    \\
\hline
H$_2$O &  3257 & 3.070  &  2.2$\times$10$^{-16}$  & O--H stretch & \cite{Bouilloud2015}\\
$^{13}$CO$_2$ &  2283&   4.380 &  1.2$\times$10$^{-16}$ & $^{13}$C--O stretch & \cite{Bouilloud2015}\\
OCN$^-$ & 2170   &  4.608  & 1.5$\times$10$^{-16}$&  C--N stretch  & \cite{Gerakines2025}\\
CO & 2138 & 4.676 & 1.0$\times$10$^{-17}$ & $^{12}$C--O stretch  & \cite{Gerakines2025}\\
OCS &  2031  & 4.924  & 1.2$\times$10$^{-16}$& C--N stretch  & \cite{Yarnall_2022}\\
H$_2$CO&  1723 & 5.804 &  9.6$\times$10$^{-18}$& C=O stretch  & \cite{Bouilloud2015}\\
H$_2$O &  1657 & 6.035  &  1.0$\times$10$^{-17}$& H--O--H bend  & \cite{Bouilloud2015}\\
CH$_3$OH& 1459 & 6.854 &   7.7$\times$10$^{-18}$&  C--H deformation & \cite{Hudson_2024}\\
NH$_4^+$ &  1436 & 6.964  &  3.6$\times$10$^{-17}$ & NH$_4^+$ asym. bend & \cite{GERAKINES2024116007}\\
CH$_4$:H$_2$O & 1302   &  7.681  & 1.0$\times$10$^{-17}$ & C--H deformation & \citet{Karteyeva2026_2}\\
CH$_4$:CO$_2$ & 1301   &  7.686  & 1.2$\times$10$^{-17}$& C--H deformation  & \citet{Karteyeva2026_2}\\
CH$_4$:CH$_3$OH & 1300    &  7.692 & 1.1$\times$10$^{-17}$& C--H deformation  & \citet{Karteyeva2026_2}\\

\hline
\end{tabular}
\end{table*}

\subsubsection{MIRI MRS}\label{PAH}
MIRI MRS observations have the same data structure but differ from NIRspec by the mosaic field of view (6.1$^{\prime\prime}$ × 7.0$^{\prime\prime}$ of channel~1 and 7.6$^{\prime\prime}$ × 9.1$^{\prime\prime}$ of channel~2 compared to 5.7$^{\prime\prime}$ × 5.7$^{\prime\prime}$ of NIRspec) and overlapping spectrum ranges of channel bands: 1-SHORT (4.90–5.74~$\mu$m, pixel size of 0.13$^{\prime\prime}$), 1-MEDIUM (5.66–6.63~$\mu$m, 0.13$^{\prime\prime}$), 1-LONG (6.53–7.65~$\mu$m, 0.13$^{\prime\prime}$), 2-SHORT (7.51–8.77~$\mu$m, 0.17$^{\prime\prime}$). We regridded the spectral cubes of channel 1 to the spatial resolution and field of view of channel 2 at 8.3~$\mu$m with FWHM of 0.3$^{\prime\prime}$ and a pixel size of 0.17$^{\prime\prime}$. 
We then applied linear interpolation where the channels overlapped (by 0.10--0.15~$\mu$m) to merge the spectra with each other. The intensity change was less than 1\%. Through this procedure, we produced a single spectral cube with a range of 4.90-8.77~$\mu$m. %Interpolation does not introduce uncertainties or distort the data when intensity is converted to optical depth. % 3.2x3.7, 4.0x4.8, 3.0x3.0 frame sizes of MIRI MRS ch~1 and ch~2 and NIRspec

\begin{figure}
\centering
\includegraphics[scale=0.42]{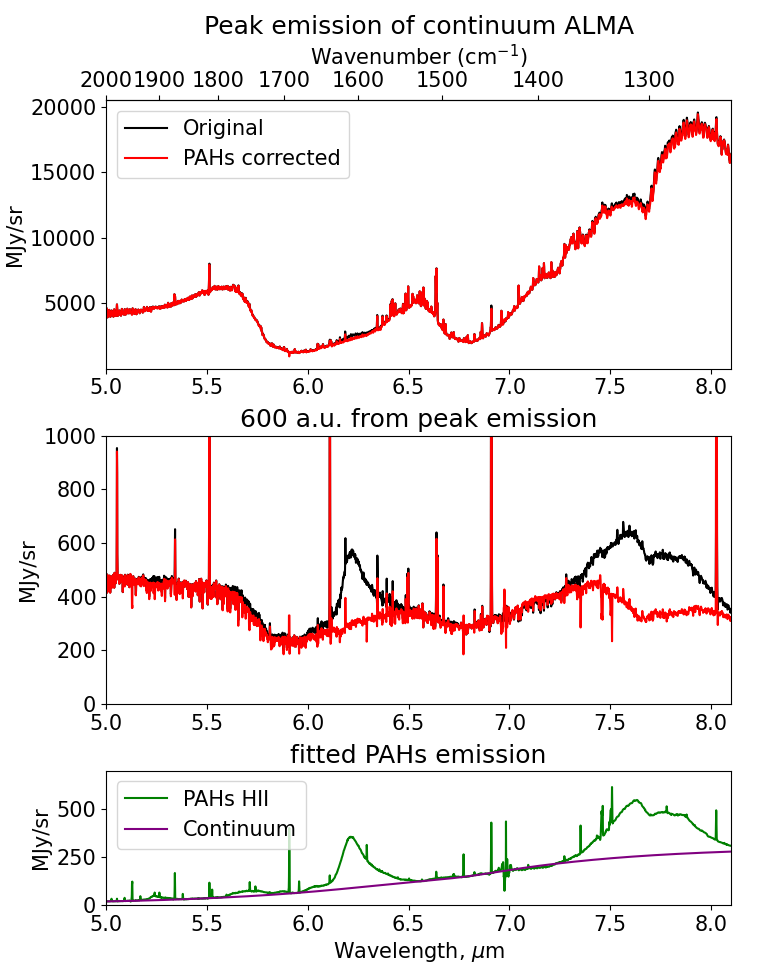}
\caption{PAHs correction for HOPS-60. Black lines show observed spectra, red lines show corrected PAHs spectra, green line shows PAHs HII spectrum from \citet{Chown2024}. {\it Top panel:} pixel at the peak of ALMA continuum emission. {\it Middle panel:} pixel 600 AU away from the peak of ALMA continuum emission. {\it Bottom panel:} HII PAHs emission from \citet{Chown2024}.} 
\label{pic:PAHs}
\end{figure}

We performed an additional correction for polycyclic aromatic hydrocarbons (PAHs) emission bands in each pixel of MIRI MRS in the 4.90–8.77 $\mu$m range using the PAH spectra obtained for Orion in \citet{Chown2024}. The HOPS-56, HOPS-60, and HOPS-108 cubes show stable, uniform PAHs emission, regardless of pixel location. We assume that this emission is located in the background or in the foreground, since it does not depend on the position in the mosaic. We subtracted the PAH emission to facilitate continuum subtraction (see Sect.~\ref{sec:cont_sub} for details) as follows. We sampled the pixels located away from the sources where ice absorption was negligible ($\sim$20 pixels) for each of the sources, except for HOPS-91 and HOPS-96, and used them to choose the most suitable of the observed PAHs spectra. HOPS-91 and HOPS-96 have PAH emissions at the noise level, since they are located farther (more than 15 arcmin in the plain of sky) than other sources from a young OI type star, NU Ori, a local source of strong UV radiation in Orion~A. The best match for the two PAHs peaks at 6.2 and 7.6 $\mu$m are the PAH spectra HII (corresponding to Orion HII region) and DF3 (corresponding to edge-on dissociation front) from \citet{Chown2024}. We adjusted the intensity of the PAHs HII spectrum for our maps and used one spectrum for each pixel within a source (see Fig. \ref{pic:PAHs}). The methane feature at 7.7 $\mu$m is corrupted for analysis due to the complex structure of PAHs emission at 7.6 $\mu$m, thus we focus only on analyzing the methane spectrum in the pixel toward the peak of ALMA continuum emission where it is not contaminated by PAHs.

\subsection{Continuum fitting}\label{sec:cont_sub}
The observed spectra are the result of emission from the protostar passing through the protostellar envelope and the disk. It is necessary to isolate ice absorption from continuum emission. We assume the continuum to be the sum of the background radiation and the protostar blackbody emission, with the protostars located in the centers of the protostellar cores. We used the \textsc{Python.specutils} and \textsc{Python.spectral\_cube} packages to further process the spectra in each pixel. The local continuum was established separately for each pixel using the following unified scheme. The pixel-wise local continuum was selected as a third-degree polynomial based on anchor points (see Fig. \ref{pic:spec_filter} in Appendix~\ref{Append}) from five spectral windows between the absorption features for NIRspec and in two spectral windows for MIRI MRS (see Table \ref{Tab:window} in Appendix~\ref{Append}), fitted separately for the NIRspec and MIRI MRS data cubes. We established uniform local continuums for the absorption bands in NIRspec and MIRI MRS in order to obtain spectra in each pixel with the same method. The local continuum fitted in each pixel differs from the global continuum fitted across the entire range from 5 to 28~$\mu$m for the aperture-extracted spectra \citep[as was done, e.g., in ][]{Chen2024} and may lead to an underestimation of the optical depth. The column density uncertainty due to the continuum selection was estimated  using a Monte Carlo method. The positions of left and right spectral window edges were varied by 2--4\% for 30,000 times and the third-degree polynomial continuum was fitted each time. The additional uncertainty of the column density due to our automated continuum selection is 5-7\% for CO$_2$, OCN$^-$, CO and OCS and 7-10\% for H$_2$O, NH$_4^+$, CH$_4$, H$_2$CO, CH$_3$OH.

\subsection{Conversion to optical depth}\label{sec:opt_dep}
\begin{figure*}
\centering
\includegraphics[scale=0.9]{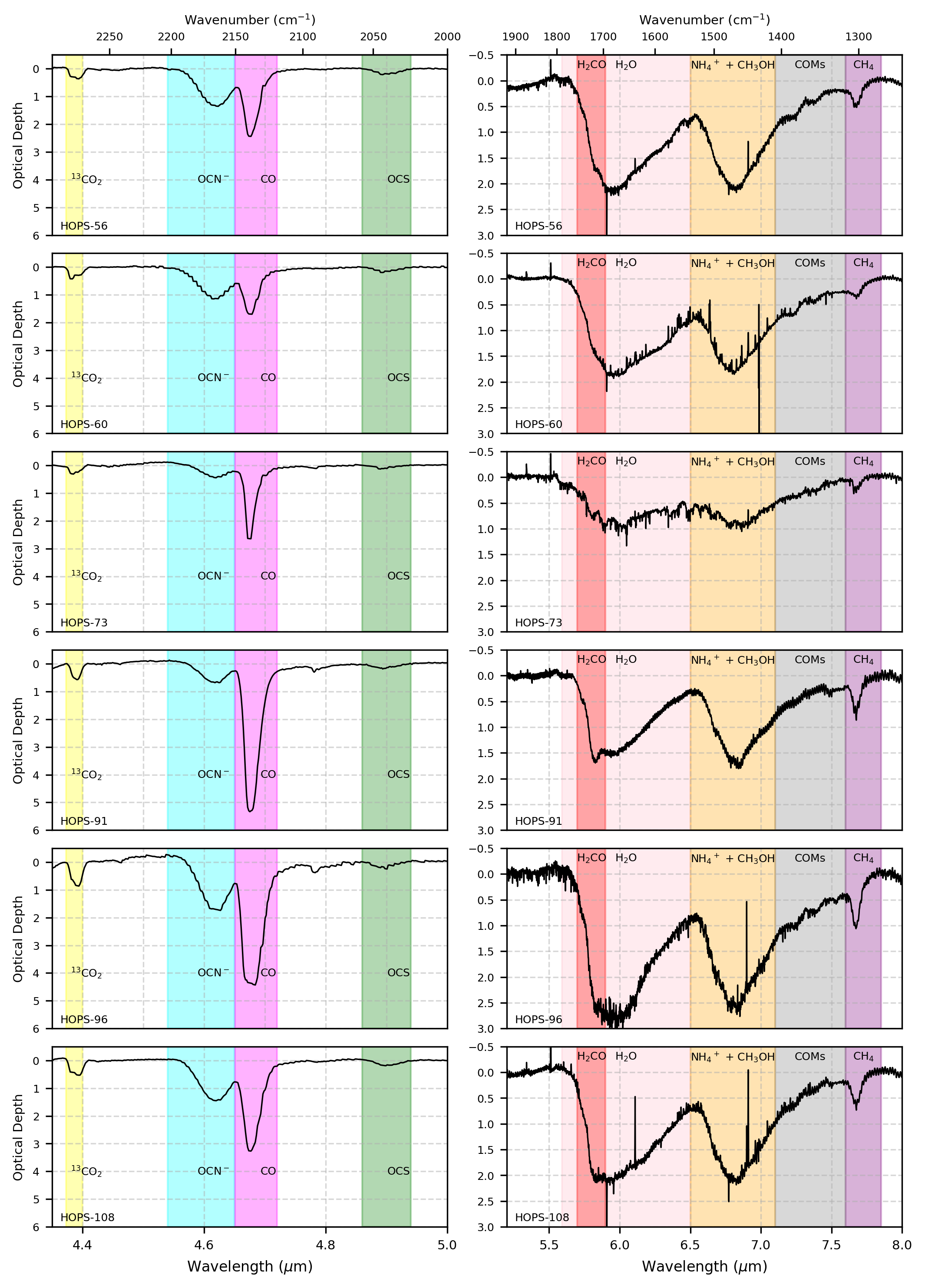}
\caption{Optical depth spectra toward the ALMA continuum intensity peak of the sources. Left column shows NIRspec spectra and right column shows MIRI MRS spectra. The colors indicate the absorption bands of various molecules.} 
\label{pic:spec_opt}
\end{figure*}

We use the continuum (see Section~\ref{sec:cont_sub}) to convert the intensity to the optical depth \citep{Tyagi_2025} separately in the ranges of 4.3--4.9~$\mu$m (NIRspec) and 5.0--8.1~$\mu$m (MIRI MRS):
\begin{equation}
\tau_{\lambda} = -\text{ln}\left(\frac{I_{\lambda,\rm obs}}{I_{\lambda,\rm cont}}\right),
\end{equation}
where $\tau_{\lambda}$ is optical depth, $I_{\rm\lambda,obs}$ is the observed intensity and $I_{\rm\lambda,cont}$ is fitted continuum intensity. We note that the MIRI MRS spectra have different continuums toward the protostar and in the background  emission. Figure \ref{pic:spec_opt} shows the optical depth spectra from NIRspec (left column) and MIRI MRS (right column) toward the ALMA continuum intensity peaks. Figure \ref{pic:spec_filter} (in Appendix~\ref{Append}) shows the continuums in NIRspec and MIRI MRS toward the peak intensity of the ALMA continuum.

We selected the most intense identified absorption bands of most abundant species in the studied range \citep[4.3-8.0 $\mu$m, ][]{Oberg_2011, Boogert2015, McClure2023}: $^{13}$CO$_2$, OCN$^-$, CO, OCS, H$_2$O, NH$_4^+$, CH$_4$, H$_2$CO and CH$_3$OH. Table \ref{tab:bands} shows the positions of the absorption band peaks and band strengths in this range; Fig.~\ref{pic:spec_opt} shows the bands in the spectrum. 

Absorption bands of $^{13}$CO$_2$ and OCS have distinct profiles and are not overlapped with other absorption bands. The OCN$^-$ absorption bands at 4.4~$\mu$m and 7.6~$\mu$m partially overlap with the CO and CH$_4$ absorption bands, respectively. We excluded NIRspec pixels with SNR of optical depth spectra $<3$ for the $^{13}$CO$_2$ and OCS absorption feature peaks and MIRI MRS pixels with SNR $<3$ for the NH$_4^+$ absorption feature peak. SNR for other absorption bands is much higher. For further study, we use the central square with sides of $4.6^{\prime\prime}\times4.6^{\prime\prime}$ that covers $\sim2000\times2000$~AU at 430~pc. Thus, for each source, we obtained two optical depth spectral maps ready for fitting in the ranges of 4.3--4.9~$\mu$m (NIRspec) and 5.0--8.1~$\mu$m (MIRI MRS) covering up to 1000 AU around the ALMA continuum intensity peak. We distinguished the CO from the OCN$^-$ band at 4.4~$\mu$m with a Gaussian profile and the bands at 7.6~$\mu$m with comparison spectra. The remaining absorption bands (H$_2$O, NH$_4^+$, H$_2$CO and CH$_3$OH) in the range from 5.0 to 7.4 $\mu$m have complex profiles, they overlap with each other and other absorption bands in this range. Therefore, to study these bands, we use the laboratory spectra of pure and mixed ices. We excluded NH$_3$ at 6.3~$\mu$m \citep[][]{McClure2023} from consideration because this absorption band is contaminated by overlapping PAHs emission. $^{13}$CO at 4.6~$\mu$m and COMs (e.g. HCOOH, CH$_3$CHO and HOCH$_2$CHO at 5.5-7.5~$\mu$m) have a low SNR of 1.0-1.5 and also cannot be analyzed in individual pixels.

\section{Laboratory ice comparison spectra}\label{sec:lab_spectra}

We used reference ice spectra obtained in our laboratory with the Ice Spectroscopy Experimental Aggregate \citep[ISEAge, ][]{Ozhiganov_2024}, publicly available spectra from the Leiden Ice Database for Astrochemistry \citep{Rocha2022} and The Cosmic Ice Laboratory\footnote{The Cosmic Ice Laboratory: \url{https://science.gsfc.nasa.gov/691/cosmicice/spectra.html}} databases to fit the observations (the details on the spectra are given in Table~\ref{tab:mix}). 
Spectrum H$_2$O:CO:CO$_2$ was selected to study the effect of the mixture on the water absorption band. We used the laboratory spectra of CH$_3$OH (at 10~K) and NH$_4^+$OCN$^-$ (at 10~K and 120~K) obtained with the ISEAge in transmission mode, since these spectra with high resolution were not available in any public database. The new spectra are available at Zenodo\footnote{Zenodo: \url{https://doi.org/10.5281/zenodo.20346324}}. 

The ISEAge is an ultra-high vacuum experimental setup that allows for growing of the interstellar ice analogues and their study via the infrared transmission spectroscopy. The ices are obtained by depositing gases or gaseous mixtures through two separate all-metal dosing lines onto a cooled substrate. The substrate temperature can be held within a 6.7--305 K. The base pressure in main chamber that contains the substrate is maintained at 2\texttimes10$^{-10}$ mbar outside of the experiments. The ratio and flux of the deposited components are controlled via the calibration curves that link the ion current on a quadrupole mass spectrometer (Stanford Research Systems RGA200) located in a close vicinity of the substrate and the column density of the deposited ice. The calibration curves are obtained individually for each pure specie used. Components of a binary mixture are introduced through separate leak valves. In case of a three-component mixture two components are introduced through a single leak valve as a gaseous mixture with a pre-calibrated ratio. The IR spectra are obtained with Thermo Scientific Nicolet iS50 Fourier Transform Infrared spectrometer in the range  between 4000 and 630~cm$^{-1}$ (2.5 and 15.9 $\mu$m) with 1~cm$^{-1}$ resolution. The spectra are continuously collected during the experiments every 45 seconds with an averaging of 32 scans. At the end of each deposition a single spectrum is obtained with an averaging of 128 scans.

\begin{figure}
\centering
\includegraphics[scale=0.53]{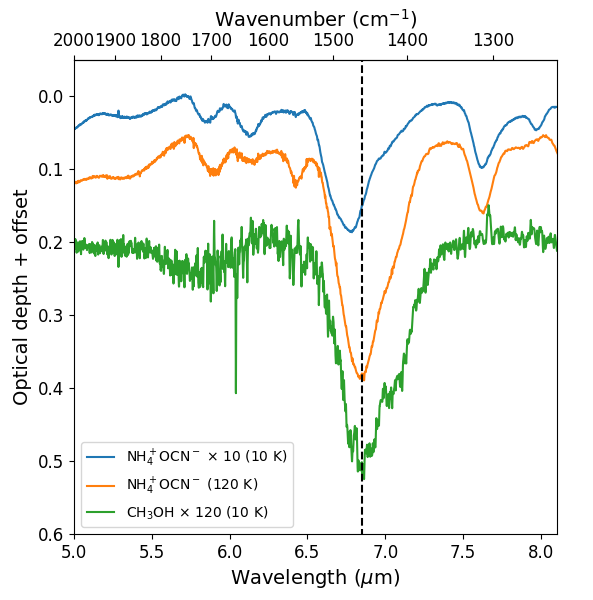}
\caption{Comparison spectra of NH$_4^+$OCN$^-$ at 10~K (blue) and at 120~K (orange) and CH$_3$OH at 10~K (green). Black line shows the peak of 1460 cm$^{-1}$. } 
\label{pic:mix_all}
\end{figure}

The pure CH$_3$OH was deposited at a rate of 1.43$\times$10$^{13}$ molecules cm$^{-2}$ s$^{-1}$ for 60 minutes onto a germanium substrate cooled to 10~K, resulting in a total column density of 5.15$\times$10$^{16}$ cm$^{-2}$. Following the deposition, the IR spectrum of the CH$_3$OH ice was recorded with 128 scans. The NH$_4^+$OCN$^-$ spectra obtained earlier in \citet{Novozamsky2001} did not resolve the NH$_4^+$OCN$^-$ absorption peak {at 6.8~$\mu$m}. The spectra of the NH$_4^+$OCN$^-$ salt (the product of the NH$_3$ + HNCO = NH$_4^+$OCN$^-$ reaction) were obtained by co-depositing ammonia (NH$_3$) and HNCO onto a 10 K substrate. Deposition rates were 5.50$\times$10$^{13}$ molecules cm$^{-2}$~s$^{-1}$ for NH$_3$ and 2.37$\times$10$^{13}$ molecules cm$^{-2}$~s$^{-1}$ for HNCO, supported for 30 minutes. The resulting ice contained NH$_3$, HNCO, and NH$_4^+$OCN$^-$ in a molecular ratio of 6:1:3, determined from the NH$_3$ umbrella band \citep[with the band strength of \textit{A'}=1.95$\times$10$^{-17}$~cm, ][]{hudson2022ammonia}, the HNCO band \citep[\textit{A'}=1.29$\times$10$^{-16}$ cm, ][]{hudson2024infrared}, and the OCN$^-$ band \citep[\textit{A'}=1.5$\times$10$^{-16}$ cm, ][]{Gerakines2025}. Subsequently, we heated the ice at a rate of 5 K min$^{-1}$ to 200~K while continuously collecting IR spectra with 32 scans per spectrum, which corresponds to an averaging of 45~seconds. 
Figure~\ref{pic:mix_all} shows the laboratory spectra obtained with ISEAge, used in the range of 5.0 to 7.4~$\mu$m. The black dashed line shows the peak at 6.85~$\mu$m (1460~cm$^{-1}$), which matches the NH$_{4}^{+}$OCN$^{-}$ salt at 120~K and the CH$_3$OH ice at 10~K. The spectrum of NH$_{4}^{+}$OCN$^{-}$ at $\sim$120~K was appropriate for fitting the observed 6.85~$\mu$m band. The temperature of salt ice does not reflect the possible temperature of interstellar ice, but rather indicates the crystalline structure of salt. Therefore, the temperature of the salt differs from that of the selected mixtures, whose absorption bands depend directly on temperature. In this work we treat the NH$_4^+$ and OCN$^-$ absorption bands separately to estimate the contributions of all salts containing NH$_4^+$ and OCN$^-$.

The compounds used for this study are CH$_3$OH ($\ge$~99.8~\%, Vekton), NH$_3$ (NPK Selivanenko, $>$99.8~\%), cyanuric acid (C$_3$H$_3$N$_3$O$_3$, Sigma Aldrich, 98~\%). Gaseous HNCO was obtained via thermal decomposition of cyanuric acid, following the procedure described in \cite{raunier2003reactivity}.

\begin{table}
\begin{center}
\caption{Spectra of the laboratory ices used to fit the observed spectra.}\label{tab:mix}
\begin{tabular}{ l  c  c  c  }\hline
Mixtures & Ratio & $T$ & Reference \\ 
 &  & (K) &  \\ 
\hline
CH$_3$OH& -- &   10  & This work\\
NH$_4^+$OCN$^-$ &  --  &  10, 120  & This work\\
H$_2$O &  --  &  15  & \citet{Oberg2007}\\
H$_2$O:CO$_2$:CO &  100:20:23  & 20 &  \citet{Ehrenfreund1997}\\
H$_2$O:H$_2$CO &  8:1  &  15  & \citet{Moore1998}\\
CH$_4$:CO$_2$ &  1:5  &  10  & \citet{Karteyeva2026_2}\\
CH$_4$:H$_2$O &  1:10  &  10  & \citet{Karteyeva2026_2}\\
CH$_4$:CH$_3$OH &  1:3  &  10  & \citet{Karteyeva2026_2}\\
OCN$^-$ &  --  &  12  & \citet{Gerakines2025}\\

\hline
\end{tabular}
\end{center}
\end{table}

\section{Results}\label{sec:results}
\subsection{Fitting optical depth spectra}

\subsubsection{4.3 -- 4.9 $\mu$m region of NIRspec: $^{13}$CO$_2$, OCN$^-$, CO, OCS}\label{sec:tau_43}
\begin{figure*}
\centering
\includegraphics[scale=0.85]{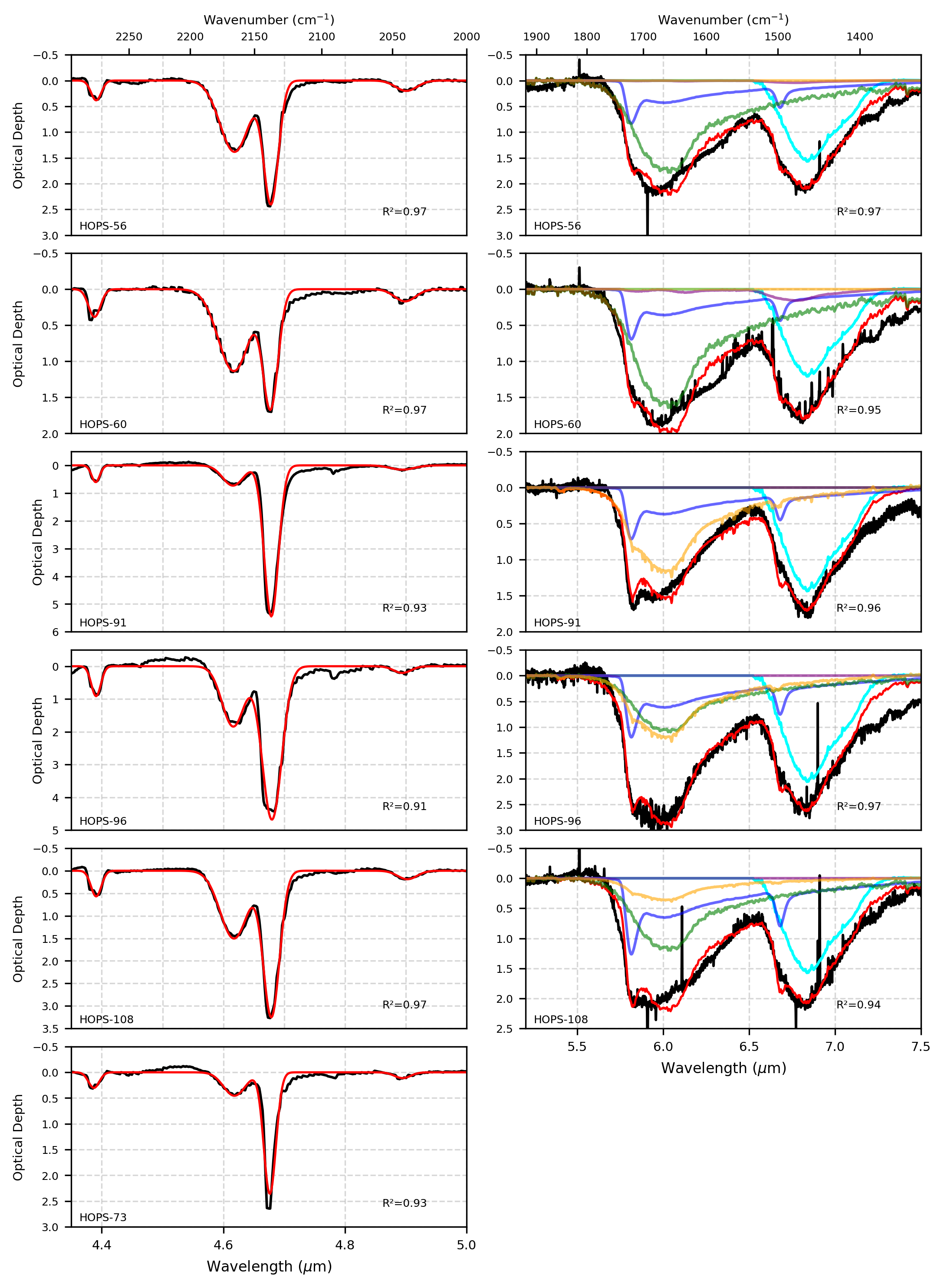}
\caption{Optical depth spectra toward the ALMA continuum intensity peak of the sources. The red line shows the best fit for each spectrum, the black line shows the optical depth spectrum. The R$^2$ of the fits is shown in the lower right corners. {\it Left panels:} Gaussian fits of $^{13}$CO$_2$, OCN$^-$, CO and OCS.
{\it Right panels:} Color lines show the components of the laboratory spectra of H$_2$CO:H$_2$O (blue), H$_2$O (green), NH$_4^+$ (15~K, purple), H$_2$O:CO:CO$_2$ (orange) and NH$_4^+$(120~K, cyan).} 
\label{pic:spec_fit}
\end{figure*}

The absorption bands of OCN$^-$, CO, and OCS have symmetrical absorption profiles in this range. In contrast, $^{13}$CO$_2$ forms a double peak that contains blended peaks of the pure ice and $^{13}$CO$_2$ mixed mostly with H$_2$O (see left panels of Fig.~\ref{pic:spec_opt}), also observed in other Class~0 sources \citep[e.g.,][]{Brunken2024,Tyagi_2025}. The low resolution of the spectra does not allow for detailed study of the structure of each absorption band, thus we use Gaussian profiles for the fits. We apply four Gaussian profiles $G(\lambda)$ for $^{13}$CO$_2$, OCN$^-$, CO and OCS to simultaneously fit all NIRSpec species in each pixel:
\begin{equation}
G(\lambda) = \tau_{\rm peak}\exp\left(-\frac{(\lambda -\lambda_0)^2}{2\sigma^2}\right) ,
\end{equation}
where $\tau_{\rm peak}$ is the peak of fitted optical depth, $\lambda_0$ is the peak intensity wavelength (cm$^{-1}$), $\sigma$ is line width (FWHM/2$\times$(2$\times$ln(2))$^{1/2}$, cm$^{-1}$). We used linear gradient fitting to obtain the parameters of each absorption band in each pixel:  $\tau_{\rm peak}$, $\lambda_0$ and $\sigma$. The best fit was obtained by minimizing $\chi^2$:
\begin{equation}
\chi^2(\lambda) = \sum^{N_{\rm points}}_{i=1}{{(\tau_{i, {\rm obs}}(\lambda)-G_{i, {\rm model}}(\lambda))^2}\over{\sigma_{i, {\rm obs}}^2}},
%R^2 = 1 - \frac{ \sum_{i=1}^{N_{\rm point}} w_i (\tau_{\rm obs}^i - G^i_{\rm model})^2 }
%{ \sum_{i=1}^{N_{\rm point}} w_i (\tau_{\rm obs}^i - \bar{\tau}_{\rm obs})^2 },
\end{equation}
where $N_{\rm point}$ is the number of analyzed data points, $\tau_{\rm obs}$ is observed optical depth and $\sigma_{\rm obs}$ is standard deviation of the observational spectrum, $G$$_{\rm model}$($\lambda$) = $G$$_{\rm CO}$($\lambda$)+ $G$$_{\rm ^{13}CO_2}$($\lambda$) + $G$$_{\rm OCN^-}$($\lambda$)+ $G$$_{\rm OCS}$($\lambda$). The goodness of the fit after processing was estimated using the standard coefficient of determination ($R^2$), weighted by the average noise value of the spectrum. 
%\begin{equation}
%R^2 = 1 - \frac{ \sum_{i=1}^{N_{\rm point}} w_i (\tau_{\rm obs}^i - G^i_{\rm model})^2 }
%{ \sum_{i=1}^{N_{\rm point}} w_i (\tau_{\rm obs}^i - \bar{\tau}_{\rm obs})^2 },
%\end{equation}
%where $w_i$ is noise weight (1/$\sigma_{\rm obs}^2$) and $\bar{\tau}_{\rm obs}$ is weighted median optical depth. 
We used the covariance matrix returned by the fitting routine \citep[curve fit and {\sc SciPy},][]{Virtanen2020} and propagated the uncertainties of the parameters as linear combinations of individual spectral components to derive the fitted optical depth and its uncertainty. Figure \ref{pic:spec_fit} (left panels) shows the best fit in pixels toward the ALMA continuum emission peak. We separated the OCN$^-$ and CO absorption bands for separate analysis as was done in \cite{Tyagi_2025}.

Absorption feature of $^{13}$CO$_2$ has a complex profile that may include the bands of pure $^{13}$CO$_2$, $^{13}$CO$_2$:CO, $^{13}$CO$_2$:CH$_3$OH, $^{13}$CO$_2$:H$_2$O \citep[e.g.,][]{Ehrenfreund1999,Brunken2024}. Fitting several components gives too ambiguous fit (with an uncertainty $>50$\%); the $^{13}$CO$_2$ single-component fit has an uncertainty that is 2-5\% larger than those of the fits of OCN$^-$, CO and OCS. We note that in most pixels toward HOPS-56, HOPS-91, HOPS-96 and HOPS-108, the optical depth peak of pure $^{13}$CO$_2$ ice is 2-3 times lower than the peak of $^{13}$CO$_2$:H$_2$O mixture, so the fitting of the feature with a single Gaussian causes negligible uncertainty. To quantify the difference between the pure and mixed $^{13}$CO$_2$ ice, we plot the ratios of the peak optical depths of the two main contributors, at 4.381~$\mu$m (pure $^{13}$CO$_2$, 2283~cm$^{-1}$) and 4.392~$\mu$m ($^{13}$CO$_2$:H$_2$O mixture, 2278~cm$^{-1}$, see Fig.~\ref{pic:CO2_clean} in Appendix~\ref{Append}). The maps of HOPS-56, 73, 91 and 96 are mostly noisy except for the central area, so we additionally calculate the median $\tau_{\rm ^{13}CO_2}/\tau_{\rm ^{13}CO_2:H_2O}$ over the central 1$^{\prime\prime}$. The median in HOPS-60 and 73 (1.36 and 1.22) is a factor of 2--3 higher then in HOPS-56, 91 and 96 (0.63, 0.43, 0.49); the median in HOPS-108 is somewhere in between (0.87). Additionally, HOPS-60 shows an increase in $\tau_{\rm ^{13}CO_2}/\tau_{\rm ^{13}CO_2:H_2O}$ by a factor of 2 ($\sim$0.7--1.4) toward the central area while in HOPS-108 the ratio is relatively uniform within 0.8--1.0 (see discussion in Sect.~\ref{discussion:evolution_stage} for more details).

% H56 0.63 
% H60 1.36
% H73 1.22
% H91 0.43
% H96 0.49
% H108 0.87

%Optical depth spectra show that HOPS-96 is the source with the highest absorption. 
The CO absorption bands are optically thick at a distance of $>$400~AU from the ALMA intensity peaks of HOPS-91 and HOPS-96 possibly because of lack of background IR emission. The CO bands in HOPS-56, HOPS-60 and HOPS-73 have peak optical depths by a factor of 2-3 smaller than in the other sources. The $^{13}$CO$_2$ optical depth values toward the ALMA intensity peaks of HOPS-56, HOPS-60 and HOPS-73 are comparable within the uncertainty. In addition, we observe that the optical depth of CO ice decreases as the intensity of the CO gas lines increases in these sources. We also note the presence of CO gas absorption lines and the most intense CO ice absorption band in HOPS-96.

\subsubsection{5.0 -- 7.4 $\mu$m region of MIRI MRS: H$_2$CO, H$_2$O, NH$_4^+$, CH$_3$OH}\label{fit}
The spectral region of 5.0 -- 7.4 $\mu$m has a complex structure containing many absorption molecular bands (see right panels of Fig.~\ref{pic:spec_opt}). We consider molecules that contribute most to the absorption of this range: H$_2$O, NH$_4^+$, H$_2$CO and CH$_3$OH \citep[][]{McClure2023,Rocha2024,Rayalacheruvu2025}.  We exclude the range of 6.25-6.35~$\mu$m to avoid the possible effect of PAHs emission. We used the following laboratory comparison transmission IR spectra to obtain a total fit: H$_2$O$_{\rm pure}$ (15~K), H$_2$O:CO$_2$:CO (20~K), NH$_4^+$ (15~K), NH$_4^+$ (120~K), H$_2$CO:H$_2$O (15~K). The laboratory spectra were interpolated with wavelength sampling of the observational spectra and baseline corrected with third-degree polynomial using the same spectral windows as for  the observational data (see Table \ref{Tab:window} in Appendix~\ref{Append}). 

We use linear gradient with minimization of $\chi^2$ to derive the coefficients ($k_{\rm mix}$) of the linear combination of laboratory spectra \citep[see][for the details]{Nakibov_2025,Karteyeva2026}:
\begin{equation}
\tau_{\rm model}(\lambda) = \sum_{\rm mix}k_{\rm mix}\tau_{\rm mix}(\lambda),
\end{equation}
\begin{equation}
\chi^2(\lambda) = \sum^{N_{\rm points}}_{i=1}{{(\tau_{i, {\rm obs}}(\lambda)-\tau_{i,{\rm model}}(\lambda))^2}\over{\sigma_i^2}},
\end{equation}
where mix indicates the list of mixtures (H$_2$O$_{\rm pure}$ at 15~K, H$_2$O:CO$_2$:CO at 20~K, NH$_4^+$ at 15~K, NH$_4^+$ at 120~K, H$_2$CO:H$_2$O at 15~K), $k_{\rm mix}$ is the fitted linear coefficient of the laboratory comparison spectrum, $\tau_{\rm mix}(\lambda)$ is the optical depth of the comparison spectrum, $N_{\rm points}$ is the number of analyzed data points, $\tau_{\rm obs}$ is observed optical depth and $\sigma$ is a standard deviation of the observational spectrum.

The absorption band at 6.85 $\mu$m is analyzed similarly to \citet{McClure2023}. The spectra we obtained for NH$_4^+$ at 120~K and CH$_3$OH at 10~K coincide and similarly describe the absorption band at 6.85 $\mu$m (see Fig. \ref{pic:mix_all}). Previous studies of  methanol ice absorption show that the influence of methanol absorption in this band is low compared to NH$_4^+$ \citep[e.g., ][]{Rayalacheruvu2025}. We assume that the absorption of methanol in this absorption band is at the level of uncertainty of the local continuum (see Sect.~\ref{sec:cont_sub}). We also tested mixtures H$_2$O:CH$_3$OH 5:1 at 15~K \citep{Luna2018} and H$_2$O:CH$_3$OH:CO$_2$ 9:1:2 at 15~K
\citep{Ehrenfreund1999} for possible separation through water mixtures, but we could not separate the absorption bands of NH$_4^+$ and CH$_3$OH this way either.

The NH$_4^+$ bending feature represents a sum of contributions from different salts, such as formate, hydrosulfide, cyanide, cyanate and others \citep[][]{Slavicinska2025_NH4}. We treated this NH$_4^+$ feature only with the NH$_4^+$OCN$^-$ spectrum, considering the feature independent of the counter-ion type of NH$_4^+$. This approximate approach introduces additional uncertainty into the column density of the NH$_4^+$ cation, due to notable differences in the band strengths of its bending feature when NH$_4^+$ originates from different salts. For example, the band strength of the NH$_4^+$ feature for formate and hydrosulfide differs by 20-40~\% \citep[][]{schutte2003origin, Slavicinska2025_NH4}. Since the observed profile at 6.85~$\mu$m is well fitted using only the absorption band of NH$_4^+$OCN$^-$, the profile of the NH$_4^+$ feature appears similar for different salts. The contribution of the salts to NH$_4^+$ absorption is not fully established: only 15-20~\% of the observed NH$_4^+$ abundance is compensated by the observed and expected anions \citep[][]{Slavicinska2025_NH4}. Future investigations of the charge balance in ices and the identification of the main contributors to the interstellar NH$_4^+$ deformation band will allow for more accurate column density determinations for this cation.

The right panels of Fig.~\ref{pic:spec_fit} show the pixel fits toward peaks of the ALMA continuum emission. Figure \ref{pic:spec_fit_b} (in Appendix~\ref{Append}, right panel) shows the pixel fits toward the pixel 400 AU down along declination axis from the ALMA continuum intensity peak. We assume that at 5.9-6.4~$\mu$m, there are also absorption bands of COMs, HCOO$^-$ and ammonia \citep[corresponding to COMs absorption bands,][]{rocha2025ice,Rayalacheruvu2025} not taken into account in our work. We fixed the fitted spectra for all mixtures except for pure H$_2$O (15~K) and varied the linear coefficient of the pure H$_2$O spectrum to correct the peak overfit at 6.0 $\mu$m for the pixel toward ALMA continuum peak. Thus, we got that our H$_2$O overestimation is 10\% for HOPS-56, 10\% for HOPS-60, and 6\% for HOPS-108, which contributes less than the H$_2$O column density uncertainty. However, we do not observe such an overestimation in the HOPS-91 and HOPS-96 spectra. We could not fit the HOPS-73 spectra correctly with the selected comparison spectra due to possibly high temperature of the ice mixtures.

\subsubsection{7.7 $\mu$m region of MIRI MRS: CH$_4$ and COMs}\label{methan}

\begin{figure*}
\centering
\includegraphics[scale=0.39]{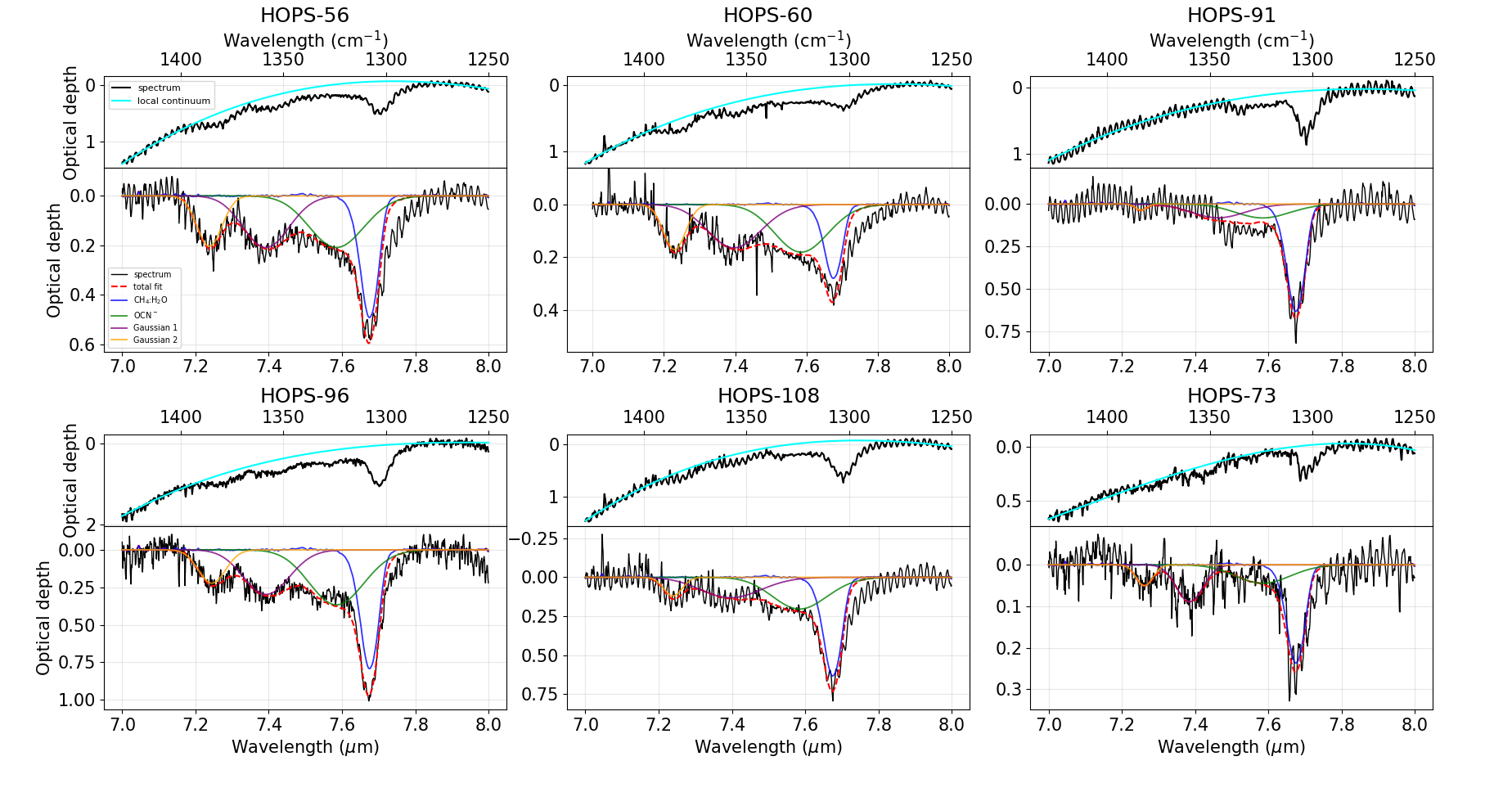}
\caption{Optical depth spectra toward the ALMA continuum intensity peaks. The red line shows the best fit for each spectrum, the cyan line shows the best fit for local continuum. Color lines show the components CH$_4$:H$_2$O (blue), OCN$^-$ (green), the Gaussian at 1380 cm$^{-1}$ (purple) and the Gaussian at 1350 cm$^{-1}$ (orange). Table \ref{tab:COMs} shows the Gaussian components reflecting the COMs absorption bands.} 
\label{pic:spec_fit_CH4}
\end{figure*}

The absorption band at 7.7 $\mu$m is associated with the CH deformation mode of methane (CH$_4$). Most JWST data are described by laboratory spectra of methane ices in the matrix of water \citep[e.g., ][]{Rocha2024,Chen2024,Rayalacheruvu2025}. However, \citet{Boogert1996} and \citet{Nakibov_2025} showed that observational data can be described by non-water methane-bearing ices. We selected the most likely mixtures for analysis, as shown in Table \ref{tab:mix}. The laboratory spectra were taken from \citet{Karteyeva2026_2}. 

We subtracted additional local continuum (third-degree polynomial with the spectral windows of 7.0--7.1~$\mu$m and 7.8--8.0~$\mu$m, see Fig.~\ref{pic:spec_fit_CH4}), as was done by \citet{Slavicinska2023_COMs}, to study this peak independently of the COMs and NH$_4^+$ absorption bands placed at 7.0-8.0 $\mu$m. We used the approach that is similar to COMs fitting (with linear coefficients applied to the spectra of laboratory ice mixtures, see Sec. \ref{fit}) for the methane absorption band. The OCN$^-$ spectrum from \cite{Gerakines2025} was also included in the fit for the 1320 cm$^{-1}$ (7.58~$\mu$m) feature, considering the OCN$^-$ column density obtained through the 4.6~$\mu$m band. We used the fitting result of 4.6~$\mu$m to describe band  OCN$^-$ at 7.58~$\mu$m. 
Also, in this region PAH emission bands overlap methane absorption. Therefore, we subtracted the multi-component PAH emission similarly to the main section (5.0--7.4~$\mu$m, see Sect.~\ref{PAH}). However, due to the uncertainty in the shape of the adapted PAH emission spectrum, the analysis of the CH$_4$ 7.7~$\mu$m band is hindered in low-intensity pixels at distances larger than 400 AU from the center. Therefore, methane maps were not constructed for these sources; we focused on analyzing the methane band and COMs toward the ALMA continuum intensity peak where the PAH emission is negligible compared to the backlight from the protostar.

We detected CH$_4$ peak in each source. According to the fit, the most probable environment for CH$_4$ is H$_2$O. Figure \ref{pic:spec_fit_CH4} shows the additional local continuum and the best fit toward the ALMA continuum intensity peaks. Column density estimates of the OCN$^-$ absorption peak are in line with values derived from the analysis of 4.4~$\mu$m band. In addition to absorption bands of CH$_4$, OCN$^-$, and COMs, this region also contains absorption band of N$_2$O with an abundance larger than CH$_4$ with respect to water \citep[][]{rocha2025ice}. However, \citet{Karteyeva2026} shows detection of N$_2$O in the main absorption band at 4.45 $\mu$m, which clearly identifies the contribution of N$_2$O in 7.7 $\mu$m region as insignificant at the noise level. We note that for HOPS-96 there is an excess absorption near 7.63 $\mu$m that can be attributed to SO$_2$ \citep[e.g., ][]{Chen2024}. However, we did not explore this feature due to low signal-to-noise ratio and continuum uncertainty.

We analyzed the absorption of NH$_4^+$HCOO$^-$ and NH$_2$CHO similarly to \citet{Slavicinska2023_COMs}. The NH$_4^+$HCOO$^-$:H$_2$O (14~K) mixture has the most suitable peak positions. As a result of fitting, this laboratory spectrum incompletely described the absorption peak at 1380 cm$^{-1}$ (7.25~$\mu$m), which led to an underestimation by $\sim$50\% of the area. It is possible that this component is the cumulative contribution of several COMs \citep[e.g., C$_2$H$_5$OH and HCOO$^-$, ][]{Chen2024,Rayalacheruvu2025}. For this reason, we evaluated the absorption peaks of COMs using Gaussian profiles (see Table~\ref{tab:COMs}). The positions of the peaks and the widths of the absorption bands coincide within the uncertainty in all sources except for HOPS-91, which indicates that they belong to the same COMs. 

% продолжаем про табличку комов
\begin{table}
\begin{center}

\caption{Maximum intensity, peak position of the band, and band width according to the Gaussian fit of COMs toward the ALMA continuum intensity peaks.}\label{tab:COMs}
\begin{tabular}{ l   c   c  c  c  c   c  c  c  c }\hline
Source & $A_{\rm peak}$ & $\lambda_{\rm peak}$ & $\sigma_{\rm peak}$ \\ 
    &           &     (cm$^{-1}$)    &    (cm$^{-1}$)\\ \hline
\multicolumn{4}{c}{band at 1350 cm$^{-1}$} \\\hline
HOPS-56 & 0.209$\pm$0.005 & 1352.8$\pm$0.3  & 11.5$\pm$0.4\\
HOPS-60 & 0.165$\pm$0.004 & 1352.0$\pm$0.4  & 13.6$\pm$0.6\\
HOPS-73 &0.089$\pm$0.005 & 1354.0$\pm$0.4  & 6.7$\pm$0.5\\
HOPS-91 & 0.081$\pm$ 0.007 & 1340.0$\pm$1.3  & 13.1$\pm$1.3\\
HOPS-96 & 0.297$\pm$0.005 & 1352.6$\pm$0.4 & 11.5$\pm$0.4\\
HOPS-108 & 0.133$\pm$0.006 & 1352.0$\pm$0.8 & 15.1$\pm$1.3\\\hline
\multicolumn{4}{c}{band at 1380 cm$^{-1}$}\\\hline
HOPS-56 &0.204$\pm$0.006 & 1381.6$\pm$0.3 & 6.5$\pm$0.3\\
HOPS-60 &0.171$\pm$0.006 & 1382.2$\pm$0.3 & 5.9$\pm$0.3\\
HOPS-73 &0.051$\pm$0.006 & 1377.1$\pm$0.6 & 4.3$\pm$0.6\\
HOPS-91 &0.036$\pm$0.012 &1378.8$\pm$1.2  &2.9$\pm$1.2\\
HOPS-96 &0.223$\pm$0.009 & 1381.4$\pm$0.4 & 7.0$\pm$0.4\\
HOPS-108 &0.120$\pm$0.010 & 1381.3$\pm$0.5 & 4.8$\pm$0.5\\
\hline
\end{tabular}

      \small
      \begin{flushleft}
      \item \textbf{Note:} Figure \ref{pic:spec_fit_CH4} shows the applied local continuum and the multi-component fit. 
      \end{flushleft}

\end{center}
\end{table}

\subsection{Ice column density maps}
\subsubsection{Conversion to column density}

We use the fitted linear coefficients and Gaussian parameters to estimate the ice column density ($iN_{\rm tot}$) of the molecules. We assume that the dust is finely dispersed, with a size of less than 1 $\mu$m, which allows us to use the Beer–Bouguer–Lambert (BBL) extinction law (see Section \ref{app:model}). $^{13}$CO$_2$, OCN$^-$, CO and OCS column densities were estimated using the obtained Gaussian fit parameters:
\begin{equation}
iN_{\rm tot}^{\rm mol} = \frac{\tau_{\rm peak}^{\rm mol}\sigma_{\rm mol}\sqrt{2\pi}}{A'_{\rm mol}},
\end{equation}
where $\tau_{\rm peak}$ is the peak of optical depth, $\sigma$ is line width, mol indicates the molecule and $A'$ is band strength. We applied the $^{12}$C/$^{13}$C=66 isotopic ratio for the local ISM \citep{Yan2023} to estimate $^{12}$CO$_2$ column density from $^{13}$CO$_2$.
The column densities of NH$_4^+$, CH$_4$ and H$_2$CO were estimated using the area of the corresponding bands in reference ice spectra:
\begin{equation}
iN_{\rm tot}^{\rm mol} = \frac{k_{\rm mol}S_{\rm mol}}{A'_{\rm mol}}, 
\end{equation}
where $k$ is the spectrum linear coefficient obtained from the fit, $S$ is the absorption peak area of the laboratory comparison spectrum and  $A'$ is the band strength. Table \ref{tab:bands} shows the band strengths and peak absorption wavelengths. OCS column density has high uncertainty with low SNR across all maps, thus we do not analyze the OCS distribution and only provide the OCS column densities toward the ALMA continuum peaks (see Table~\ref{tab:coln_den}).

%The H$_2$O column density was obtained through fitting of 6 $\mu$m band. Due to strong silicate absorption at  $\sim$9 $\mu$m the right wing of broad asymmetric H$_2$O band is underestimated near the 7.7 $\mu$m CH$_4$ band.  To manage this, we subtracted the continuum in H$_2$O laboratory spectra similarly to observational data. This procedure yields more correct fitted linear coefficients for H$_2$O but distorts the laboratory spectrum in this range.  Therefore, we multiplied each laboratory H$_2$O spectrum by its fitted linear coefficient and calculated the column density from the 3~$\mu$m H$_2$O band, which is unaffected by continuum subtraction. 

The H$_2$O column density is obtained through fitting of the $\sim$6~$\mu$m band. Due to strong silicate absorption at $\sim$9~$\mu$m the right wing of broad asymmetric H$_2$O band is underestimated near the 7.7 $\mu$m CH$_4$ band and this complex profile may bring a 10-15\% overestimation to the H$_2$O column density \citep{Keane2001}. To manage this, we subtracted the continuum in H$_2$O laboratory spectra similarly to observational data in 5.0-8.1~$\mu$m range. This procedure yields more correct fitted linear coefficients for H$_2$O. However, such subtraction affects the laboratory spectra, making us unable to calculate the column density from 6.6~$\mu$m H$_2$O absorption band. Therefore, we multiplied each laboratory H$_2$O spectrum by its fitted linear coefficient (from the fit at the 6~$\mu$m band) and calculated the column density from the area of the 3~$\mu$m H$_2$O band (of the laboratory spectrum), which is unaffected by continuum subtraction. In addition to the pure water, we took into account the column density of the ice mixtures obtained during the experiments:

\begin{multline}
iN_{\rm tot}({\rm H_2O}) = \left(\frac{k_{\rm H_2O:H_2CO}S_{\rm H_2O:H_2CO}}{A'_{\rm H_2O}}\right)+ \left(\frac{k_{\rm H_2O}S_{\rm H_2O}}{A'_{\rm H_2O}}\right) + \\ 
+\left(\frac{k_{\rm H_2O:CO:CO_2}S_{\rm H_2O:CO:CO_2}}{A'_{\rm H_2O}}\right), %+0.67*k_{H_2O:CO:CO_2}N_{tot}(H_2O:CO:CO_2), 
\end{multline}
where $k_{\rm H_2O:H_2CO}$ is the linear coefficient of the H$_2$O:H$_2$CO mixture from fitting, $S_{\rm H_2O:H_2CO}$ is the H$_2$O band area at 3~$\mu$m from the H$_2$O:H$_2$CO mixture ($S_{\rm H_2O:H_2CO}$ = 346~cm$^{-1}$), $A'_{\rm H_2O}$ is the band strength of H$_2$O at 3~$\mu$m, $k_{\rm H_2O}$ is the linear coefficient for pure H$_2$O from fitting, $S_{\rm H_2O}$ is the H$_2$O band area at 3~$\mu$m from pure H$_2$O ($S_{\rm H_2O}$ = 75~cm$^{-1}$), $k_{\rm H_2O:CO:CO_2}$ is the linear coefficient the H$_2$O:CO:CO$_2$ mixture from fitting, $S$(H$_2$O:CO:CO$_2$) is the H$_2$O band area at 3~$\mu$m from the H$_2$O:CO:CO$_2$ mixture ($S_{\rm H_2O:CO:CO_2}$ = 53~cm$^{-1}$). The water column density may be still overestimated by $<$10~\% due to contribution of COMs absorption (see Sect.~\ref{fit}) that was not analyzed here.

The uncertainty of the column density is 10--30\%, it includes the uncertainty caused by the continuum selection uncertainty, by the fit of the comparison spectra and the Gaussian, and by the noise in the optical depth spectra. The uncertainty increases with decreasing background emission and ice column density in the areas of ice evaporation. The pixels with the uncertainties $>30$\% are excluded from our analysis.

\subsubsection{Molecular hydrogen column density}\label{col_H2}
\begin{figure*}
\centering
\includegraphics[scale=0.4]{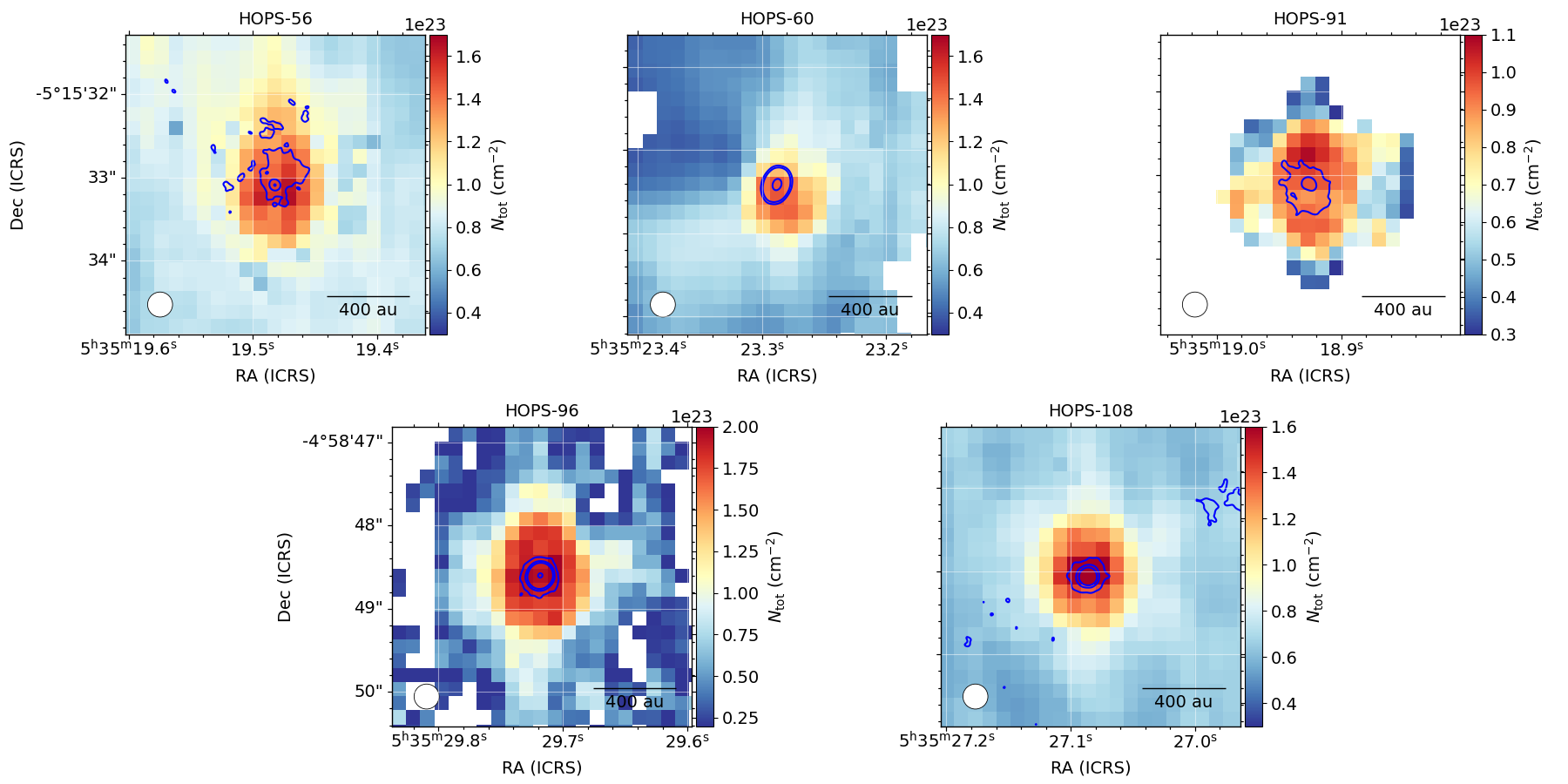}
\caption{Maps of molecular hydrogen column density. The contours show the ALMA continuum emission (like in Fig.~\ref{pic:continuum}).} 
\label{pic:N_H2}
\end{figure*}

The H$_2$ column density estimates obtained in earlier studies of this region are not suitable to analyze the JWST observations due to their low spatial resolution \citep[2--36.3$^{\prime\prime}$, e.g.,][]{Furlan2016}. Our estimates of solid H$_2$O column density enable the analysis of the column density of molecular hydrogen at $\sim0.3^{\prime\prime}$ scale. We determined the optical depth of the water OH stretch band at 3 $\mu$m with laboratory spectra using the linear coefficients ($\tau_{\rm 3\mu m}\times k$). Next, visual extinction ($A_V$) was obtained through the observational dependence $A_V$ = $\tau_{\rm 3\mu m}/q + A_{\rm th}$, with the empirical coefficients $q = 0.072\pm0.002$ and $A_{\rm th} = 3.2\pm0.1$, obtained in \citet{Whittet_2001}. Finally, we converted visual extinction into the column density of molecular hydrogen with $N$(H$_2$) = 1.1$\times$10$^{21}\times A_V$ \citep{Guver2009}.

Figure \ref{pic:N_H2} shows the maps of molecular hydrogen column density toward HOPS-56, HOPS-60, HOPS-91, HOPS-96 and HOPS-108. HOPS-73 is missing since we could not obtain $N$(H$_2$O) there (see Sect.~\ref{fit}). Our estimates are by a factor of 1.5--10 higher than those obtained with the Spitzer, Herschel, and ground-based IR facilities data (given in Table~\ref{tab:coord}), which is due to their lower spatial resolution (see Fig.~\ref{pic:Orion}).

\subsubsection{Ice column density distribution}

\begin{figure*}
\centering
\includegraphics[scale=0.45]{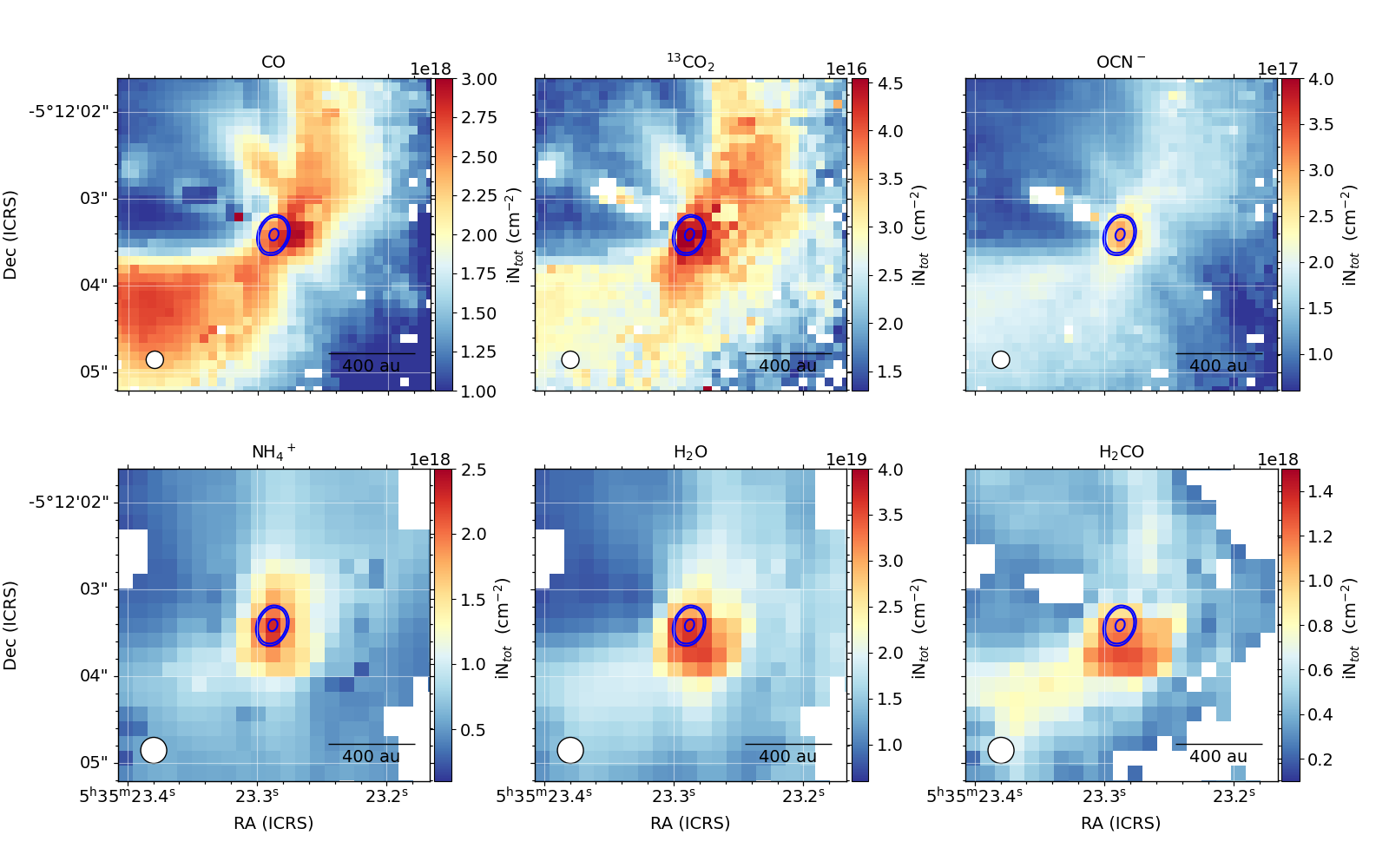}
\caption{Maps of ice column densities toward HOPS-60. The contours show the ALMA continuum emission (like in Fig.~\ref{pic:continuum}). } 
\label{pic:N_H60}
\end{figure*}

The protostellar phase includes several evolutionary stages starting with the transition from a spherically symmetric envelope to a more complex structure of envelope, the formation of an accretion disk, and jets \citep[][]{Ehrenfreund_2002,Machida2013}. Figure \ref{pic:N_H60} and Fig. \ref{pic:N_H56}, \ref{pic:N_H73}, \ref{pic:N_H91}, \ref{pic:N_H96} and \ref{pic:N_H108} (in Appendix~\ref{Append}) show the maps of the $^{13}$CO$_2$, OCN$^-$, CO, H$_2$O, NH$_4^+$ and H$_2$CO column density toward HOPS-56, HOPS-60, HOPS-73, HOPS-91, HOPS-96 and HOPS-108. Distribution of NH$_4^+$ is similar to that of H$_2$O and increases toward the ALMA continuum peak. Observed spatial distributions of column densities are consistent with the structure of late protostellar envelopes where the protostellar object has a central source, a disk connected to the inner envelope, an outer envelope, and jets that significantly influence the dust along the axes \citep[e.g., ][]{Dishoeck2025}.

The CO, $^{13}$CO$_2$, and OCN$^-$ maps of HOPS-60  show a complex structure of the inner envelope (seen as two flared wing-shaped regions with increased column densities in Fig.~\ref{pic:N_H60}). In addition to the two main wings of the inner envelope, we also observe an additional wing, which may be a streamer \citep[e.g., ][]{Dishoeck2025} located in a layer different from the main inner envelope or an element of accretion disk \citep[][]{Hartmann1999,Vorobyov2013}. The distributions of CO and $^{13}$CO$_2$ differ, indicating small temperature fluctuations within the envelope. We do not observe similar structures of CO, $^{13}$CO$_2$, and OCN$^-$ in the other sources. 

The distribution of column densities in HOPS-60 and HOPS-73 shows the effect of the outflows on icy mantles. This is indicated by the zones of decreased column density of all species due to icy mantles heating, see left top quadrant in Fig. \ref{pic:N_H60} and right top quadrant in Fig. \ref{pic:N_H73} in Appendix~\ref{Append}. Some pixels are blank since the optical depth is too low there and propagates its large uncertainty to the column density --- these positions may indicate jet knot positions. The gas emission lines of CO, H$_2$, H$_2$O and ionized metals (Fe, Ne, Ar) are strong there as well. 

In HOPS-56, we observe enhanced column densities at a distance of 500 AU from the ALMA continuum peak (see left bottom quadrant of the CO, $^{13}$CO$_2$, and OCN$^-$ maps in Fig.~\ref{pic:ab_sli}), which may indicate the initial jet activity \citep[][]{Machida2013}. Given the detected CO gas lines in that area, we can assume the presence of outflows. No such structures were observed in the other sources, and their distributions resemble a spherically symmetric structure of the envelope. We assume that this structure is related to a more massive outer envelope (see $M_{\rm env}$ in Table~\ref{tab:coord}) and to an earlier stage of a protostar. 

In HOPS-60, we observe a complete similarity between the OCN$^-$ distribution and the ALMA continuum, revealing an accretion disk. HOPS-56 and HOPS-108 show the same similarity between the OCN$^-$ column density distribution and the ALMA continuum emission. HOPS-96 has a ring-shaped H$_2$CO column density distribution around the center of the source at 200~AU, with a decrease toward the center. In the other sources, we observe an increase in the H$_2$CO column density toward the center of the sources.  

The average column densities of CO, CO$_2$ and H$_2$O obtained toward all sources are comparable to the column densities of the Chameleon~1 dense cloud within an order of magnitude \citep[][]{Smith2025}. Toward the prototsars, we observe column densities of CO and H$_2$O that are higher by a factor of 2-4 and can be explained by higher density and ice thickness.

%The following estimates of column density are toward all maps: 0.1--10.8$\times$10$^{18}$~cm$^{-2}$ of CO, 0.9--10.6$\times$10$^{18}$~cm$^{-2}$ of CO$_2$, 1.1--13.3$\times$10$^{16}$~cm$^{-2}$ of $^{13}$CO$_2$, 0.8--6.1$\times$10$^{16}$~cm$^{-2}$ of OCS, 0.5--3.4$\times$10$^{17}$~cm$^{-2}$ of OCN$^-$, 0.4--5.2$\times$10$^{19}$~cm$^{-2}$ of H$_2$O, 0.1--2.9$\times$10$^{18}$~cm$^{-2}$ of H$_2$CO and 0.1--4.1$\times$10$^{18}$~cm$^{-2}$ of NH$_4^+$; with uncertainty of 8-22\%.

\subsubsection{Peaks of ice column density}

%\subsubsection{Analyzes of column density distribution}

% пик лучевой концентрации в  HOPS-56 b HOPS-73 смещен
%  HOPS-73 низкое значения льда 
\begin{table*}
\begin{center}
\caption{Column densities toward the ALMA continuum peaks.}\label{tab:coln_den}
\begin{tabular}{ l   c   c  c  c  c   c  c  c  c }\hline
Source & H$_2$O& CO & $^{13}$CO$_2$ & OCN$^-$ & H$_2$CO & NH$_4^+$ & CH$_4$ & 
OCS\\ 
 & (10$^{19}$~cm$^{-2}$) & (10$^{18}$~cm$^{-2}$) & (10$^{16}$~cm$^{-2}$) & (10$^{17}$~cm$^{-2}$)  & (10$^{18}$~cm$^{-2}$) & (10$^{18}$~cm$^{-2}$) & (10$^{17}$~cm$^{-2}$) & 
(10$^{16}$~cm$^{-2}$)\\ \hline
HOPS-56 & 3.75$\pm$0.31  & 4.36$\pm$0.38  & 4.18$\pm$0.38  & 3.19$\pm$0.31 & 2.31$\pm$0.24  & 2.67$\pm$0.18 & 5.91$\pm$0.44 & 2.21$\pm$0.42 \\ 
HOPS-60 & 3.62$\pm$0.42 & 2.71$\pm$0.25 & 4.79$\pm$0.51 & 2.76$\pm$0.39 & 1.17$\pm$0.12  & 2.03$\pm$0.17 & 3.34$\pm$0.32 & 2.44$\pm$0.52\\ 
HOPS-73 & -- & 2.74$\pm$0.38 & 3.11$\pm$0.44 & 0.68$\pm$0.15 & --  & -- & 2.85$\pm$0.18 & 1.12$\pm$0.27\\
%HOPS-73 & 1.01$\pm$0.11 & 2.74$\pm$0.38 & 2.39$\pm$0.34 & 0.68$\pm$0.15 & 0.55$\pm$0.04 & 0.65$\pm$0.06 & 3.9$\pm$0.33 & 1.91$\pm$0.27 & 1.12$\pm$0.27\\ 
HOPS-91 & 2.72$\pm$0.22 & 8.94$\pm$0.88 & 4.48$\pm$0.79 & 1.25$\pm$0.17 & 1.18$\pm$0.15  & 2.33$\pm$0.23 & 7.56$\pm$0.51 & -- \\ 
HOPS-96 & 5.16$\pm$0.46 & 10.78$\pm$1.24 & 9.34$\pm$1.35 & 3.25$\pm$0.45 & 1.59$\pm$0.18 & 4.11$\pm$0.39 & 9.51$\pm$1.12 & -- \\ 
HOPS-108 &3.98$\pm$0.38 & 5.77$\pm$0.69 & 5.18$\pm$0.75 & 3.11$\pm$0.37 & 2.08$\pm$0.25 & 3.39$\pm$0.19 & 7.59$\pm$0.25 & 2.71$\pm$0.64    \\ \hline
%NIR38$^a$ & 0.69$^{1.25}_{0.37}$ & 2.96$^{4.66}_{1.86}$ &  2$^{3}_{2}$ & 0.2 & --  & 0.57 & 1.8$^{2.3}_{1.4}$ & 1 \\
%J110621$^a$ & 1.34$^{1.73}_{0.78}$ & 3.68$^{5.46}_{2.48}$&  2$^{4}_{2}$ & 0.3 & --  & 0.78 & 2.5$^{2.8}_{1.6}$ & 2\\
IRAS 2A$^a$ & 3.24$^{5.04}_{2.05}$ & -- &  -- & 4.90$^{6.50}_{2.68}$ & 1.51$^{1.99}_{0.83}$  & 3.73$^{5.52}_{2.31}$ & 4.55$^{6.30}_{2.43}$ & --\\
B1-c$^b$ & 2.50$\pm$0.17 & -- &  -- & 4.1$\pm$2.2 &  0.45$\pm$0.13  &  -- &   9.3$\pm$5.1 & --\\

\hline
\end{tabular}
\begin{flushleft}
      \small
      \begin{flushleft}
      \item \textbf{Note:} The values of CO, CO$_2$, OCN$^-$ and OCS column density were taken from the column density maps regridded to the field of view and pixel size of the H$_2$O column density map. Table \ref{tab:coln_den_non} in Appendix~\ref{Append} shows the non-regridded column density values.
%      $^a$ -- column densities obtained toward Chameleon 1 (NIR38, $A_V$=60 mag and J110621, $A_V$=95 mag) from \cite{McClure2023}. 
$^a$ The column densities obtained toward low-mass protostar of Class~0, IRAS 2A from \cite{Rayalacheruvu2025} and $^b$ the column densities obtained toward the low-mass protostar of Class~0, B1-c from \cite{Chen2024}.
      \end{flushleft}
\end{flushleft}
\end{center}
\end{table*}
%OCN$^-$ column densities are obtained with the CH$_4$: 3.6$\times$10$^{17}$cm$^{-2}$ for HOPS-56, 3.1$\times$10$^{17}$cm$^{-2}$ for HOPS-60, 1.7$\times$10$^{17}$cm$^{-2}$ for HOPS-73, 2.3$\times$10$^{17}$cm$^{-2}$ for HOPS-91, 5.6$\times$10$^{17}$cm$^{-2}$ for HOPS-96, and 3.7$\times$10$^{17}$cm$^{-2}$ for HOPS-108.[]

% сравнение
Comparing the peak positions of the ALMA continuum emission and of the ice column densities provides information about chemical evolution and the impact of different physical conditions on the dust surface. The column density peaks of $^{13}$CO$_2$, OCN$^-$, H$_2$O and NH$_4^+$ coincide with the peaks of ALMA continuum emission in HOPS-56, HOPS-60, HOPS-91, HOPS-96 and HOPS-108. We observe a shift in the peak of CO ice column density in HOPS-60 from the peak of ALMA continuum intensity, which may be related to heating \citep[][]{Herbst2009,Chen2024} or to non-axisymmetric perturbation in the disk \citep[e.g., ][]{Soker2022}. The assumption of heating is also supported by the fact that the band of pure $^{13}$CO$_2$ prevails over the mixed $^{13}$CO$_2$ toward the ALMA continuum peak in HOPS-60 and HOPS-73 (see the spectra in Fig.~\ref{pic:spec_opt} and the ratios of the optical depths in Fig.~\ref{pic:CO2_clean}, discussed below in Sect.~\ref{discussion:evolution_stage}),  with a median ratio of maximal optical depths over the central 1$^{\prime\prime}$ $\tau_{\rm CO_2}:\tau_{\rm CO_2:H_2O}>1$, see Sect.~\ref{sec:tau_43}. These factors indicate the evolution of the chemical composition of dust \citep[][]{Boogert2015}. In HOPS-73, the peaks of column densities are shifted from the ALMA continuum emission peak, which might be caused by the outflow projecting on the continuum peak. %which suggests a more complex structure of outflows in beam in these source.} % The assumption of heating is also supported by the fact that the component/band of pure 13CO2 prevails over the mixed 13CO2 toward the ALMA continuum peak in HOPS-60.

We derived the column densities obtained toward the ALMA continuum intensity peaks, as shown in Table~\ref{tab:coln_den}. The column densities increase with $A_V$ (given in Table~\ref{tab:coord}). As we mentioned in Section \ref{methan}, CH$_4$ was detected in a mixture with H$_2$O, which is characteristic of low-mass protostars \citep[e.g., ][]{Rocha2024,Chen2024}. In addition to the estimates of OCN$^-$ by 4.45~$\mu$m band, we also estimated the OCN$^-$ column densities from the blue wing (7.6~$\mu$m) of CH$_4$ band and obtained the following estimates for the ALMA intensity peaks in Table \ref{tab:coln_den}. %3.6$\times$10$^{17}$cm$^{-2}$ for HOPS-56, 3.1$\times$10$^{17}$cm$^{-2}$ for HOPS-60, 1.7$\times$10$^{17}$cm$^{-2}$ for HOPS-73, 2.3$\times$10$^{17}$cm$^{-2}$ for HOPS-91, 5.6$\times$10$^{17}$cm$^{-2}$ for HOPS-96, and 3.7$\times$10$^{17}$cm$^{-2}$ for HOPS-108. 
The uncertainty of the OCN$^-$ column density is greater than 50\% for this absorption band due to the uncertainty in the local continuum. Continuum selection uncertainty, however, has a low impact on CH$_4$ column density estimates (7-8\%). The OCN$^-$ estimates are consistent with those obtained via the 4.45~$\mu$m band.

We compared the column densities with the values obtained toward the low-mass protostar of Class~0, IRAS 2A from \cite{Rayalacheruvu2025}. Our estimates of H$_2$O, NH$_4^+$ and H$_2$CO are within the uncertainty limits of IRAS~2A. The column densities of CH$_4$ agree within the uncertainties in sources HOPS-56, HOPS-60 and HOPS-73. Sources HOPS-91, HOPS-96, and HOPS-108 have the column densities of CH$_4$ $\sim$1.5 times higher than that of IRAS 2A (taking the uncertainties into the account). We also compared our peak column densities with the column densities toward Class~0 protostar B1-c \citet{Chen2024}. Our column densities of OCN$^-$ and CH$_4$ are consistent with their estimates within the uncertainties. Our column densities of H$_2$O and H$_2$CO are 1.1-2 and 2.6-5.7 times higher, respectively. Their low H$_2$CO value (also compared to IRAS 2A) may be due to the use of a single absorption band (while we use two H$_2$CO bands).

The obtained similar column densities indicate the similar envelope in different star-forming regions in low-mass sources at the Class 0 stage. This is also confirmed by our estimates of extinction (see Table \ref{tab:coord}) and the column density of molecular hydrogen (see Fig. \ref{pic:N_H2}), which indicates high density with variations in the mass of the protostellar core.
We compared our column densities to those toward the outflow of a low-mass Class~0 protostar HOPS-370 from \cite{Tyagi_2025}. The estimates of the CO$_2$ column density in HOPS-60 and HOPS-73 are larger then the column density toward the HOPS-370 outflow by a factor of 1.7--2.5. At the same time, our estimate of HOPS-60 extinction along the outflows direction is $\sim$2.5 times higher than in HOPS-370. This suggests that ice column density may show that the outflows are shielded by the envelope.

\subsection{Ice abundances with respect to H$_2$O}

%Наши оценки
We regridded the column density maps of CO, $^{13}$CO$_2$, CO$_2$ and OCN$^-$ to the pixel size and the field of view of the H$_2$O column density map for HOPS-56, HOPS-60, HOPS-91, HOPS-96 and HOPS-108. We calculated the CO$_2$ (via $^{13}$CO$_2$), OCN$^-$, CO, NH$_4^+$ and H$_2$CO abundance maps with respect to H$_2$O as follows: 
\begin{equation}
iX_{\rm H_2O}=iN_{\rm tot}/iN_{\rm tot}(\text{H}_2\text{O}),
\end{equation}
where $iN_{\rm tot}$ is the ice column density of the species and $iN_{\rm tot}$(H$_2$O) is the column density of H$_2$O. Table \ref{tab:ratio} shows the abundance of the species toward the ALMA continuum intensity peak. 

\subsubsection{Peaks of ice abundance}

We find that, in all sources except for HOPS-60, the abundance of CO is larger than that of CO$_2$ by a factor of 1.3--2.9, with an average ratio of 1.9. In contrast, HOPS-60 shows a CO$_2$ abundance that is 1.2 times higher than that of CO. This may indicate a more advanced evolutionary stage of a Class~0 source with a relatively warm envelope ($T\geq$ 20~K). As time passes, the envelopes of protostars are subject to heating from the developing central object and outflows. 
CO desorbs at temperature of ~18--40 K, while CO$_2$ desorbs at significantly higher temperature of 50--80~K \citep[e.g.,][]{Fayolle2011,Cuppen2024}. Trapping effects may additionally increase effective sublimation temperature of CO$_2$ from water ice \citep{Fayolle2011} because it is predominantly located in H$_2$O-rich ice, whereas CO is in CO-rich layer \citep{Boogert2015}. Additionally CO$_2$ entrapment in water ice is more efficient than that of CO \citep{Fayolle2011}. Thus, even considering entrapment effects, decline of CO abundance relative to CO$_2$ may indicate thermal desorption of a part of the former.

The sources HOPS-56, HOPS-60, and HOPS-108 exhibit OCN$^-$ abundances that are approximately 1.5 times higher than those observed in HOPS-91 and HOPS-96. This suggests enhanced exposure to UV emission at earlier stages of formation \citep[][]{Fedoseev2016}. Consistently, these same sources (HOPS-56, HOPS-60, and HOPS-108) show strong PAH emission (see Section \ref{PAH}), which indirectly supports the presence of an elevated high-energy emission background in outer envelopes. In contrast, the abundances of H$_2$CO, CH$_4$, and NH$_4^+$ measured toward the ALMA intensity peaks do not show a dependence on neither envelope density or the level of background emission, possible at earlier stage of star formation.

Table \ref{tab:ratio} compares our derived abundances with those reported for cold cloud Cha~1 \citep[with two background stars, NIR38 and J110621;][]{McClure2023}, and protostars IRAS 2A \citep[][]{Rayalacheruvu2025}, B1-c \citep{Chen2024} and other low-mass and massive protostars \citep[from Spitzer observations][]{Oberg_2011,Boogert2015}. The abundances of CO$_2$, $^{13}$CO$_2$, NH$_4^+$, and CH$_4$ obtained in this work fall within the ranges reported for Cha~1. If the abundance of complex species and salts in ices increases with both density and evolutionary time, one would expect NH$_4^+$ abundances to exceed those measured in Cha~1. While this is not observed for NH$_4^+$, we find OCN$^-$ abundances to be higher than those of Cha~1 by a factor of 2.5--3.5  (see Table~\ref{tab:ratio}). Finally, the CO abundances in HOPS-91 and HOPS-96 are comparable to those reported for Cha~1 and for low-mass protostars (LYSOs in Table~\ref{tab:ratio}). This indicates that ice inventory in the envelopes of HOPS-91 and HOPS-96 are similar to the initial cold dense cores. In contrast, HOPS-56, HOPS-60, and HOPS-108 exhibit systematically lower CO abundances, similar to massive protostars (MYSOs in Table~\ref{tab:ratio}), which is consistent with envelope temperatures exceeding approximately 20~K.

The abundances of CO$_2$ in HOPS-91 and HOPS-96, H$_2$CO in HOPS-108 and HOPS-91, CH$_4$ in HOPS-56 and NH$_4^+$ in HOPS-108 coincide within the uncertainty with the abundances in the low-mass protostar IRAS~2A \citep{Rayalacheruvu2025}. Our derived abundances are consistent with the ranges reported by \citet{Boogert2015}. This similarity suggests that the ice compositions in our sources are shaped by evolutionary processes comparable to those inferred for the objects studied by \citet{Boogert2015}. 

We compared our abundances to the median abundances obtained separately for high-mass and low-mass protostars by \citet[][see their Table~3]{Oberg_2011}. The abundances of CO/H$_2$O in HOPS-91 and HOPS-96 are consistent with those of low-mass protostars while CO/H$_2$O in HOPS-56, HOPS-60, HOPS-108 are consistent with those of high-mass protostars. The abundances of CO$_2$/H$_2$O in all our protostars are similar to that of high-mass protostars. The abundances of OCN$^-$ are consistent with both categories and closer to the massive protostars; the abundances of CH$_4$ are lower and the abundances of NH$_4^+$ are larger than those of both high-mass and low-mass protostars (see Table~\ref{tab:ratio}). The difference in CO$_2$ abundance is possibly caused by chosen $^{12}$C/$^{13}$C ratio, which may differ at this spatial resolution.

\begin{figure*}
\centering
\includegraphics[scale=0.5]{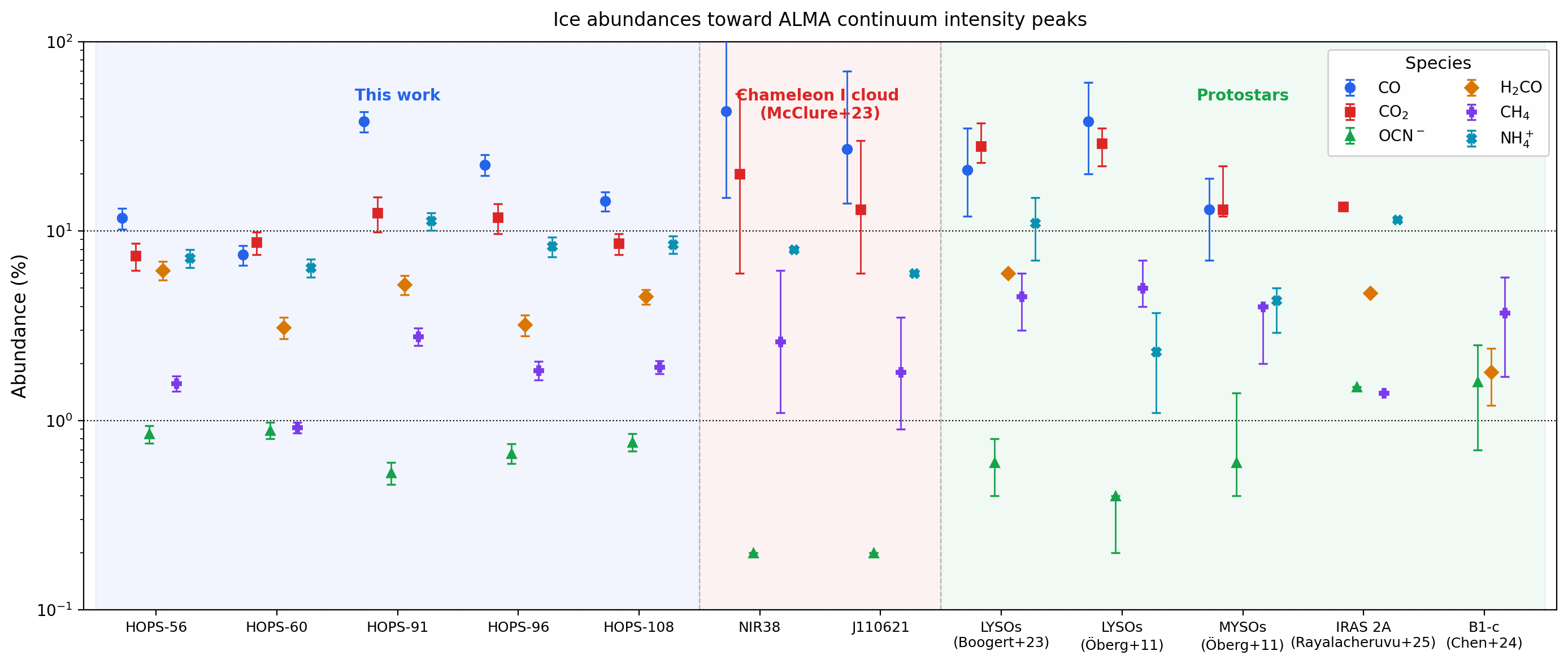}
\caption{Abundances of the species with respect to H$_2$O toward the ALMA continuum intensity peaks (see Table~\ref{tab:ratio}). } 
\label{pic:abund_all}
\end{figure*}

\begin{table*}
%\begin{center}
\caption{Abundances of the species with respect to H$_2$O toward the ALMA continuum intensity peaks.}\label{tab:ratio}
\begin{tabular}{ l  c  c  c  c   c  c   c}\hline
Source & CO/H$_2$O &CO$_2$/H$_2$O& $^{13}$CO$_2$/H$_2$O & OCN$^-$/H$_2$O & H$_2$CO/H$_2$O & CH$_4$/H$_2$O  & NH$_4^+$/H$_2$O\\ 
      &(\%) & (\%) &(\%) &(\%)  &(\%) &(\%) &(\%) \\\hline
HOPS-56 & 11.7$\pm$1.5  & 7.4$\pm$1.2& 0.11$\pm$0.02 & 0.85$\pm$0.09 & 6.2$\pm$0.7 & 1.57$\pm$0.15&  7.2$\pm$0.8\\ 
HOPS-60 & 7.5$\pm$0.9  & 8.7$\pm$1.2& 0.13$\pm$0.02& 0.89$\pm$0.09 & 3.1$\pm$0.4 & 0.92$\pm$0.06& 6.4$\pm$0.7\\ 
%HOPS-73 & 29\%$\pm$6\%  &26\%$\pm$6\%& $<$1\% &5.7\%$\pm$0.9\% & 6.8\%$\pm$0.8\%& 4.1\%$\pm$0.8\%\\ 
HOPS-91 &37.9$\pm$4.6  &12.5$\pm$2.6&  0.19$\pm$0.04& 0.53$\pm$0.07 & 5.2$\pm$0.6 & 2.78$\pm$0.30&11.3$\pm$1.2 \\ 
HOPS-96 & 22.4$\pm$2.8  &11.8$\pm$2.1 & 0.18$\pm$0.03 & 0.67$\pm$0.08 & 3.2$\pm$0.4 & 1.84$\pm$0.21 &8.3$\pm$1.0\\ 
HOPS-108 &  14.4$\pm$1.7  &8.6$\pm$1.1& 0.13$\pm$0.02 & 0.77$\pm$0.08 & 4.5$\pm$0.4& 1.91$\pm$0.15 &8.5$\pm$0.9 \\ 
\hline
NIR38$^a$ &  43$^{125}_{15}$ &20$^{53}_{6}$& 0.28$^{0.80}_{0.16}$ & 0.2 & -- &  2.6$^{6.2}_{1.1}$ & 8 \\
J110621$^a$ &  27$^{70}_{14}$ &13$^{30}_{6}$&  0.15$^{0.51}_{0.11}$ & 0.2 & -- & 1.8$^{3.5}_{0.9}$& 6 \\
LYSOs$^b$ &  21$^{35}_{12}$ &28$^{37}_{23}$& --& 0.6$^{0.8}_{0.4}$ & $\sim$6 & 4.5$^{6}_{3}$ & 11$^{15}_7$ \\
    & ($\leq$3)-85 &  12-50 &  &($\leq$ 0.1)-1.1 & -- &  1-11 &  4-13 \\ 
LYSOs$^c$ &  38$^{61}_{20}$ &29$^{35}_{22}$& -- &0.4$^{0.4}_{0.2}$ & -- & 5$^{7}_{4}$ & 2.3$^{3.7}_{2.1}$ \\
MYSOs$^c$ &  13$^{19}_{7}$ &13$^{22}_{12}$& -- &0.6$^{1.4}_{0.4}$ & -- & 4$^{4}_{2}$ & 4.3$^{5.0}_{2.9}$ \\
IRAS 2A$^d$ &  -- &13.4&-- & 1.51& 4.7 & 1.4 & 11.5 \\
B1-c$^e$ &  -- &--&-- &  1.6$\pm$0.9& 1.8$\pm$0.6 & 3.7$\pm$2.0 & -- \\
\hline
\end{tabular}
\begin{flushleft}
      \small
      \begin{flushleft}
      \item \textbf{Note:} 
      $^a$The abundances from the column densities obtained toward Cha~1 (NIR38, $A_V$=60 mag and J110621, $A_V$=95 mag) from \cite{McClure2023}. $^b$The abundances for the low-mass protostars from \cite{Boogert2015}. $^c$The abundances for the low-mass protostars from \cite{Oberg_2011}. $^d$The abundances for the low-mass protostar of Class~0, IRAS 2A from \cite{Rayalacheruvu2025}. $^e$ The abundances obtained toward the low-mass protostar of Class~0, B1-c from \cite{Chen2024}.
      \end{flushleft}
\end{flushleft}
%\end{center}
\end{table*}

\subsubsection{Maps of ice abundances}
%Описание карт и профилей
Figure \ref{pic:R_all} in Appendix~\ref{Append} shows ice abundance maps with respect to H$_2$O. Pixels with uncertainty larger than 30\% have been removed. The distribution of abundances differs from the distribution of column densities (increase toward the center of the protostar), although the zones of ice evaporation are also visible in the abundance maps. Since the desorption temperature of water \citep[100--110~K at the heating rate of 0.01~K~yr$^{-1}$, typical for protostars;][]{Collings2004} %\citep[$\sim$140~K, ][]{Fraser2001} 
is higher than that of other molecules, such as CO and CO$_2$, we observe a decrease in the abundances of CO, CO$_2$, H$_2$CO, and OCN$^-$ around the peaks of ALMA intensity in HOPS-56, HOPS-60, and HOPS-108. HOPS-91 and HOPS-96 do not have pronounced abundance peaks, which indicates the influence of a cold young envelope. We observe a peak in the abundance of NH$_4^+$ toward HOPS-56, HOPS-60 and HOPS-108. The difference between the two groups of sources may indicate different evolutionary ages of the envelopes and different contributions of the outer and inner envelopes. As the protostar evolves, its envelope becomes denser, heats up from the central source, and undergoes structural changes. All of this affects the chemical composition of the ice on the dust surface. CO abundance in HOPS-91 and HOPS-96 is larger compared to the other sources, which also demonstrates more significant contribution of the external envelope in absorption. The pixels at the edges of the maps show high abundances due to uncertainty of $\sim$29\%. %HOPS-73 has peaks of abundance in all types, which is associated with low column density of H$_2$O compared to near pixels. In our fit, we do not take into account the small contributions of formaldehyde and ammonia groups at the 6.3 $\mu$m absorption peak. At the same time, in this source, we note a broadening of the water peak, which indicates a temperature >20K.

\begin{figure*}
\centering
\includegraphics[scale=0.38]{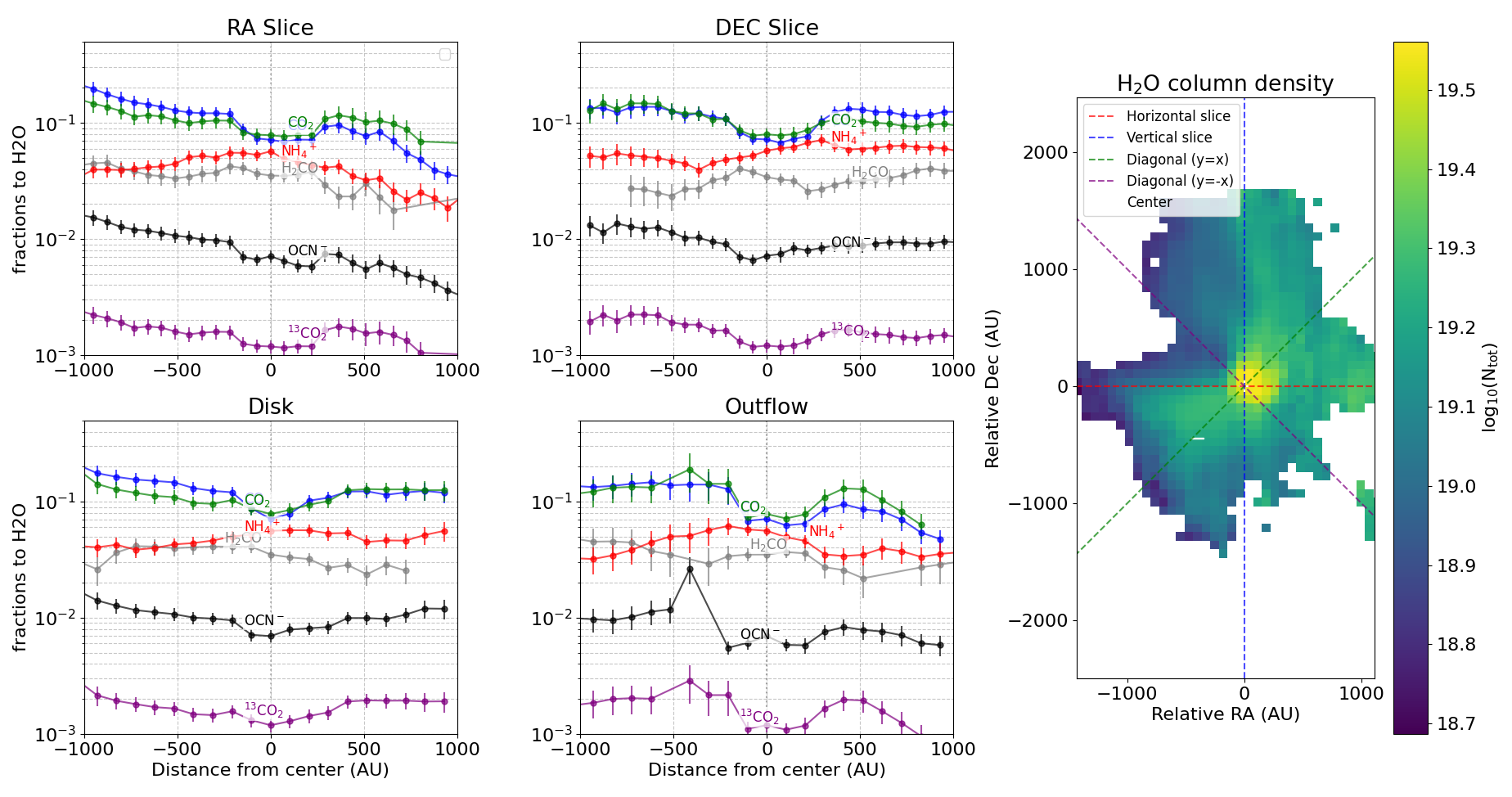}
\caption{Slices of species abundance in four directions of the protostar HOPS-60.} 
\label{pic:ab_H60}
\end{figure*}

We plotted the abundance slices with respect to H$_2$O along the RA and DEC axes and with slices along the outflows and the disk plane of HOPS-60 (Fig. \ref{pic:ab_H60}) and the abundance slices along the RA and DEC axes for HOPS-56, 91, 96 and 108 (Fig. \ref{pic:ab_sli} in Appendix~\ref{Append}). We observe a decrease in the abundances around the ALMA intensity peaks compared to the positions 400~AU away from it in the slices of HOPS-56, HOPS-60, and HOPS-108 in all plotted directions. Presumably, this decrease in the abundances is associated with the sublimation of the molecules, with the exception of water \citep[][]{Boogert2015}. Moreover, HOPS-91 and HOPS-96 do not have such decreases in the abundances, which indicates prevalent contribution from the outer envelope. The greatest change in the abundances is seen in HOPS-60, where they decrease by a factor of $\sim$2, while for HOPS-56 and HOPS-108 the abundances decrease by a factor of $\sim$1.7. We note a similar decrease in H$_2$CO abundances in HOPS-96. The HOPS-91 and HOPS-96 slices have peaks of CO abundance toward the centers of ALMA intensity, which show the highest density. At the same time, we observe a peak in CO in HOPS-108, which may indicate the perpendicular position of the disk. There is also an increase in the CO$_2$, OCN$^-$, NH$_4^+$ abundances caused by a decrease in H$_2$O column density towards the continuum peak. This might indicate strong water evaporation caused, for example, by an accretion burst. Although we do not fully resolve the accretion disk in observations, the difference between the inner and outer envelopes near the peak intensity reflects the effect of heating of the inner envelope. Based on this, we assume that the visibility of the heating effect depends on the visibility and differences between the inner and outer envelope during evolution. In our observations, we cannot distinguish between the inner envelope (closer to the protostar) and the outer envelope (farther from the protostar) in the line of sight. Although $A_V$ may be the same in two different sources, the contribution of the warmer inner envelope and the colder outer envelope may differ. It is precisely this effect that we can observe in our sources in the example of the abundance differences. This difference also directly indicates different stages of evolution within Class~0. This effect of ice evolution was described in \citet{Herbst2009} and \citet{Chen2024} based on chemical models and pointing gas and ice observations, and now we have contributed the picture with the ice maps. 

%As shown in the column density maps, HOPS-73 has a more complex structure of dust and outflows. The slices show two peaks of abundance decline on both sides of the accretion disc. As we have already mentioned, we do not take into account the absorption of less common molecules in this range. Due to these species, we overestimate water in our other sources. Thus, HOPS-73 has a hotter environment in which volatile molecules have left the dust surface. We also do not rule out the possibility that there may be an additional source of emission in the form of outflows on the beam. According to observations \cite{Furlan2016}, this source has low luminosity ($\sim$1.5 L$\bigodot$) and temperature ($\sim$42K) compared to our other sources, which may indicate that we are seeing the absorption of light received from outflows. At the same time, the slice directed to the outflow indicates a rapid increase in CO$_2$, which we also see in other slices in places subject to outflows.

\subsection{Radiative transfer model}\label{app:model}

\begin{figure*}
\centering
\includegraphics[scale=0.25]{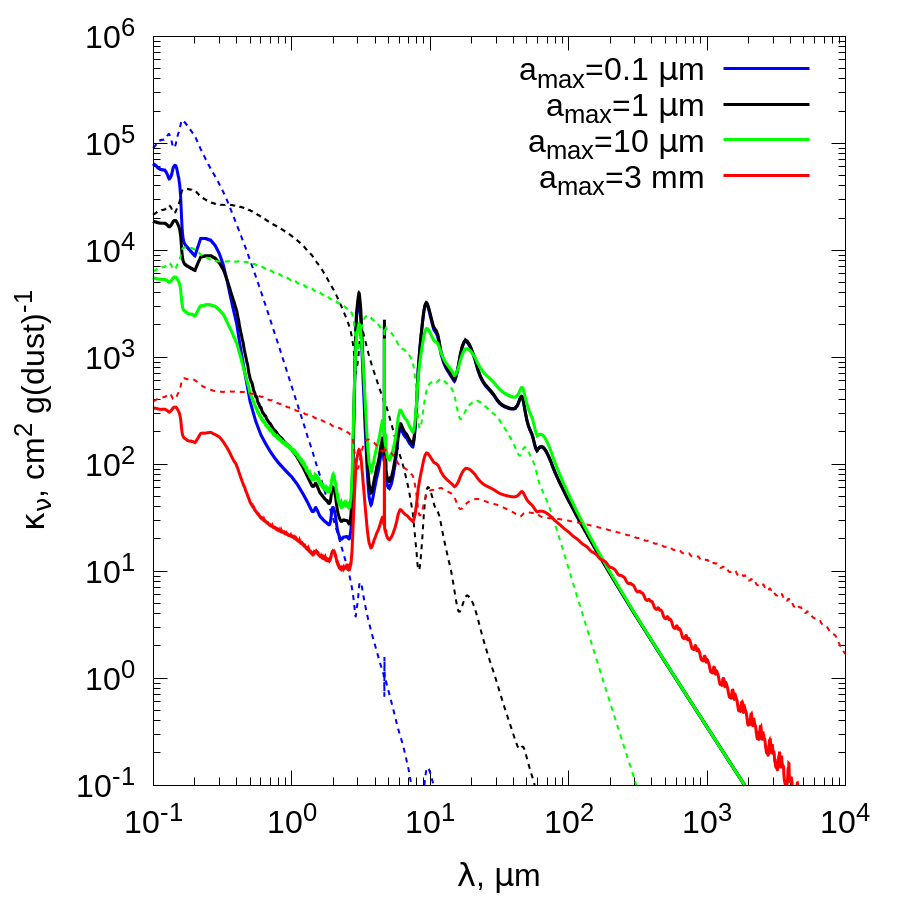}
\includegraphics[scale=0.25]{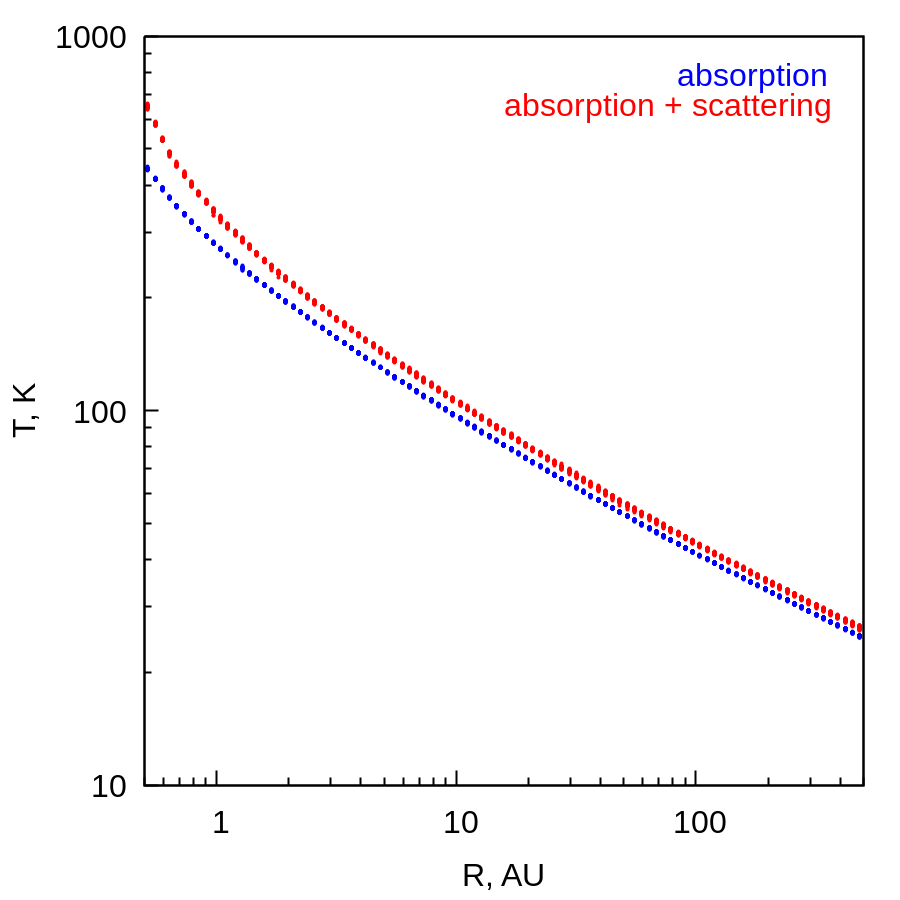}
\caption{Left: absorption and scattering coefficients calculated using \textsc{OPTOOL} code for a number of grain dust distributions with different maximal grain sizes $a_\text{max}$. Absorption and scattering coefficients are shown with solid and dashed curves, correspondingly. Right: temperature distributions calculated for the model where both absorption and scattering are accounted (red line), and for the model with only absorption (blue).} 
\label{pic:Appendix-B1}
\end{figure*}

\begin{figure*}
\centering
\includegraphics[scale=0.14]{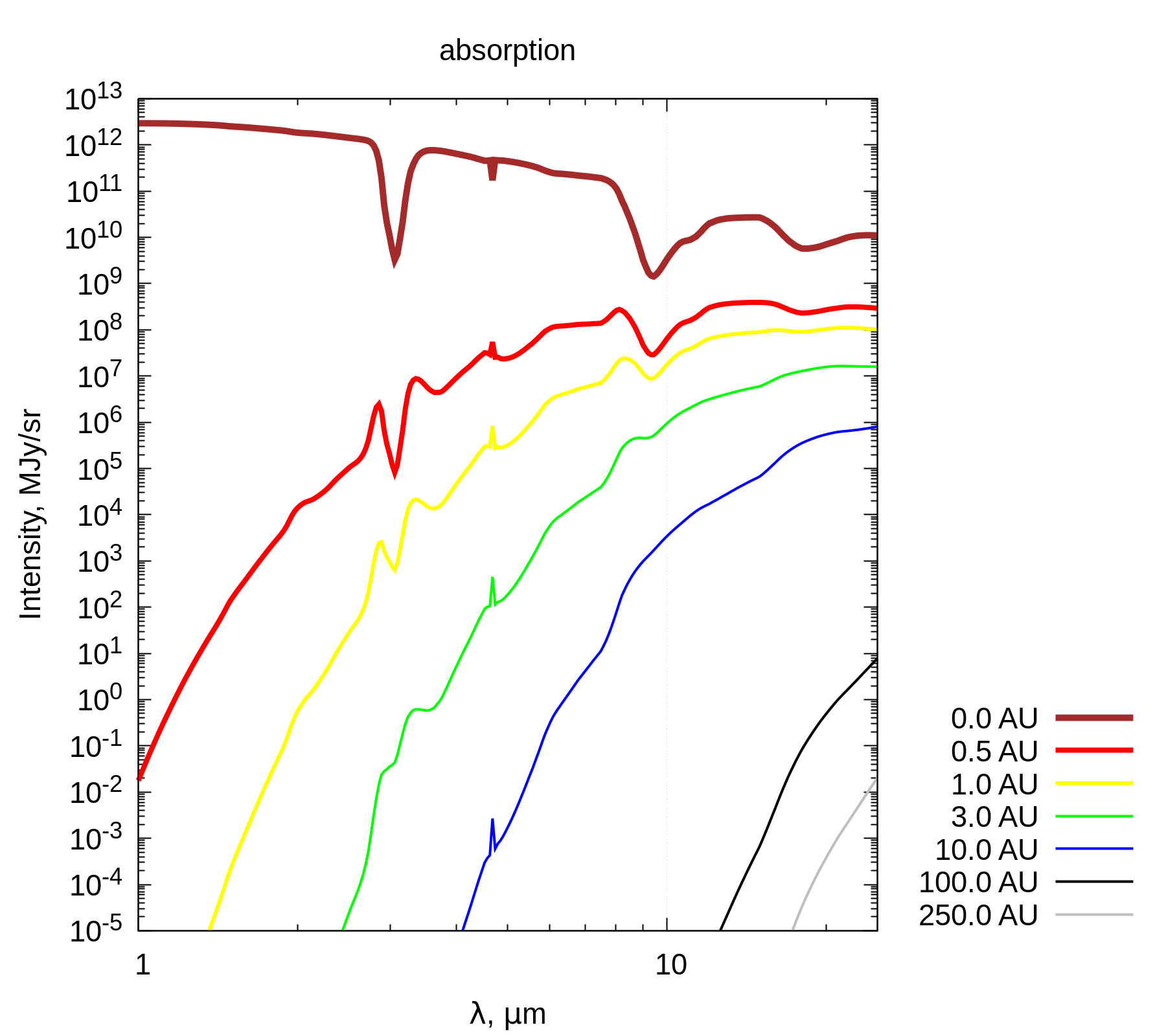}
\includegraphics[scale=0.14]{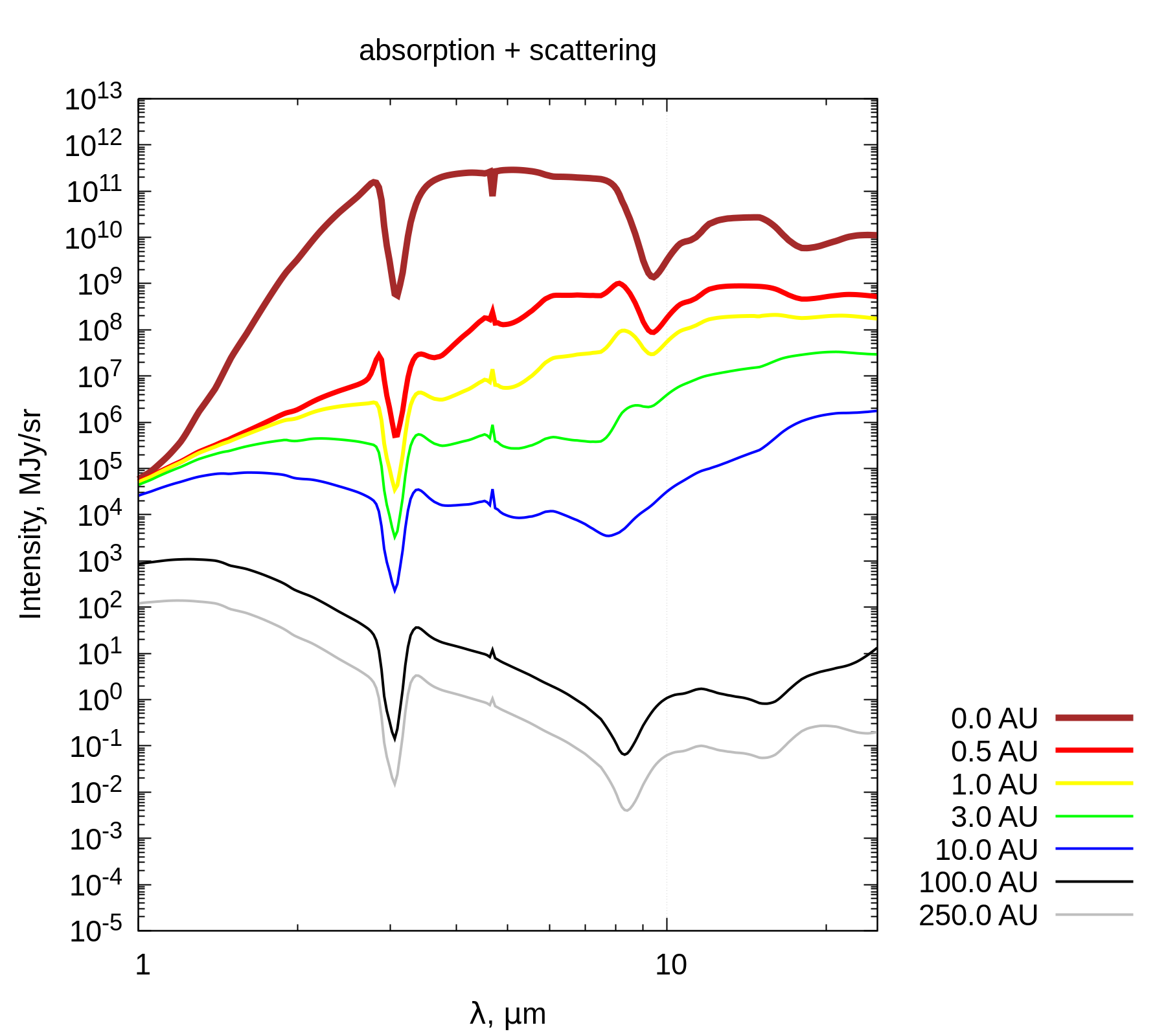}\\
\includegraphics[scale=0.14]{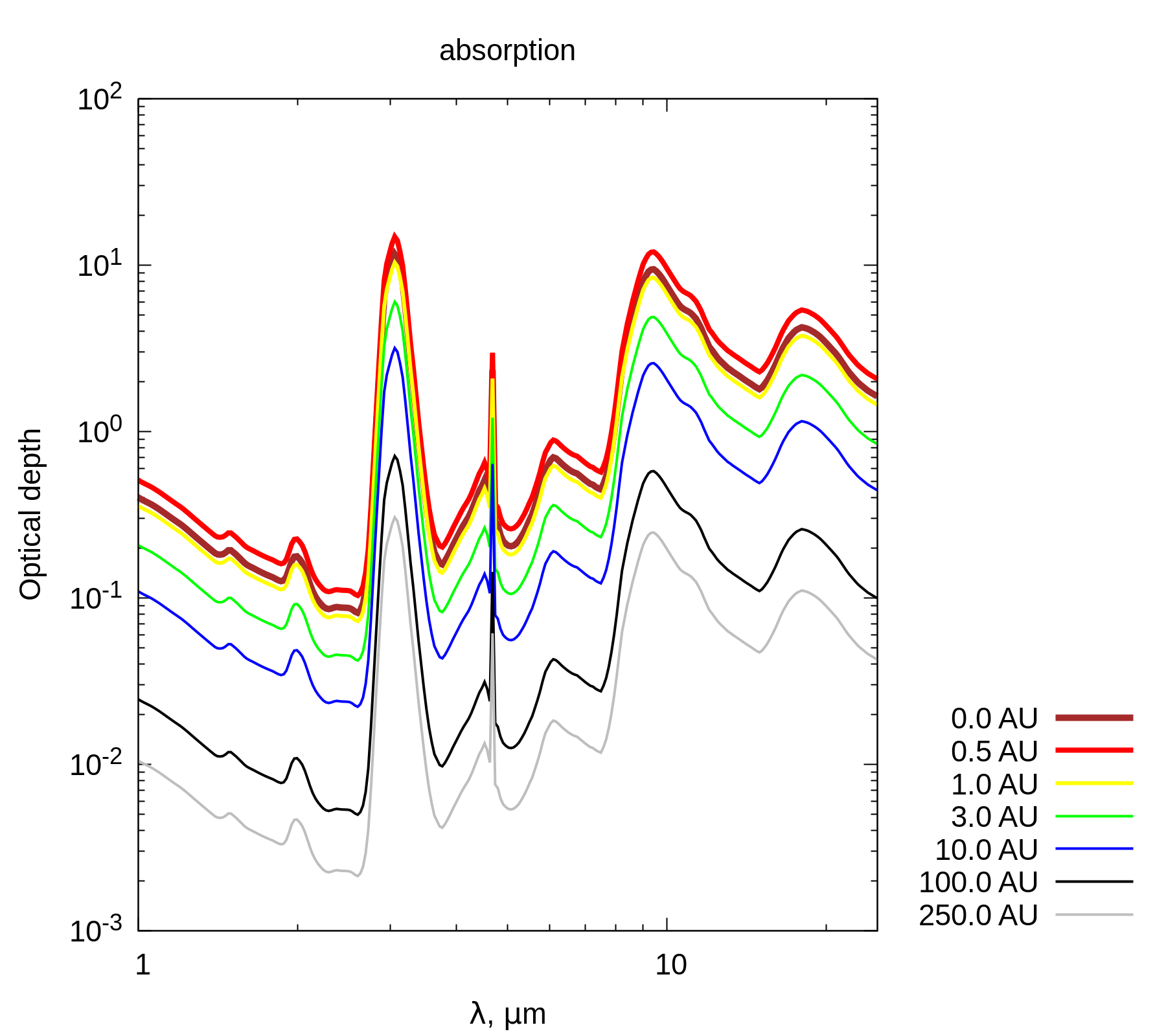}
\includegraphics[scale=0.14]{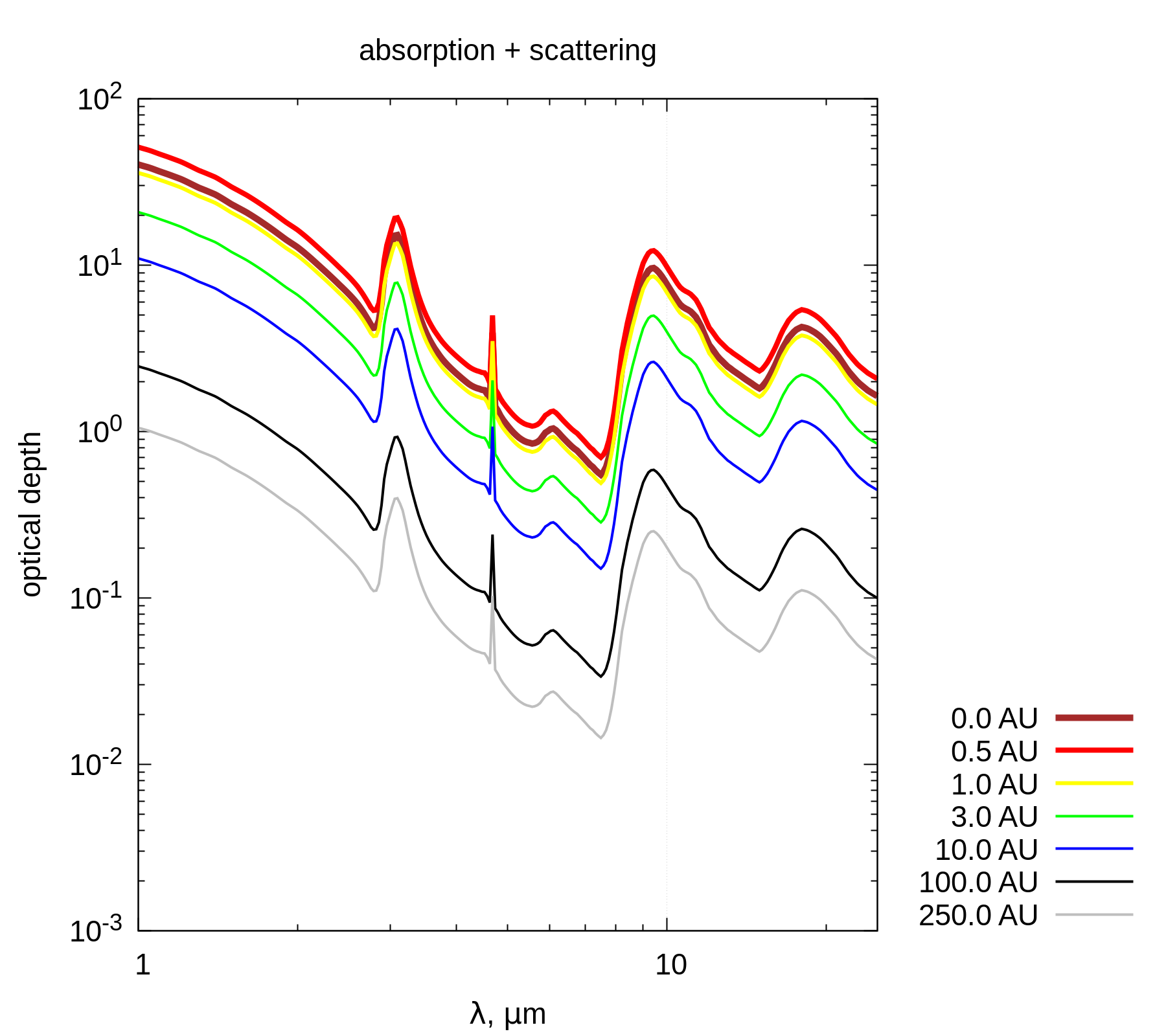}\\
\caption{{\it Top line:} Spectral energy distributions at different impact distance for the model with only absorption (left panel) and for the model with both absorption and scattering (right panel). {\it Bottom line:} Spectral distributions of optical depth at different impact distance for the true absorption model (left panel) and full model (right panel). The corresponding impact distances are shown in color legend.} 
\label{pic:Appendix-B2}
\end{figure*}

The calculation of the ice column density performed in this study is based on several key assumptions. Namely, we assumed the absorption to take place over some background emission. However, the emission behind and outside the absorption peaks is likely to originate in the cloud itself. Second, we completely neglected light scattering which may dominate over absorption in the considered wavelength range. Finally, the convolution with telescope beam (which we did not consider explicitly) may strongly affect conclusions on radial structure of the cloud. 

To illustrate all these factors and to inspect the restrictions of our conclusions, we provide here the results of radiative transfer simulations for a model spherically symmetric protostellar core. The core has the power-law density distribution $n(\text{H}_2) = 10^{9} \text{cm}^{-3}(r/\text{AU})^{-3/2}$ between 0.5 AU and 500 AU. The corresponding H$_2$ column density from the star to the core border is $4.3\times 10^{22}$~cm$^{-2}$, the mass of the envelope is $5\times 10^{-4}$ $M_{\odot}$.

The cloud is heated by central star with solar radius and black body spectrum with $T_\text{eff}=5780$~K. The adopted dust opacities are produced using \textsc{OPTOOL} package\footnote{OPTOOL: https://github.com/cdominik/optool?ysclid=mj6xzo7lg227325696}~\citep{2021ascl.soft04010D} for a mixture of dust grains including silicate, water and carbon monoxide given the MRN-like \citep[][]{MRN1977} size distribution $n(a)\propto a^{-3.5}$ with  $a_\text{min}=0.05$~$\mu$m and $a_\text{max}=1~\mu$m\footnote{The opacities were produced with the following command: optool pyr-mg70 1 -m h2o-w 0.2 -m co-a 0.04 -a 0.05 1.0 3.5 40 -l 0.1 1e4 -nlam 100000 -mie}. These opacities are shown with black color in the left panel of Fig.~\ref{pic:Appendix-B1}. Note that for the selected dust model the scattering dominates over absorption in the short wavelengths up to $8$~$\mu$m. The dust opacities are kept constant over the cloud i.e. we do not take any effects of dust processing. The dust-to-gas mass ratio is 0.01.

We use \textsc{RADMC-3D} code\footnote{RADMC-3D: https://www.ita.uni-heidelberg.de/~dullemond/software/radmc-3d/} \citep[][]{Dullemond2012} to calculate the self-consistent dust temperature and spectral energy distributions. The obtained results have been also double-checked using the \textsc{NATALY} code \citep[][]{2012MNRAS.421.2430P}. The temperatures and SEDs are calculated for two cases: 1) when both absorption and scattering are taken into account; 2) when scattering is neglected.  The corresponding temperature distributions are shown in the right panel of Fig.~\ref{pic:Appendix-B1}. We see that temperature profiles are close to the power-law while scattering does not strongly affect the temperature.

The calculated spectral energy distributions between 1 and 25 $\mu$m toward different positions in the core are shown in Fig.~\ref{pic:Appendix-B2}. In fact, the spectra look different for the cases with and without taking scattering into account (left and right panels of Fig.~\ref{pic:Appendix-B2}, respectively). The intensity of the SED toward the star surface (labeled as ``0.0~AU'') for $\lambda<8~\mu$m is significantly higher in the case when scattering is turned off (top left panel of Fig.~\ref{pic:Appendix-B2}) since the optical depth to scattering at these wavelengths is few times higher than to absorption ($\tau_\text{tot}=40$ vs $\tau_\text{abs}=0.3$ at $\lambda=1$~$\mu$m), see bottom panels of Fig.~\ref{pic:Appendix-B2}. In the non-scattering case, the intensity of the SED decreases rapidly with impact distance (shown with color and line thickness in Fig.~\ref{pic:Appendix-B2}). In the full (scattering + absorption) model, the SED variations are much smoother, whereas the profiles themselves look flatter. In the full case, the absorption features are present for all the impact distances, while in the non-scattering case, they disappear outside 1~AU. Physically, these differences are explained by the diffusion of the stellar and intrinsic emission via (multi-)scattering over extended parts of the cloud. Thus, the scattered emission acts as a ``background''\, for the absorption features to appear.

\begin{figure*}
\centering
\includegraphics[scale=0.28]{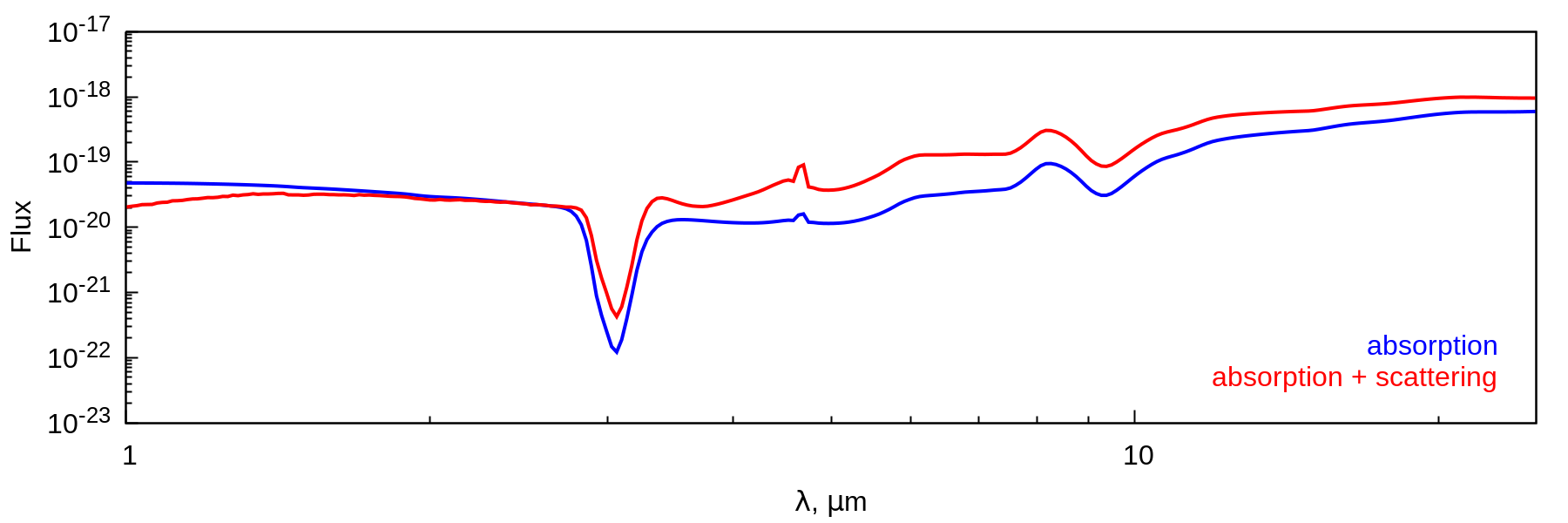}
\caption{Total (integrated over the entire core of 500~AU, equivalent to our 1.16$^{\prime\prime}$ which is $\sim$4 of our FWHM) spectral energy distributions for the models with and without scattering.} 
\label{pic:Appendix-B4}
\end{figure*}

Despite the tremendous differences in the SEDs toward specific impact distance, the total fluxes (SEDs integrated over the entire source) for the non-scattering and scattering cases look similar, see Fig.~\ref{pic:Appendix-B4}. In both models, the profiles are relatively flat, and absorption feature at 3$~\mu$m is prominent.  
The similarity between the total SEDs can be understood in terms of energy conservation. In the non-scattering case, the direct stellar emission which relatively freely escapes the core provides the significant (dominant up to 8~$\mu$m for our model) contribution into the total SED. In the model with scattering, the stellar light does not freely escape the core but diffuses over it and leaves the core via multi-scatters from more distant positions. So, the procedure of beam convolution in the non-scattering model would have the same effect as diffusion of scattered emission in the full model.

Given the illustrated importance of the light scattering and convolution, it would be useful to derive optical depth and H$_2$ column density from the obtained SEDs using the fitting procedure adopted in the paper. Below we perform it for the ideal (non-convolved) SED toward the core center calculated for non-scattering model (see brown profile in the top left panel of Fig.~\ref{pic:Appendix-B2}), and for the total SED calculated for the model with both absorption and scattering (see red profile in Fig.~\ref{pic:Appendix-B4}). The first case is more consistent with the assumptions of the fitting model, while in the second case the applicability of the fitting is not obvious a priori.

\begin{figure*}
\centering
\includegraphics[scale=0.4]{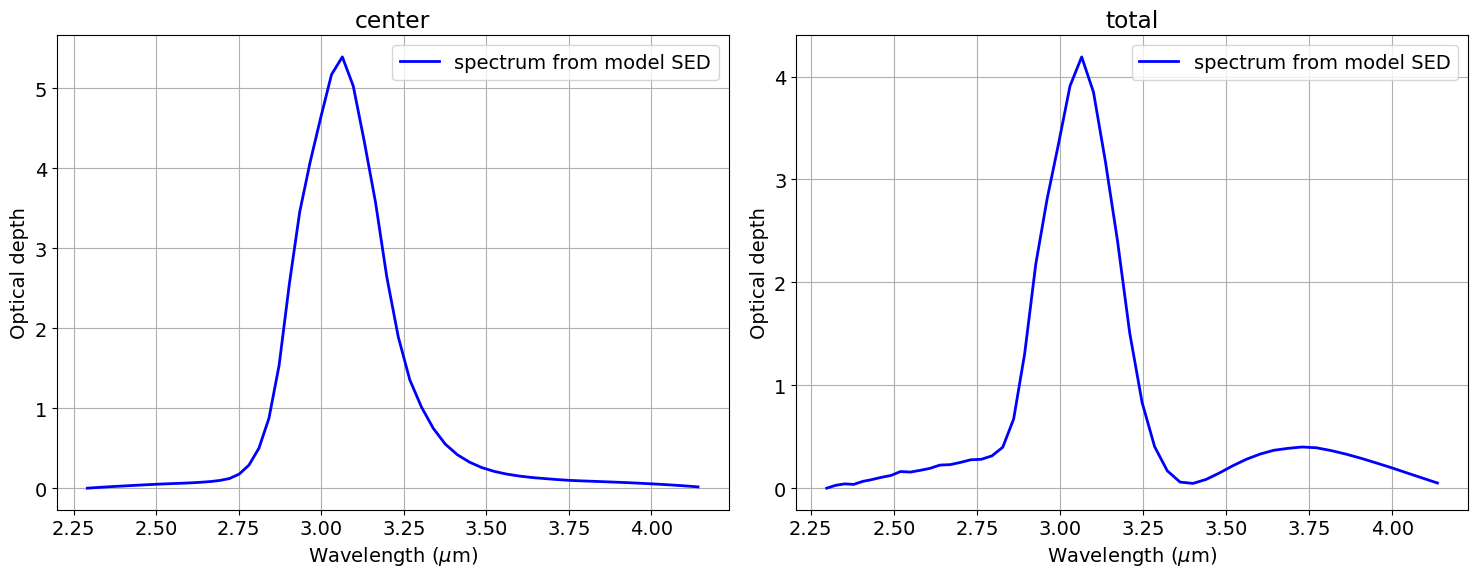}
\caption{The optical depth spectra derived by the fitting procedure. {\it Left panel:} for the central non-convolved SED without scattering. {\it Right panel:} for the total SED with absorption and scattering.} 
\label{pic:Appendix-B5}
\end{figure*}

We perform all the steps of our fitting procedure for the water absorption band at 3~$\mu$m treating the calculated SEDs as observed. The ``continuum''\ curve around 3~$\mu$m is derived using the linear fit through reference points at 1.5, 2.7, 3.6, and 4.5 $\mu$m. The derived optical depth profiles for the central and the total SEDs are shown in Fig.~\ref{pic:Appendix-B5}. The peak values ($\tau\approx 5$ for the central SED and $\tau\approx 4$ for the total SED) are about two times lower than the optical depth toward the cloud center ($\tau\approx 12$) obtained in the radiative transfer simulation, see brown profile in the bottom left panel of Fig.~\ref{pic:Appendix-B2}. Note, however, that the radiative transfer optical depth $\tau\approx 12$ is calculated through the entire cloud (from the back side to the front of the core) while the absorption of stellar radiation takes place over half of this way. So, both derived optical depths are consistent with the original radiative transfer value.
The column densities of water ice associated with the derived optical depths are $8.3\times10^{18}$~cm$^{-2}$ and $6\times10^{18}$~cm$^{-2}$ for the central and total SEDs, respectively. These values are obtained using the band strength of 2.2$\times$10$^{-16}$~cm~molecule$^{-1}$. Using the mass fraction of water ice (0.16) from the original \textsc{OPTOOL} opacities and dust-to-gas mass ratio (0.01), we obtain $N$(H$_2$)~= $4.8\times10^{22}$~cm$^{-2}$ for the central SED, and $3.5\times10^{22}$~cm$^{-2}$ for the total SED. These values are close to the input value of $4.3\times10^{22}$~cm$^{-2}$.

We suggest that relatively good agreement between the input and derived hydrogen column density is a consequence of using the dust opacities calculated for the MRN-like distribution ($n(a)\propto a^{-3.5}$) with relatively small maximal grain size, $a_\text{max}<\lambda/2\pi$. For small grains the absorption efficiency $Q_\text{abs}(a)\propto a$, and absorption coefficient weakly depends on the minimal and maximal grain sizes (see blue, green and black curves in Fig.~\ref{pic:Appendix-B1}). In this case, the absorption cross section of the dust grain is proportional to its volume. This case corresponds to the conditions at which the band strength is determined calculating the absorption by dust slices in laboratory experiments. If the maximal dust size is high enough ($a_\text{max}>\lambda/2\pi$, such as the absorption efficiency of large grains turns into $Q_\text{abs}(a)\approx 1$)
the absorption coefficient starts to deviate from the low-frequency limit, see red curve in Fig.~\ref{pic:Appendix-B1}. In the latter case, the transformation of optical depth into the dust column density $\Sigma(\text{dust})$ using the band strength will underestimate $\Sigma(\text{dust})$.
The effects of scattering also depend on the assumed dust size distribution. Thus, the increase of the maximum grain size expands the area where the scattering dominates absorption, see Fig.~\ref{pic:Appendix-B1}.

Given this example, we conclude that the adopted fitting procedure allows us to derive the appropriate column densities of the ice components for the entire cloud but should be used with caution to study their spatial distributions. The obtained result is consistent with other studies on radiation transfer in protostars \citep[e.g.,][]{Dartois2022} and does not contradict the influence of scattering on absorption bands \citep[e.g., ][]{Ehrenfreund1997,Dartois2024}.

\section{Discussion}\label{sec:discussion}
\subsection{Global chemical inventory}
Our results can be directly compared with recent astrochemical models of ice formation in dense cores and protostellar envelopes, such as those presented by \citet{Jimenez-Serra2025} and by \citet{Borshcheva2025}. These studies predict that interstellar ices are dominated by H$_2$O, with CO and CO$_2$ as the main carbon-bearing species, contributing $\sim$30-45\% and $\sim$13-20\% relative to H$_2$O, respectively, while CH$_4$ as well as CH$_3$OH and NH$_3$ provide secondary contributions at the level of a few percent. %Grain-surface chemistry, particularly nondiffusive reactions in prestellar cores, is critical for the formation of CH$_3$OH and NH$_3$ prior to protostellar heating. 
When expressed as fractions of the total ice budget, this corresponds to $\sim$55-70\% for H$_2$O and $\sim$30-45\% in all other species.

In our sample, the non-H$_2$O species account for $\sim$20-40\% of the total ice. This places our sources within the range predicted by the models, although generally toward the H$_2$O-rich end of the distribution.
The models presented in \citet{Jimenez-Serra2025} suggest that our observations recover a substantial fraction ($\sim$50-90\%) of the non-water ice reservoir, particularly the dominant carbon-bearing species CO and CO$_2$, and water. The remaining discrepancy likely reflects the absence of key ice components such as CH$_3$OH and NH$_3$ in our analysis. As \citet{Borshcheva2025} show, these molecules are formed efficiently via nondiffusive grain-surface reactions in dense cores, representing a significant portion of the total ice budget and the nitrogen reservoir. Overall, the observed ice composition is consistent which the bulk of the ice is established during the prestellar phase and subsequently inherited by the protostellar envelope. Our data estimate 90\% of all ice on the surface is dust including H$_2$O, CO and CO$_2$. %Про 90 проц всего льда и камету 
%Our data capture most of the major carbon reservoirs but only part of the full chemical complexity predicted by contemporary astrochemical models.

\subsection{Possible scattering effects}
In addition to the discussed above, the emission observed near the center of a protostar is a subject to scattering. To estimate the effects of scattering, we perform radiation transfer modeling (see Section~\ref{app:model}). The model that includes scattering shows that the scattering effects are significant. However, our angular resolution resembles diffusion of the scattered light, and the model with scattering when considering the entire emission is consistent with the model without scattering (see Section~\ref{app:model} for details). The spatial resolution of our maps does not allow us to resolve the accretion disk and the central object of the protostar. Switching to the envelope, we observe an increase in the source size with long wavelengths (NIRspec and MIRI MRS), similar to the increase in the beam.

Background emission is one of the factors that allows the study of protostellar envelopes. Earlier, in other sources discussed in \citet{Brunken2024,Chen2024,Rocha2024,Rayalacheruvu2025}, where the background emission depended on the distance to the center of the source and was caused by scattering. \citet{Slavicinska2025_NH4} and \citet{Tyagi_2025} show sources where the background emission comes from the outflow. In our case, the background emission is the re-emission of dust or PAHs under the influence of UV background emission.

In the current paper, we have restricted ourselves to use the Beer-Bouguer-Lambert extinction law as a main instrument to derive the ice column densities. The radiative transfer model was only used to inspect the restrictions of our approach by considering the spherically symmetric core as a simplest source prototype. The outcome of our radiative transfer modeling is that the adopted fitting procedure allows us to derive the appropriate column densities of the ice components for the entire source but caution is required when drawing conclusions about the actual ice spatial distributions. The only way to make the conclusions (e.g. about the dominant regions contributing to the observed ice absorption) more reliable is to construct the detailed model of the source including the likely presence of a star, accretion disk, envelope, polar cavities, and considering the effects of ice sublimation, radiative scattering, spatial orientation, etc. The construction of such a model and its examination would be too heavy in the frame of this paper but start to be addressed by other studies, see e.g. \citet{Thompson2026}.

%Добавить про H2o:CO2:CO
\subsection{Possible overestimates of H$_2$O and H$_2$CO and underestimates of CO column densities}
One of the difficult problems in our work was fitting the region of 5.0-7.4 $\mu$m. In addition to the molecules we studied, there is a large number of molecules with low absorption contributions, such as NH$_3$, HCOOH, etc. We did not take them into account in the fit because it is difficult to distinguish these molecules without additional absorption peaks. In our case, with pixel processing and source mapping, we are limited by two conditions that prevent us from separating water for the analysis of the other molecules in this absorption range. Class~0 protostars have low emission in the range below 3.5~$\mu$m, which makes the H$_2$O band at 3~$\mu$m unavailable for analysis due to its optical depth. Analysis of the water spatial distribution through 15~$\mu$m water band is not possible due to the low spatial resolution of the channels. 
% Кеани 2001 и 

As \cite{Chen2024} and \cite{Slavicinska2025} show, the approach using different absorption bands allows for more accurate determination of the column density of H$_2$O, but without spatial distribution. The absorption bands from COMs and HCOO$^-$, which affect the fitting of the water band, are present toward the ALMA continuum peak (see Fig. \ref{pic:spec_fit}). Our overestimate of H$_2$O column density in all pixels is less than 10\% for HOPS-56, 10\% for HOPS-60, and 6\% for HOPS-108. The uncertainty of the abundances with respect to water takes into account the overestimation of H$_2$O.

%Our approach overestimates H$_2$O column density by up to $\sim$10\% due to the additional absorption by other molecules, but allows us to construct a spatial distribution of the column density. We tried to minimize the overestimate by considering H$_2$CO, which has a significant contribution to the water absorption band \citep[][]{Boogert2015,Rocha2024}. We observe its significant contribution to the absorption band in the sources with a dense envelope, HOPS-91 and HOPS-96.

The absorption peaks of CO in HOPS-91 and HOPS-96 have optically thick lines. This effect is enhanced at distances $>$400~AU from the center of the ALMA intensity peak. Consequently, our estimates of column density and abundance of CO in this sources represent the lower limit for pixels located more than 400~AU away. In addition, the chosen $^{12}$C/$^{13}$C ratio of 66 may underestimate CO$_2$ values due to possible fractionation and a decrease in $^{13}$C in the species in the sources \citep[][]{Brunken2024_1}. %Therefore, the underestimation will be 5-10\% for the pixels closer then 400~AU to the ALMA peak and 20-40\% for the other pixels. For this estimation, we performed an additional fitting using the wing method for a Gaussian of the same width described in \cite{Smith2025}.

\subsection{The evolution stage or different envelope?}\label{discussion:evolution_stage}
Our maps and abundance slices with respect to H$_2$O show a decrease in abundances of CO, CO$_2$ and H$_2$CO and increase in NH$_4^+$ toward the ALMA intensity peaks in HOPS-56, HOPS-60, and HOPS-108. HOPS-91 and HOPS-96 do not have a significant decrease in the abundance. Based on  this difference we split our sources in two groups. For our protostars, we assume certain temperatures, luminosities (see Table~\ref{tab:coord}) and structures of the envelope. All these factors indirectly indicate different stages of Class~0 protostars. The large discrepancies in the parameters of the envelope, disk, and central object obtained through the Spizer/Herschel and ALMA SED modeling do not provide specific information about the status within Class~0 \citep[parameters of the HOPS sources are taken from][]{Furlan2016,Tobin2020,Sheehan2022}. The ALMA data provides information only about the central parts of the protostellar source -- accretion discs, while the IR Spitzer/Herschel SEDs having lower resolution include emission from the outflows. We observe an increase in the absorption of pure $^{13}$CO$_2$ compared to $^{13}$CO$_2$ mixed with H$_2$O toward HOPS-60 and HOPS-73, in its central area and the zones where the column densities of all ices are decreased possibly due to ice mantle heating by the outflow (see left top quadrant for HOPS-60 in Fig.~\ref{pic:N_H60} and right top quadrant for HOPS-73 in Fig.~\ref{pic:N_H73} for the column densities and Fig.~\ref{pic:CO2_clean} for the $\tau_{\rm ^{13}CO_2}/\tau_{\rm ^{13}CO_2:H_2O}$). The ratio in HOPS-108 is elevated but uniform (see Fig.~\ref{pic:CO2_clean}). Such segregation of $^{13}$CO$_2$ indicates the impact of protostars heating on ice \cite{Tyagi_2025}.  We assume that there is a different ratio of warm ($T\sim$20~K) inner envelope and cold ($T\sim$10~K) outer envelope on the line of sight. Figure \ref{pic:CO_CO2} shows the maps of the ratio of column densities of CO and CO$_2$ toward all sources. The ratio varies from 0.5 to 2.5 and generally increases towards the ALMA continuum peaks. We suggest that the increase in the CO/CO$_2$ ratio toward the protostars in all sources except HOPS-60 is associated with active CO freeze-out \citep[][]{Boogert2015,Caselli2022}. At the same time, HOPS-60 has the CO/CO$_2$ ratio $<$ 1 that increases toward the flared wing-shaped regions of the envelope structure, which is associated with heating. That is, we do not rule out the heating and desorption processes of the inner regions of HOPS-91 and HOPS-96, but due to the high density of the outer envelope, this abundance decrease is masked by the absorption of the outer envelope. This is also indicated by the large difference in CO abundance between the two groups of protostars (factor of $\sim$2--4), which is most sensitive to  the moderate heating.

Based on the assessment of the column density of molecular hydrogen, we distinguish two groups of sources that differ within Class~0. The first group is HOPS-56, HOPS-60, HOPS-73, and HOPS-108, in which signs of temperature influence on ice are observed, resulting in a noticeable decrease in CO abundance and an increase in the salt (NH$_4^+$) abundance. The second group is HOPS-91 and HOPS-96, in which no changes in abundances are observed and the distribution is spherically symmetric. That is, we do not rule out the heating processes of the inner envelope in HOPS-91 and HOPS-96, but due to the high density of the envelope, this decreased abundance is masked by the absorption of the outer envelope. This is consistent with the theory of protostellar evolution, where the mass of the envelope decreases and its temperature increases over time. Therefore, the visibility of this effect is directly related to the column density ratio of the inner and outer envelopes and, consequently, to the time of evolution.

\subsection{NH$_4^+$ and OCN$^-$ ice abundances}

We observe a distinct increase in the NH$_4^+$ abundance and decrease in the OCN$^-$ abundance along the slices of HOPS-56, HOPS-60 and HOPS-108. This seemingly leads to an increase in the negative charge deficit in the ices and contradicts the previous consideration of the OCN$^-$ anion as a counter-ion to NH$_4^+$ \citep{McClure2023, Slavicinska2025_NH4}. Additionally, the OCN$^-$ ice abundance is significantly lowered near the center of the slices toward HOPS-108 and HOPS-60, associated with heating of the ices, whereas the NH$_4^+$OCN$^-$ salt evaporates at $\sim$~200~K \citep{cottin2001stability}, which is significantly higher than the desorption temperature of H$_2$O ice. These two facts are consistent with the plausible origin of a portion of the OCN$^-$ anion from the dissociative ionization of HNCO in H$_2$O ice due to solvation effects -- mechanism supported by both laboratory experiments and quantum chemical calculations \citep{raunier2003reactivity, theule2011kinetics}. The reaction equation for this process is as follows: HNCO + H$_2$O $\to$ OCN$^-$ + H$_3$O$^+$. The species formed are more volatile than H$_2$O ice, when comparing the corresponding parameters of desorption kinetics: the binding energy of H$_2$O ice is $\sim$~5700~K \citep{minissale2022thermal} while that of dissociated HNCO is $\sim$~4800~K \citep{theule2011kinetics}. Taking this into account, lowered OCN$^-$ abundances in heated regions can be caused by desorption of dissociated HNCO. The observed decrease in OCN$^-$ abundance toward the center may also be caused by CO evaporation from the H$_2$O environment and a reduction in the contribution of the corresponding 2152 cm$^{-1}$ band \citep[][]{Al-Halabi_2004,Bouwman_2007}. This band although previously observed only in laboratory ice, may be present in HOPS due to the high abundance of CO in the polar environment. The latter is evidenced by the large width of the CO bands ($\sim$17~cm$^{-1}$).

In its turn, NH$_4^+$ abundance peaks in the heated regions can be assigned to the products of other effective acid-base reactions between icy NH$_3$ and acids, such as HCOOH, producing NH$_4^+$HCOO$^-$, CH$_3$COOH, producing NH$_4^+$CH$_3$COO$^-$ \citep[e.g.,][]{Rocha2024}, which are less volatile than H$_2$O ice \citep[][]{kruczkiewicz2021ammonia}. NH$_4^+$OCN$^-$ and NH$_4^+$HCOO$^-$ account for up to 20\% of NH$_4^+$, as shown by e.g., \citet{Boogert2015,Rayalacheruvu2025,Slavicinska2025_NH4}. However, the observed abundance growth by a factor of $\sim$2 can hardly be attributed by these reactions alone. The NH$_4^+$ band, generally, remains one of open questions: 80-85 $\%$ of this cation remain uncompensated by observed and expected anions \citep[][]{Slavicinska2025_NH4}. Future investigation of NH$_4^+$ counter-ions may better explain observed peaks of its ice abundance.  %\citep[][]{Slavicinska2025_NH4}. % The increase in band NH$_4^+$ may also contain the contribution of methanol (see Sec.~\ref{fit}), which is supported by the presence of bands of other COMs.

\subsection{PAHs and UV background}
The Orion A star-forming region is located near the young OI type star NU Ori, which creates elevated background UV emission. The UV background is estimated to be $G_0$ = (2.2 - 7.1)$\times$10$^4$~erg~cm$^{-2}$~s$^{-1}$~sr$^{-1}$ with the median of 5.9$\times$10$^4$~erg~cm$^{-2}$~s$^{-1}$~sr$^{-1}$ \citep[][]{Peeters2024}, which is several times higher than in other star-forming regions. Even taking into account the high embedding of the protostar, the UV background is very strong, which has an impact on dust and ice. The the presence of background re-emitted IR PAHs emission in HOPS-56, HOPS-60, and HOPS-108 is an indicator of background UV emission. We assume that UV emission has the greatest impact on the outer parts of the envelope and molecular cloud. We do not observe PAHs emission in the less dense parts of the protostellar structure, it is uniform across the entire frame. At the same time, the PAHs emission allows us to study ice absorption without saturating the absorption bands.

\subsection{Structure of the envelope of HOPS-56, HOPS-60 and HOPS-73}
Unlike the other sources, HOPS-56 (Fig. \ref{pic:N_H56} in Appendix~\ref{Append}), HOPS-60 (Fig. \ref{pic:N_H60}), and HOPS-73 (Fig. \ref{pic:N_H73} in Appendix~\ref{Append}) have additional dust structures outside the accretion disks and CO gas emission lines. HOPS-60 and HOPS-73 also have pronounced mantle evaporation in the outflows (see Fig. \ref{pic:N_H60} and \ref{pic:N_H73} in Appendix~\ref{Append}) coinciding with increased IR continuum emission (Fig. \ref{pic:continuum}). As shown in \cite{Dishoeck2025}, the structure of a protostar has inner envelopes that transit into accretion disks. We associate these dust structures with our ice distributions; CO and CO$_2$ maps best reveal these structures. At the same time, we note that these cold envelope structures are visible when studying ice absorption bands, rather than the emission of continuum. Therefore, ice distribution maps is a tool for studying the internal structures of the envelopes.

\cite{Tobin2020} show that HOPS-56 may have a more complex, unusual structure possibly consisting of three protostellar components. Here we do not see pronounced outflows and mantle evaporation, instead, we see the CO gas emission toward the area of enhanced CO ice column density outside the ALMA continuum (see Fig.~\ref{pic:N_H56} in Appendix~\ref{Append}), which indicates a different mechanism for the formation of CO emission lines. Therefore, we assume that this source either has a more complex kinematic structure of gas and dust or possibly marks the beginning of the CO outflow stage.

\section{Conclusions}\label{sec:conclusions}
We present spatial distribution of the ice chemical inventory of six Class~0 protostars in an active star-forming region, Orion~A. 
We processed the spectra for pixel-by-pixel analysis in the ranges from 4.3 to 4.9 and from 5.0 to 8.1~$\mu$m of IFU observations of NIRspec and MIRI MRS from JWST toward HOPS-56, HOPS-60, HOPS-73, HOPS-91, HOPS-96, and HOPS-108. The NIRspec observations were filtered to avoid distortion by CO emission lines. MIRI MRS observations were regridded to the spatial resolution and field of view of channel 2 to construct a single cube of spectra. We identified PAH emission in HOPS-56, HOPS-60 and HOPS-108 (foreground or background) that was uniform within the frames that allowed us to study the ices farther from the protostar. We showed that the CH$_3$OH contribution to the absorption band at 6.8~$\mu$m is negligible. We compared our abundances obtained toward the protostars to those obtained for the molecular cloud Cha~1 and other protostars observed with JWST and other IR facilities and analyzed through aperture extraction
\citep{Oberg_2011,Boogert2015,McClure2023,Chen2024,Rayalacheruvu2025}. 
We showed the applicability limits of the column density calculation using a radiation transfer model and model spectra of silicates and ice.
%We mapped column densities and abundances with respect to H$_2$O for $^{13}$CO$_2$, OCN$^-$, CO, H$_2$O, NH$_4^+$ and H$_2$CO and analyzed OCS, CH$_4$ and COMs absorption toward the ALMA continuum peaks. 
We analyzed the abundances of Class~0 protostars with different evolution stages and estimated the abundance distributions in the envelopes near accretion disks. Our conclusions are as follows:

\begin{enumerate}

\item We present maps of $^{13}$CO$_2$, OCN$^-$, CO, H$_2$O, NH$_4^+$ and H$_2$CO ice column densities, with the ice column density peaks mostly coinciding with the ALMA continuum peaks. We also present the column densities of OCS and CH$_4$ and analysis of COMs absorption bands toward the ALMA continuum peaks. The column densities are similar to those toward protostars in the Perseus star-forming region \citep[IRAS 2A and B1-c, ][]{Chen2024,Rayalacheruvu2025} and to other protostars and the cloud \citep[][]{Oberg_2011,Boogert2015}. This indicates that the chemical composition of the envelopes is the same in different star-forming regions.

\item We present the first $\sim$100~AU scale $^{13}$CO$_2$, CO$_2$, OCN$^-$, CO, NH$_4^+$ and H$_2$CO abundance maps with respect to H$_2$O toward Class~0 protostars. The obtained abundances are consistent with model and observed estimates for Class~0 protostars. At the same time, we observe that the abundance of CO is comparable to that of CO$_2$ in HOPS-56, HOPS-60, and HOPS-108, while in HOPS-91 and HOPS-96, the abundance of CO is 2--3 times larger than that of CO$_2$.

\item The abundance maps of HOPS-56, HOPS-60, and HOPS-108 show a decrease in the abundance of CO, $^{13}$CO$_2$ and H$_2$CO and increase in the abundance of NH$_4^+$ toward the peak of ALMA continuum. We present slices of abundances along RA and DEC axes for HOPS-56, HOPS-91, HOPS-96, and HOPS-108 and in outflow and disk for HOPS-60. Slices HOPS-56, HOPS-60, and HOPS-108 show a deficit in abundance at the ALMA continuum peaks, associated with the heating of the inner parts of envelope by protostars. We do not rule out that HOPS-91, HOPS-96 have the same effects, but the contribution of outer envelope absorption covers the internal structures for study, which can be used to study the outer envelopes of protostars.

\item We present for the first time a verification of the Beer–Bouguer–Lambert extinction law for estimating column density of the entire source for optical depth greater than one. The radiative transfer model, taking into account scattering with density distribution $n(\text{H}_2) = 10^{9} \text{cm}^{-3}(r/\text{AU})^{-3/2}$ and the different set of ice dust \textsc{OPTOOL} spectra, shows that this approach can be implemented for fine dust, with the maximum size of the dust particle $<1\,\mu$m.

\item We observe a decrease in ice column densities of all species toward the outflows in HOPS-60 and HOPS-73 indicating mantle evaporation. Our ice distributions correspond to the internal structure of protostars \citep[][]{Dishoeck2025}.

\item We present the first maps of the CO/CO$_2$ column densities ratio toward all sources. Maps of all sources except HOPS-60 show an increase in the ratio toward the center, indicating active CO freeze-out and CO$_2$ formation \citep[][]{Boogert2015,Caselli2022}. At the same time, HOPS-60 shows CO evaporation in the ALMA continuum peak and a broadening of the pure CO$_2$ band.

\end{enumerate}
The obtained estimates of the column densities and the abundances improve our understanding of the evolution of ice and dust in protostars and can be actively used to benchmark astrochemical models \citep[e.g., ][]{Jimenez-Serra2025,Borshcheva2025} and radiation transfer models \citep[e.g., ][]{Kargaltseva2022}.

\section*{Acknowledgments}
The authors thank the anonymous reviewer for excellent discussion and strengthening the conclusions of the study. This work is based on observations made with the NASA/ESA/CSA James Webb Space Telescope. The data were obtained from the Mikulski Archive for Space Telescopes at the Space Telescope Science Institute, which is operated by the Association of Universities for Research in Astronomy, Inc., under NASA contract NAS 5-03127 for JWST. The authors acknowledge Alexander Matveev and Andrey Ostrovsky for the stimulating discussions. The authors acknowledge the financial support of the Ministry of Science and Education of Russia, the FEUZ-2025-0003 project. The work associated with radiative transfer simulations (Section 5.4) was carried out under the state assignment of the Institute of astronomy of RAS.

%%%%%%%%%%%%%%%%%%%%%%%%%%%%%%%%%%%%%%%%%%%%%%%%%%
\section*{Data Availability}
All the JWST data used in this paper can be found in MAST: \dataset[10.17909/163h-vp81]{http://dx.doi.org/10.17909/163h-vp81}

\facilities{JWST, HST(SPIRE)}

\software{ \textsc{astropy} \citep{Astropy2013,Astropy2018,Astropy2022}.  
%          Cloudy \citep{2013RMxAA..49..137F}, 
%          Source Extractor \citep{1996A&AS..117..393B}
          }


\bibliography{sample631}{}
\bibliographystyle{aasjournal}

\appendix

% Written by Pavlyuchenkov: not finished yet!
\section{Tables and plots}\label{Append}
\begin{figure*}
\centering
\includegraphics[scale=0.55]{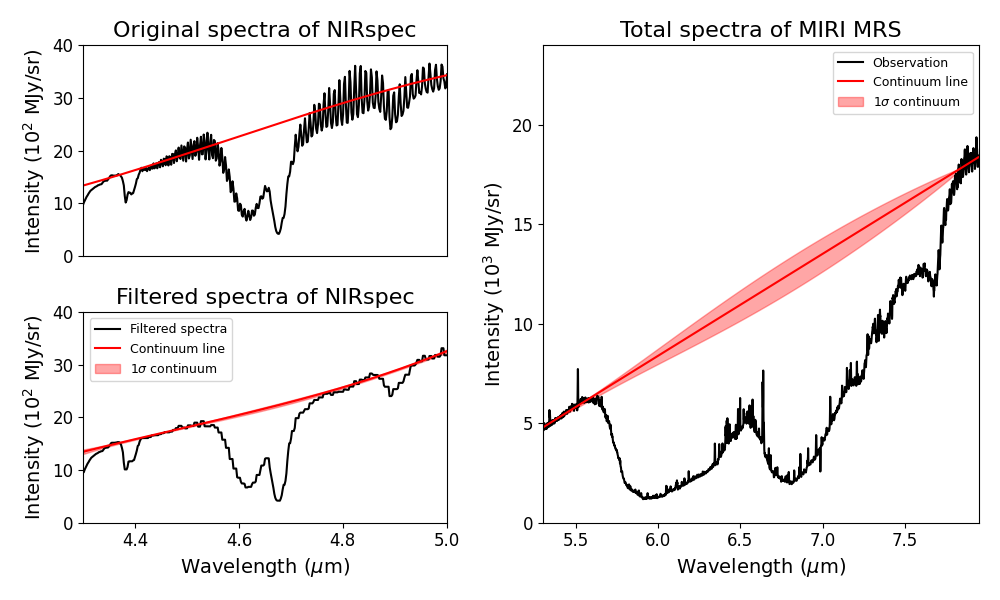}
\caption{Spectra in the pixel toward the ALMA continuum intensity peak of HOPS-60 before (top left panel) and after filter (bottom left panel) processing for NIRspec and the total spectrum for MIRI MRS (right panel). The red line shows the continuum lines for estimating the optical depth.} 
\label{pic:spec_filter}
\end{figure*}

\begin{figure*}
\centering
\includegraphics[scale=0.85]{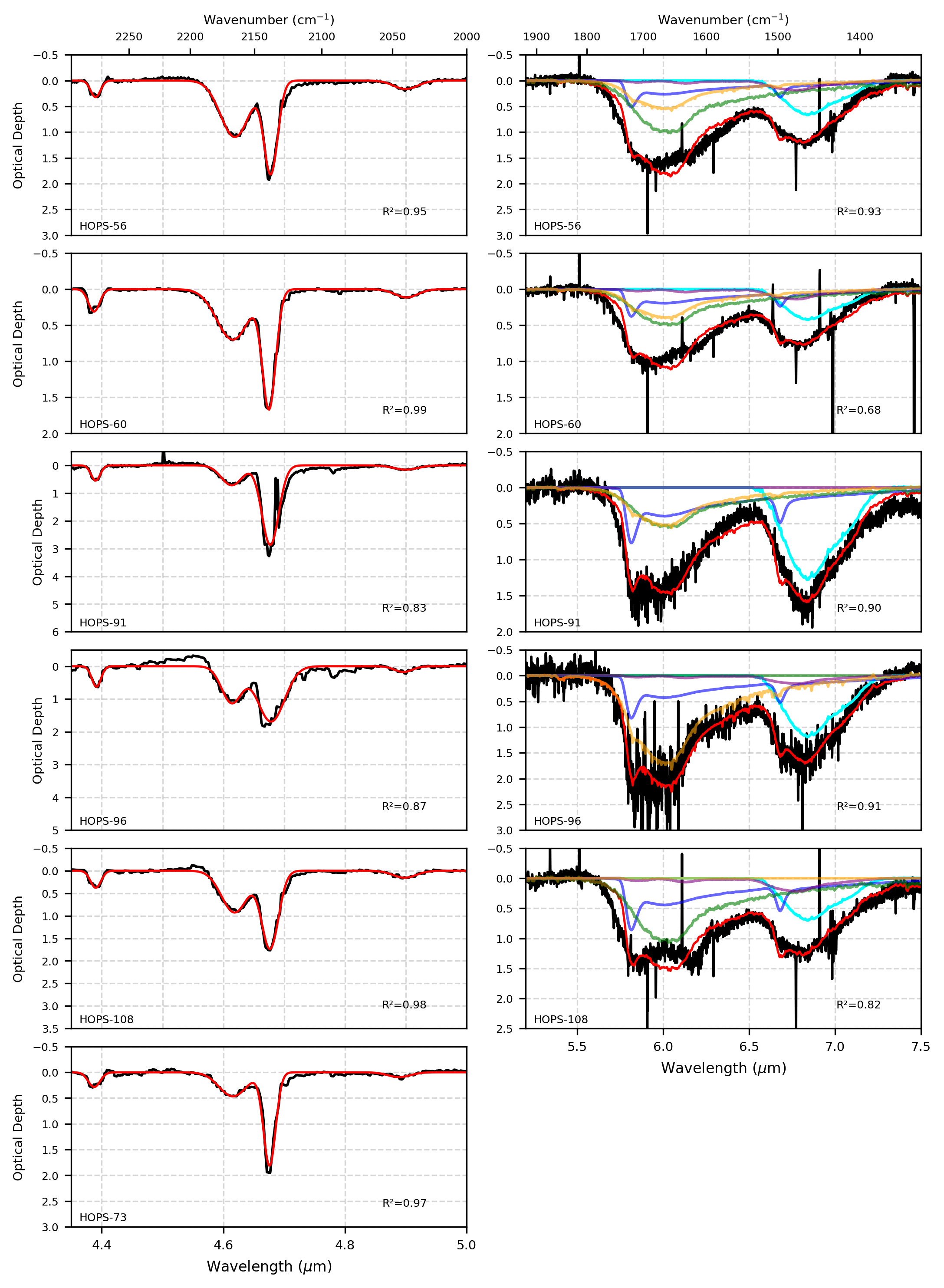}
\caption{Optical depth spectra toward the pixel 400 AU down a long declination axis from the ALMA continuum intensity peak of sources. The red line shows the best fit for each spectrum, the black line shows the optical depth spectrum. The R$^2$ of the fits is shown in the lower right corners. {\it Left panels:} Gaussian fits of $^{13}$CO$_2$, OCN$^-$, CO and OCS.
{\it Right panels:} Color lines show the components of the laboratory spectra of H$_2$CO:H$_2$O (blue), H$_2$O (green), NH$_4^+$ (15~K, purple), H$_2$O:CO:CO$_2$ (orange) and NH$_4^+$(120~K, cyan).} 
\label{pic:spec_fit_b}
\end{figure*}

\begin{figure*}
\centering
\includegraphics[scale=0.41]{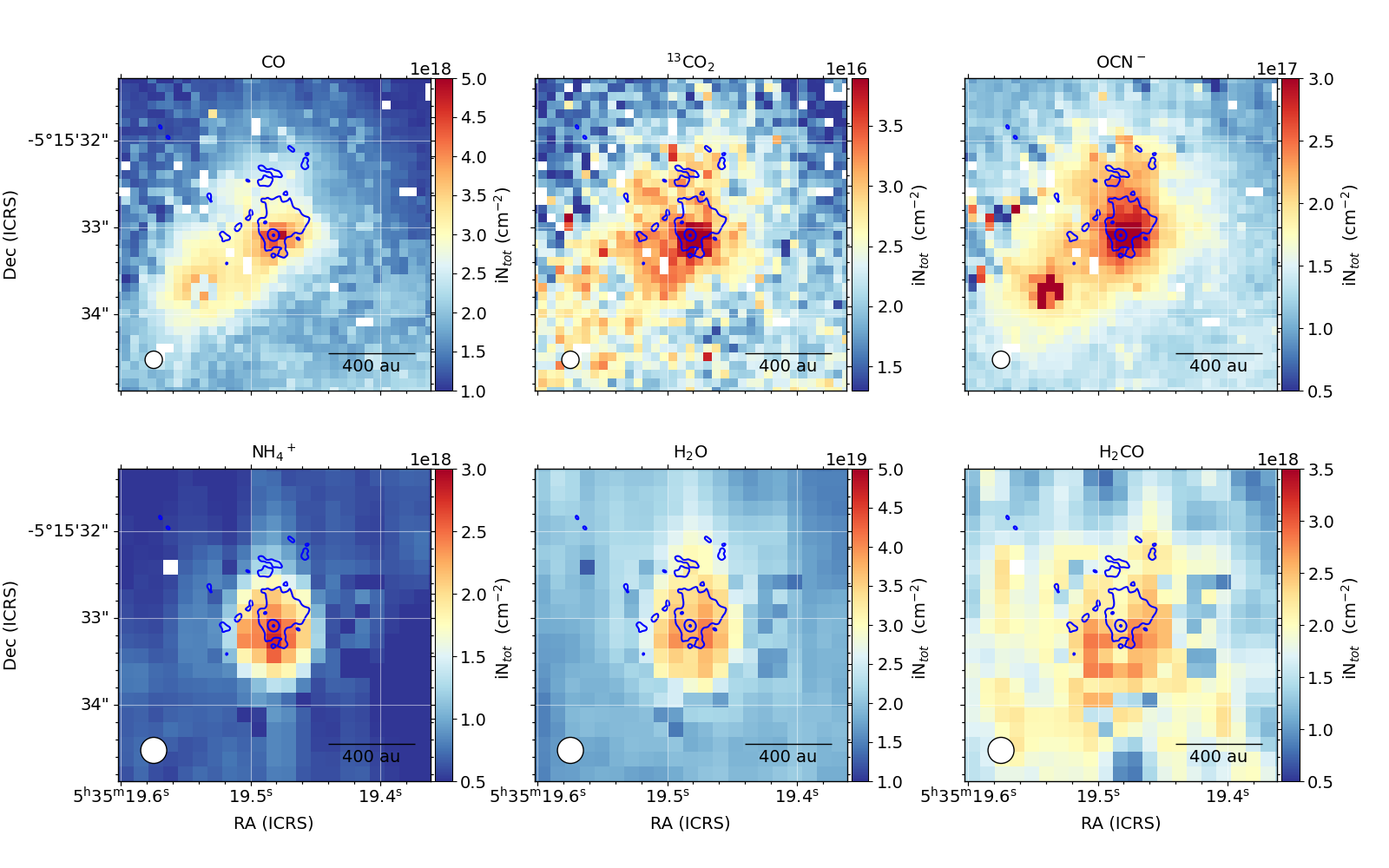}
\caption{Maps of ice column densities toward HOPS-56. The contours show the ALMA continuum emission (like in Fig.~\ref{pic:continuum}).} 
\label{pic:N_H56}
\end{figure*}

\begin{figure*}
\centering
\includegraphics[scale=0.4]{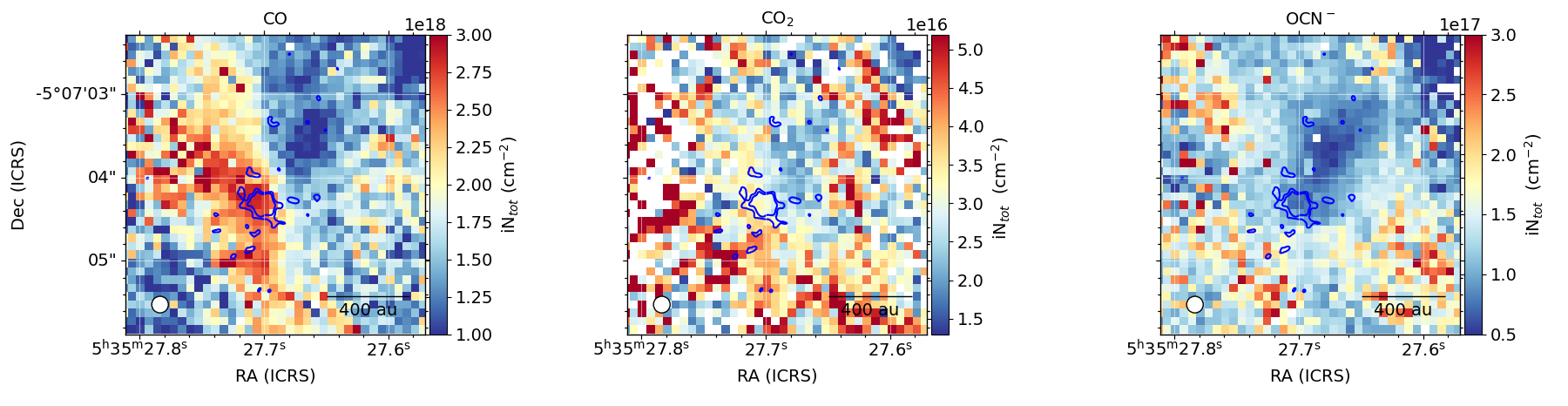}
\caption{Maps of ice column densities toward HOPS-73. The contours show the ALMA continuum emission (like in Fig.~\ref{pic:continuum}).} 
\label{pic:N_H73}
\end{figure*}

\begin{figure*}
\centering
\includegraphics[scale=0.39]{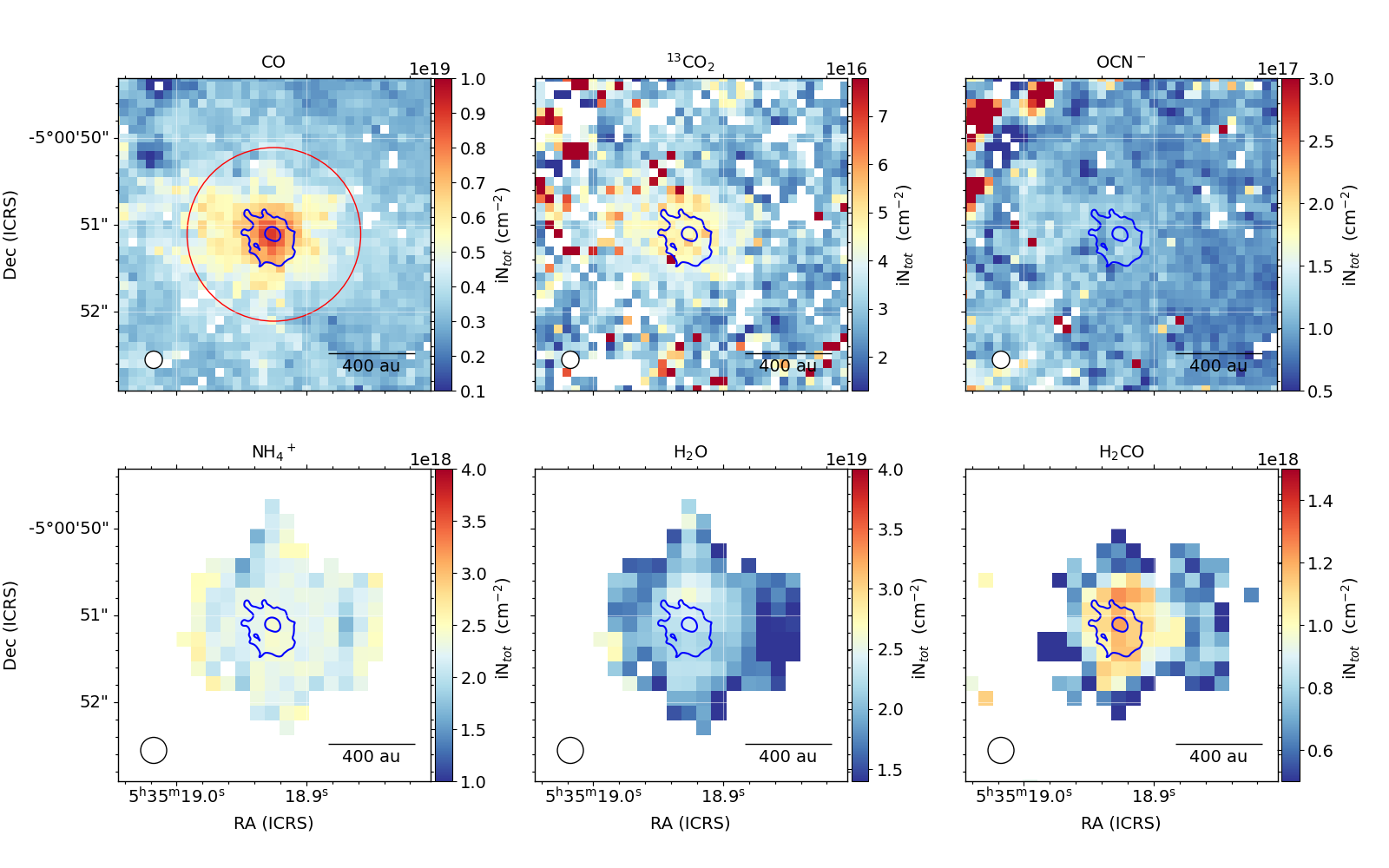}
\caption{Maps of ice column densities toward HOPS-91. The circle shows the region of the optically thin CO band. The contours show the ALMA continuum emission (like in Fig.~\ref{pic:continuum}).} 
\label{pic:N_H91}
\end{figure*}

\begin{figure*}
\centering
\includegraphics[scale=0.39]{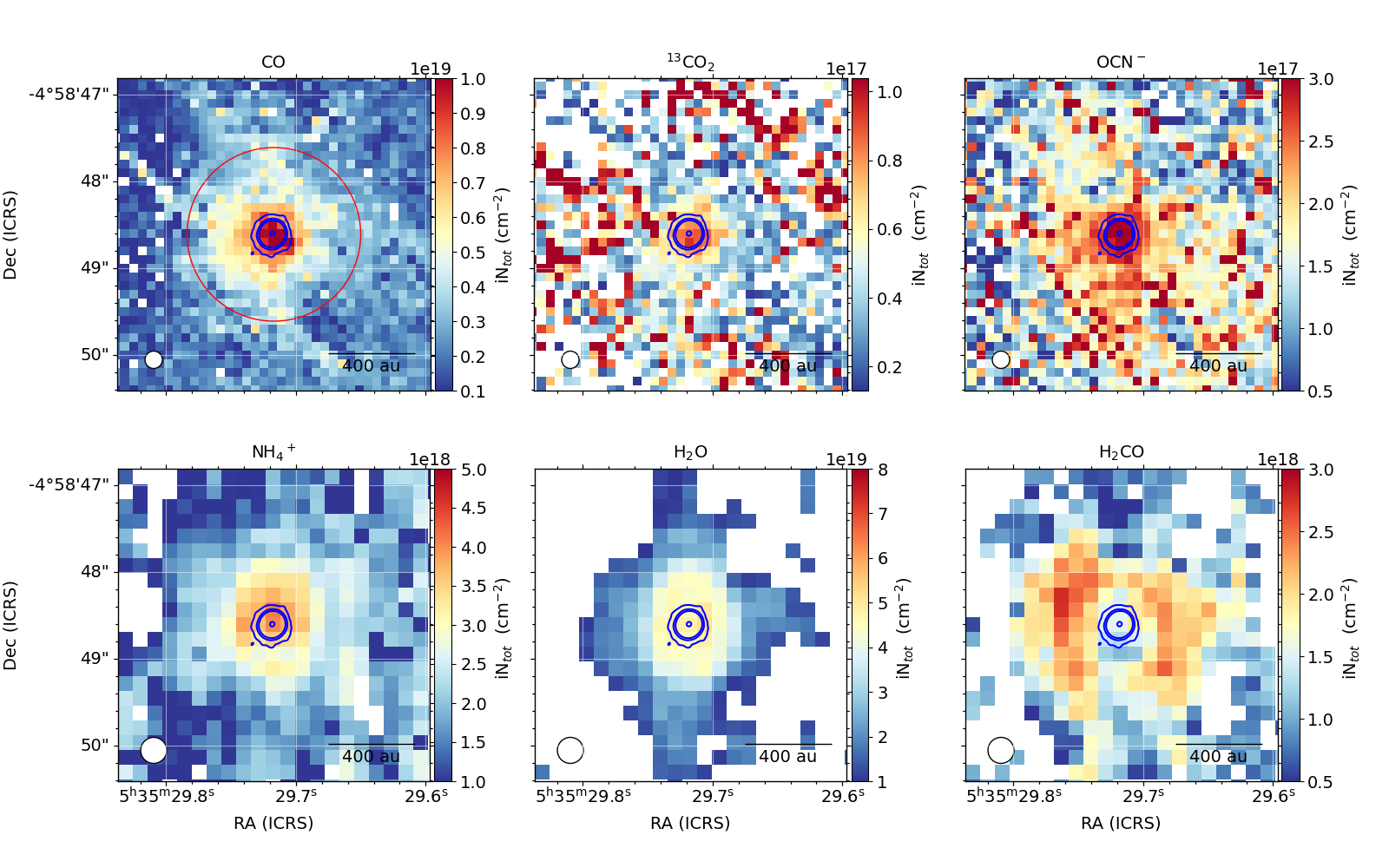}
\caption{Maps of ice column densities toward HOPS-96. The circle shows the region of the optically thin CO band. The contours show the ALMA continuum emission (like in Fig.~\ref{pic:continuum}).} 
\label{pic:N_H96}
\end{figure*}

\begin{figure*}
\centering
\includegraphics[scale=0.39]{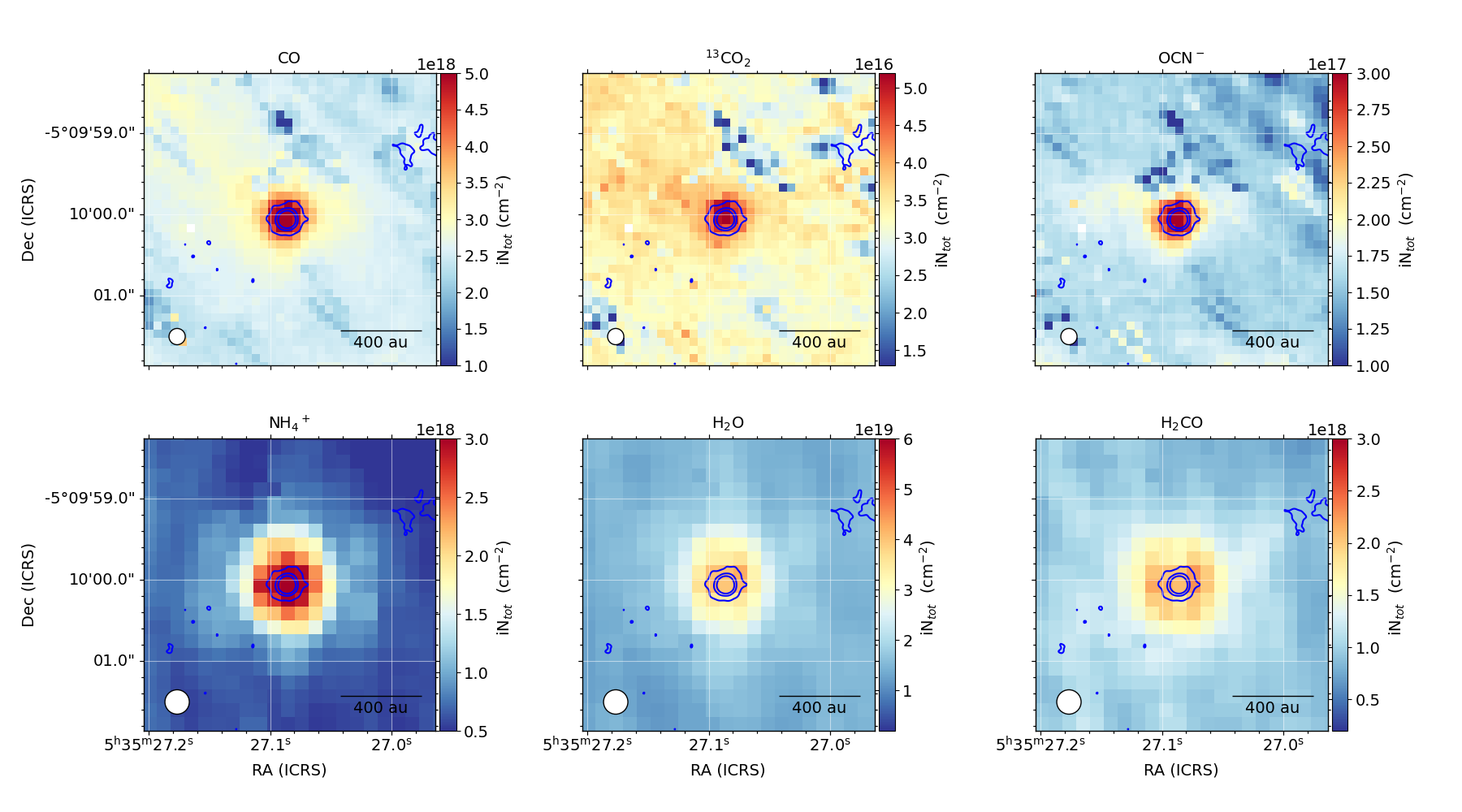}
\caption{Maps of ice column densities toward HOPS-108. The contours show the ALMA continuum emission (like in Fig.~\ref{pic:continuum}).} 
\label{pic:N_H108}
\end{figure*}

\begin{figure*}
\centering
\includegraphics[scale=0.36]{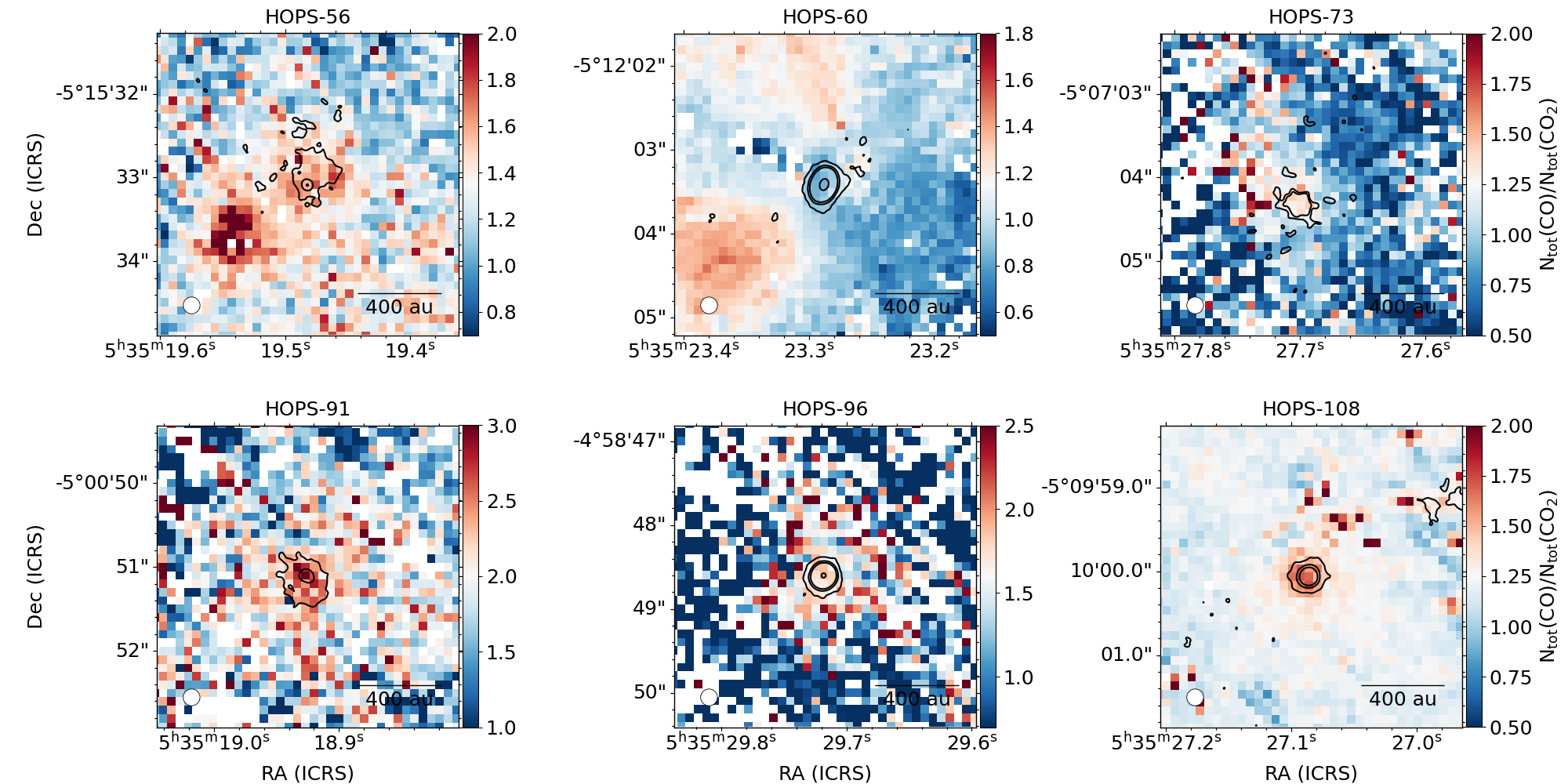}
\caption{Maps of the ratio of column densities of CO and CO$_2$ toward HOPS-56, HOPS-60, HOPS-73, HOPS-91, HOPS-96, and HOPS-108. The contours show the ALMA continuum emission (like in Fig.~\ref{pic:continuum}).} 
\label{pic:CO_CO2}
\end{figure*}

\begin{figure*}
\centering
\includegraphics[scale=0.36]{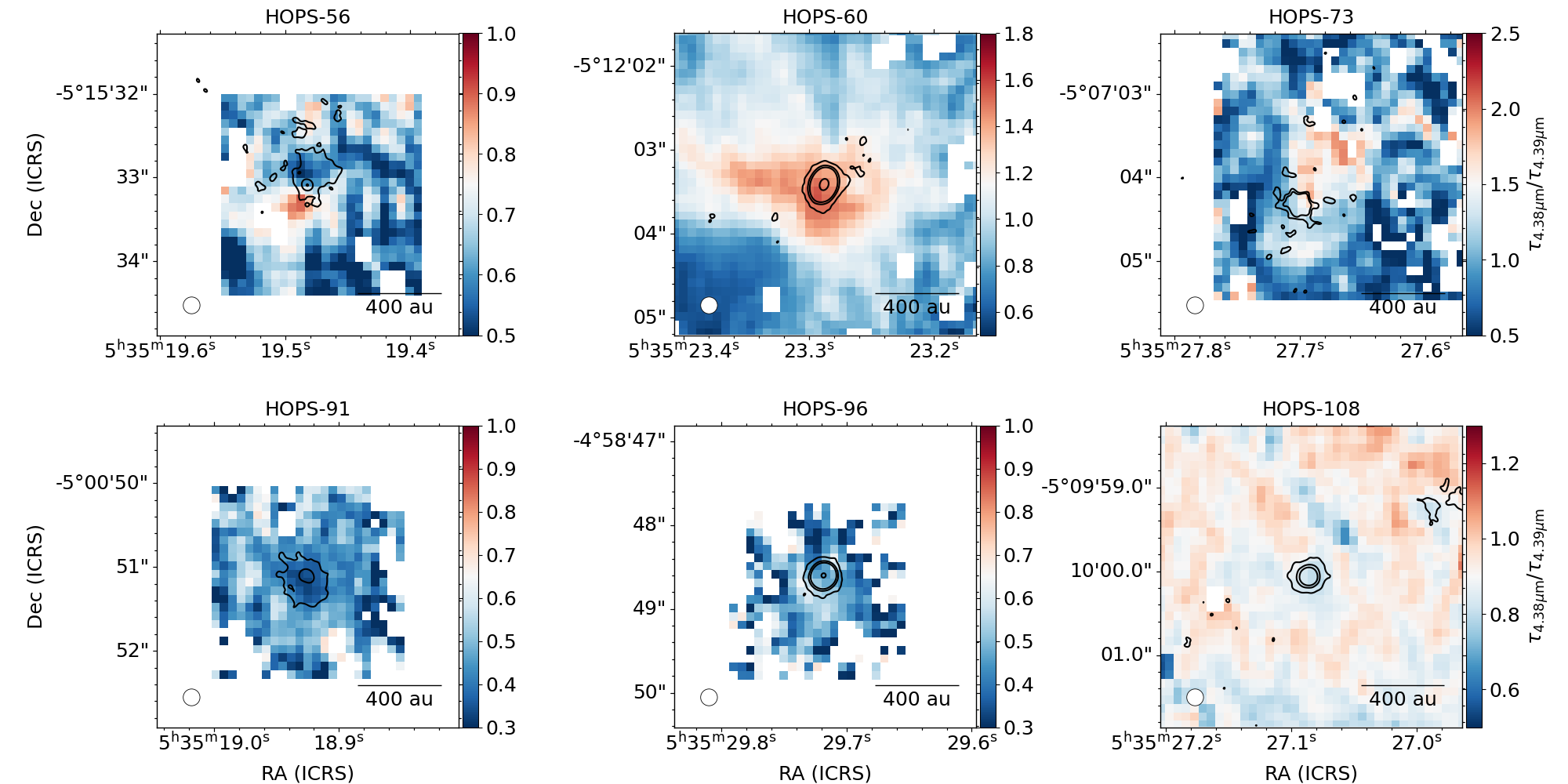}
\caption{ Maps of the ratio of optical depth of pure $^{13}$CO$_2$ (at 4.381~$\mu$m or 2283~cm$^{-1}$) and $^{13}$CO$_2$:H$_2$O (at 4.392~$\mu$m or 2276~cm$^{-1}$) toward HOPS-56, HOPS-60, HOPS-73, HOPS-91, HOPS-96, and HOPS-108. The contours show the ALMA continuum emission (like in Fig.~\ref{pic:continuum}).} 
\label{pic:CO2_clean}
\end{figure*}

\begin{sidewaysfigure}
\centering
\includegraphics[scale=0.9]{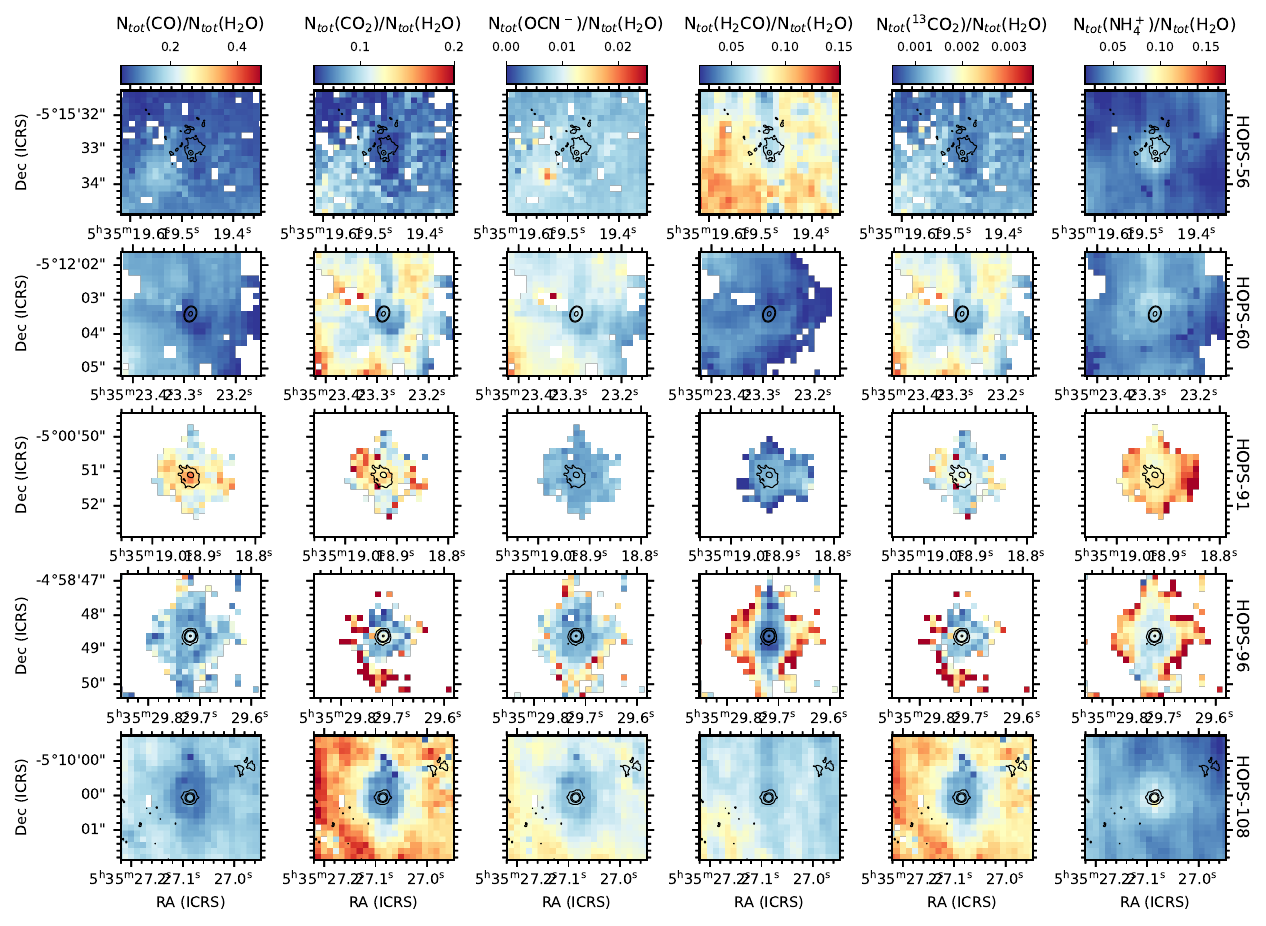}
\caption{Abundances of species with respect H$_2$O. The contours show the continuum emission at 0.87 mm according to ALMA observations. The contours show the ALMA continuum emission (like in Fig.~\ref{pic:continuum}).} 
\label{pic:R_all}
\end{sidewaysfigure}

\begin{figure*}
\centering
\includegraphics[scale=0.36]{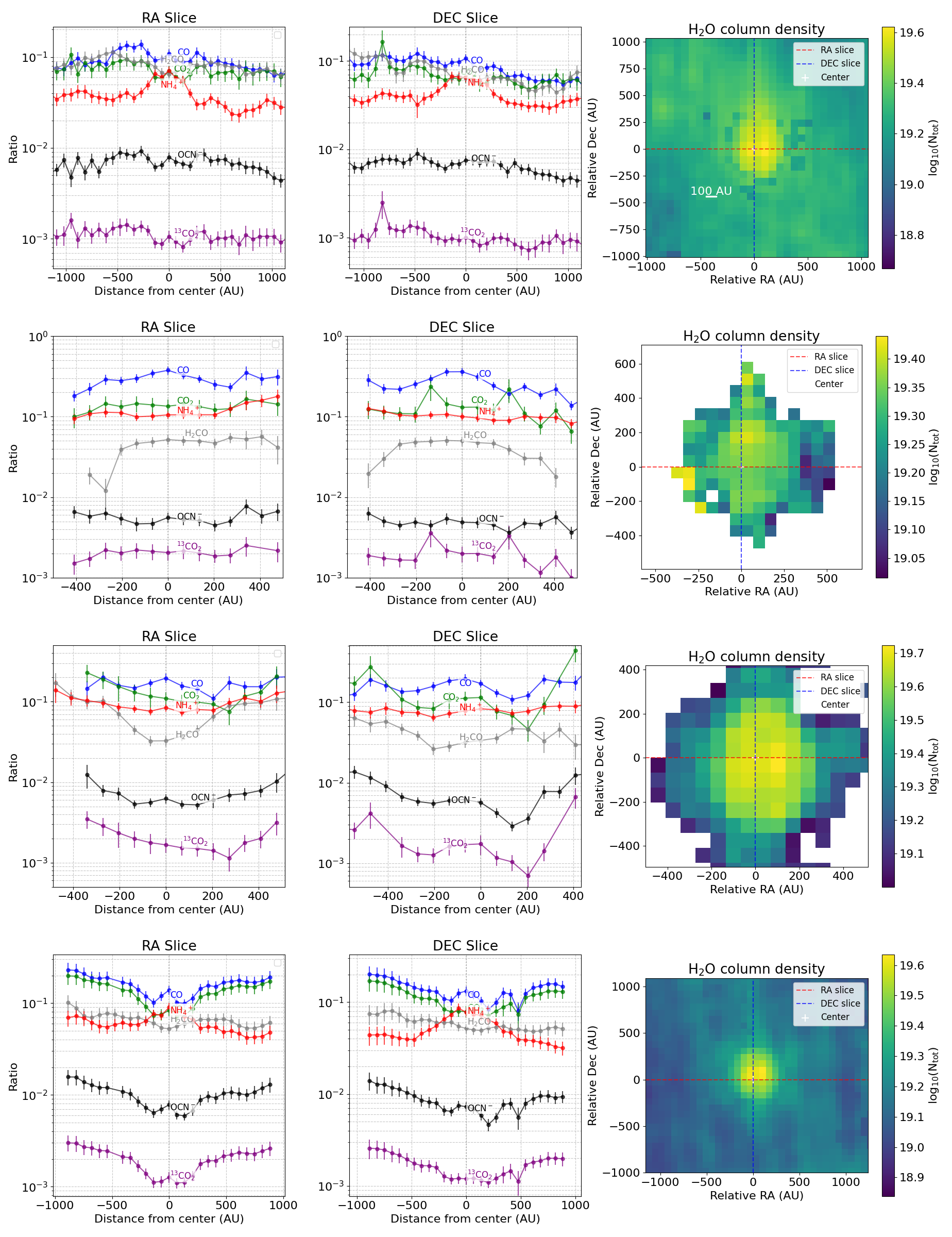}
\caption{Slices of species abundance along the RA and DEC axes toward HOPS-56 (top), HOPS-91 (middle top), HOPS-96 (middle bottom) HOPS-108 (bottom).} 
\label{pic:ab_sli}
\end{figure*}

Here we present window ranges for continuum estimation (Table \ref{Tab:window}); non-regridded column densities toward the ALMA continuum peaks (Table \ref{tab:coln_den_non}); filtered and non-filtered spectra with continuum (Fig. \ref{pic:spec_filter}); optical depth spectra toward the pixel 400 AU down a long declination axis from the ALMA continuum intensity peak with best fit (Fig. \ref{pic:spec_fit_b}); maps of ice column densities toward HOPS-56, HOPS-73, HOPS-91, HOPS-96 and HOPS-108 (Fig. \ref{pic:N_H56}, \ref{pic:N_H73}, \ref{pic:N_H91}, \ref{pic:N_H96} and \ref{pic:N_H108}); maps of the ratio of column densities of CO and CO$_2$ toward all sources (Fig. \ref{pic:CO_CO2});  maps of the ratio of optical depth of pure $^{13}$CO$_2$ and $^{13}$CO$_2$:H$_2$O (Fig. \ref{pic:CO2_clean}); maps of abundances of species with respect H$_2$O (Fig. \ref{pic:R_all}); Slices of species abundance along the RA and DEC axes toward all sources (Fig. \ref{pic:ab_sli}).

\begin{table*}
\begin{center}
\caption{Window ranges for continuum estimation.}\label{Tab:window}
\begin{tabular}{ l  c c c c c c c c  c  }\hline
Source & \multicolumn{5}{c}{windows NIRspec}  & \multicolumn{2}{c}{windows MIRI MRS}  \\ 
 & \multicolumn{5}{c}{($\mu$m)}  & \multicolumn{2}{c}{($\mu$m)}  \\ 
\hline
HOPS-56 & 4.34 - 4.37& 4.47 - 4.52& 4.75 - 4.77& 4.81 - 4.86& 4.97 - 5.15 &  5.25 - 5.46& 7.79 - 7.95\\
HOPS-60 & 4.34 - 4.37& 4.42 - 4.52& 4.75 - 4.77& 4.81 - 4.86& 4.97 - 5.13 &  5.25 - 5.46& 7.83 - 7.97\\
HOPS-73 & 4.34 - 4.37& 4.42 - 4.52& 4.75 - 4.77& 4.81 - 4.86& 4.97 - 5.10 & -- \\%5.25 - 5.46, 7.50-7.65, 7.83 - 7.97\\
HOPS-91 & 4.34 - 4.37& 4.42 - 4.55& 4.75 - 4.77& 4.81 - 4.86& 4.97 - 5.15 &  5.25 - 5.46& 7.79 - 7.95\\
HOPS-96 & 4.34 - 4.37& 4.42 - 4.52& 4.75 - 4.77& 4.80 - 4.86& 4.97 - 5.27 &  5.25 - 5.46& 7.79 - 7.95\\
HOPS-108 & 4.34 - 4.37& 4.42 - 4.52& 4.75 - 4.77& 4.81 - 4.86& 4.97 - 5.15 &  5.25 - 5.46& 7.79 - 7.95\\

\hline
\end{tabular}
\end{center}
\end{table*}

\begin{table*}
\begin{center}
\caption{Non-regridded column densities toward the peaks of intensity of the ALMA continuum.}\label{tab:coln_den_non}
\begin{tabular}{ l   c   c  c  c  c   c  c  c  c }\hline
Core & CO & $^{13}$CO$_2$ & OCN$^-$ & OCS\\ 
  & (10$^{18}$~cm$^{-2}$) & (10$^{16}$~cm$^{-2}$) & (10$^{17}$~cm$^{-2}$) & (10$^{16}$~cm$^{-2}$)\\ \hline
HOPS-56  & 4.38  & 4.18 & 3.17  & 2.55 \\ 
HOPS-60  & 2.69 & 4.78 & 3.17 & 2.44\\ 
%HOPS-73  & 2.74 & 2.39 & 0.68 &  1.12\\ 
HOPS-91  & 8.97 & 4.46 & 1.21  & - \\ 
HOPS-96  & 10.07 & 9.48 & 3.22 & - \\ 
HOPS-108 & 5.75 & 5.19 & 3.07  & 2.56    \\ 
\hline
\end{tabular}
\end{center}
\end{table*}

%\begin{figure*}
%\centering
%\includegraphics[angle=90, width=0.8\linewidth]{figure1.pdf}
%\end{figure*}

%\begin{figure*}
%\centering
%\includegraphics[scale=0.36]{piq/HOPS-56_new_last.png}
%\caption{Slices of species abundance along the RA and DEC axes toward the protostar HOPS-56.} 
%\label{pic:ab_H56}
%\end{figure*}

%\begin{figure*}
%\centering
%\includegraphics[scale=0.36]{piq/HOPS-73_all_t.png}
%\caption{Slices of species abundance in four directions of the protostar HOPS-73.} 
%\label{pic:ab_H73}
%\end{figure*}

%\begin{figure*}
%\centering
%\includegraphics[scale=0.36]{piq/HOPS-91_new_last.png}
%\caption{Slices of species abundance along the RA and DEC axes toward the protostar HOPS-91.} 
%\label{pic:ab_H91}
%\end{figure*}

%\begin{figure*}
%\centering
%\includegraphics[scale=0.36]{piq/HOPS-96_new_last.png}
%\caption{Slices of species abundance along the RA and DEC axes toward the protostar HOPS-96.} 
%\label{pic:ab_H96}
%\end{figure*}

%\begin{figure*}
%\centering
%\includegraphics[scale=0.36]{piq/HOPS-108_new_last.png}
%\caption{Slices of species abundance along the RA and DEC axes toward the protostar HOPS-108.} 
%\label{pic:ab_H108}
%\end{figure*}

\end{document}